\documentclass[11pt,a4paper]{article}
\usepackage{jheppub}
\usepackage{amsthm,mathtools,amscd,mathrsfs,booktabs,enumitem,microtype}
\newcommand{\Z}{\mathbb Z}
\newcommand{\R}{\mathbb R}
\newcommand{\Q}{\mathbb Q}
\newcommand{\C}{\mathbb C}
\newcommand{\HH}{\mathbb H}
\newcommand{\Sp}{\mathrm{Sp}}
\newcommand{\Spin}{\mathrm{Spin}}
\newcommand{\SO}{\mathrm{SO}}
\newcommand{\SU}{\mathrm{SU}}
\newcommand{\TMF}{\mathrm{TMF}}
\newcommand{\MSpin}{\mathrm{MSpin}}
\newcommand{\MString}{\mathrm{MString}}
\newcommand{\KO}{\mathrm{KO}}
\newcommand{\IZ}{I_{\Z}}
\newcommand{\Ind}{\operatorname{Ind}}
\newcommand{\ch}{\operatorname{ch}}
\newcommand{\Ad}{\operatorname{Ad}}
\newcommand{\CS}{\operatorname{CS}}
\newcommand{\tr}{\operatorname{tr}}
\newcommand{\Str}{\operatorname{Str}}
\newcommand{\LC}{\mathrm{LC}}
\newcommand{\nat}{\mathrm{nat}}
\newcommand{\eff}{\mathrm{eff}}
\newcommand{\Th}{\mathrm{Th}}

\newcommand{\hofib}{\operatorname{hofib}}
\newcommand{\cofib}{\operatorname{cofib}}

\newcommand{\Wit}{\operatorname{Wit}}
\newcommand{\Krel}{\mathsf K}
\newcommand{\tTw}{t_{\mathrm{tw}}}

\hypersetup{bookmarksdepth=3}
\newtheorem{theorem}{Theorem}[section]

\theoremstyle{definition}

\newtheorem{hypothesis}[theorem]{Hypothesis}
\theoremstyle{remark}

\makeatletter
\gdef\@fpheader{}
\makeatother

\title{Global Green--Schwarz response on quaternionic sphere bundles}
\author{Yuqi Li}
\author{and Hao-Yu Sun}
\affiliation[a]{C.N. Yang Institute for Theoretical Physics, Stony Brook University, Stony Brook, NY 11794, USA}
\affiliation[b]{Weinberg Institute, the University of Texas at Austin, 2515 Speedway, Austin, TX 78712, USA\\
SpaceXSI, USA}
\emailAdd{yuqi.li@stonybrook.edu}
\emailAdd{hkdavidsun@utexas.edu}

\abstract{
Local anomaly polynomials describe local variations, but a normalized Green--Schwarz phase also depends on global background data. We explain this dependence for ten-dimensional three-sphere bundles of quaternionic lines over closed spin seven-manifolds. The backgrounds retain two specified $\Spin(16)$ gauge bundles and an integral differential trivialization on the base, including its boundary identifications. The trivialized class is the sum of half the first Pontryagin class of the tangent bundle and the second Chern class of the quaternionic line's underlying complex bundle. For these backgrounds, the calculation combines a generally nonzero local Green--Schwarz integral with a real eta invariant evaluated by a twelve-dimensional Atiyah--Patodi--Singer filling. Together, these terms determine an integral relative characteristic number; the additive response is one third of that number, modulo integers. When the base's third and fourth rational cohomology groups vanish, the response equals eight times the Crowley--Goette invariant, without assuming that the base is simply connected. A calibration using the actual quaternionic line bundle gives additive value one third on $\Sp(2)$, recovering the known third-root phase. Distinct integral String data on one fixed ten-manifold give all three phase values, showing which global background information the response can detect. The physical identifications retain their prescribed fermion data, while the separate application to integral topological modular forms and vertex operator superalgebras requires compatible integral lifts. The response calculation constructs neither a microscopic realization nor an equivalence of complete theories.
}
\hypersetup{
  pdftitle={Global Green--Schwarz response on quaternionic sphere bundles},
  pdfauthor={Yuqi Li and Hao-Yu Sun},
  pdfsubject={Physics exposition of a normalized Green--Schwarz response}
}
\newcommand{\companionDraftNotice}{}

\begin{document}
\maketitle
\companionDraftNotice
\section{Global information in the Green--Schwarz response}
\label{sec:introduction}

Green and Schwarz showed how suitable local interactions can cancel
gauge and gravitational anomalies in ten-dimensional supergravity
coupled to Yang--Mills for the appropriate gauge groups \cite{GS84}.
Global information requires a further description. In the anomaly-line
framework, curvature measures the local anomaly, while a flat line can
retain nontrivial holonomy \cite[\S2.2]{FreedAnomalies}.
The scalar studied here compares a normalized Green--Schwarz section
with a spin reference; a nontrivial comparison does not by itself
signal a failure of anomaly cancellation.

Evaluating this normalized response on a closed background requires
global data even when the background gauge
connections on the ten-manifold are trivial. Tachikawa--Zhang identified
a nontrivial third-root response in their comparison of heterotic
Green--Schwarz terms \cite{TZ}. Quaternionic sphere bundles provide
a family of ten-manifolds on which the response can be computed and
its dependence on the background made explicit. The calculation
relates a local Green--Schwarz integral and a closed fermion eta phase
to one integral relative characteristic number. Its value can change
with the integral String datum while the underlying ten-manifold and
the fibre period remain fixed. Related worldsheet developments include the tertiary invariant of
two-dimensional minimally supersymmetric field theories
\cite{TYInvariants2025} and the study of non-supersymmetric heterotic
branes and their Green--Schwarz pairings \cite{KTYBranes2024}.
Wada--Yonekura \cite{WadaYonekura2026}, which appears simultaneously with this paper, evaluate discrete Green--Schwarz
couplings by a complementary worldsheet/tertiary-invariant method,
including an order-three response on $\Sp(2)$. The present calculation
instead uses a spacetime sphere-bundle construction and a
twelve-dimensional APS filling to evaluate the normalized response.

The mathematical derivation of the sphere-bundle formula and the
examples below are due to Li--Sun, in their unpublished working
manuscript \emph{Green--Schwarz scalars on quaternionic sphere bundles}
\cite{C1Working}. The present companion explains the background data,
calculation and physical consequences of those results for readers
familiar with anomalies but not with differential cohomology or the
Crowley--Goette invariant. Its purpose is to make the local integral,
the closed spectral contribution and their common relative pairing
visible in one calculation, while retaining the different premises of
the physical and generalized-cohomology comparisons.

For a closed ten-manifold $M$ with differential String data and
prescribed trivial gauge connections and frames, the observable is
an exponentiated scalar
$\mathcal O(M)=\exp(2\pi i\phi(M))$, with additive value
$\phi(M)\in\R/\Z$. Its normalization compares the Green--Schwarz
section to a spin reference with trivial framed gauge bundles. Let
$N$ be an eleven-dimensional filling carrying the gauge and differential
trivialization data, with three-form component $H_3$, and let
$W_{\rm ref}$ be the spin reference. The local eight-form $Y_8$ is
fixed by the full real fermion anomaly polynomial. The additive phase
combines the local integral with a closed eta term:
\begin{equation}\label{eq:introduction:normalized-response}
 \phi(M)=-a_N-\mu_*(Z),\qquad
 a_N=\int_N H_3Y_8,\qquad
 Z=\overline N\cup_M W_{\rm ref}.
\end{equation}
The orientation-reversed filling and reference form the closed
spin/gauge double $Z$. The closed term
$\mu_*$ uses a real normalization of the Atiyah--Patodi--Singer eta
invariant of the specified virtual real fermion coefficient, with
$\xi=(\eta_{\rm APS}+\dim_\C\ker)/2$ and the factor $-1/2$ multiplying $\xi$,
as well as the specified finite tangent-connection transgression.
The kernel term, reference and the two minus signs are part of the
observable. Section~\ref{sec:response-data} gives the complete
definition in \eqref{eq:response-data:observable}.

The scalar describes the specified anomaly response. Its identification
with the $\Spin(32)/\Z_2$ heterotic Green--Schwarz phase divided by the
$E_8\times E_8$ phase uses the published
fermion and Green--Schwarz prescription: the same-chirality adjoint
gauginos, identical gravitational and ghost sectors, the normalized
section and reference pairing, and no additional independent invertible
sector. Section~\ref{sec:heterotic-comparison} states this input and
compares the actual closed real Dirac systems. Equality of local
polynomials would leave precisely that global identification
undetermined. The response studied here does not specify a complete
partition function or a microscopic realization.

The geometry makes its global inputs concrete. Let $B$ be a closed
connected spin seven-manifold, with specified Spin structure, metric
and a quaternionic line $V$ with connection. Its unit sphere bundle
$M=S(V)$ has dimension ten. If $U$ is the underlying complex rank-two
bundle, write $c=c_2(U)$ and $\lambda=p_1/2$. The extra base datum is
an integral differential trivialization $\check h$, with three-form
component $h$, satisfying
\begin{equation}\label{eq:introduction:base-datum}
 d\check h=\check\lambda(TB)+\check c,
 \qquad dh=\lambda(TB)+c.
\end{equation}
The second equation is the curvature equation; the first
also retains the integral nullhomotopy and its specified morphisms. A primitive
of the four-form alone would not retain the integer information used
by the response. This datum induces the differential String datum $\check s_M$
on $S(V)$. All gauge bundles used to construct the filling lift to
the actual common cover $\Spin(16)\times\Spin(16)$.

The disk bundle $N=D(V)$ provides the induced boundary data and the
local integral. With the base-then-fibre orientation and the outward
boundary orientation used throughout, that integral is
$a_N=\frac13\int_B hc$ and is generally nonzero. Computing only the
eta term would therefore miss the dependence on $\check h$. To
evaluate the closed term in
\eqref{eq:introduction:normalized-response}, extend the actual pair
$(B,V)$ over a spin eight-manifold $W$ with a quaternionic line
$\widetilde V$. Associating the doubled four-dimensional fibre over
$W$ produces an auxiliary twelve-dimensional spin/gauge filling.
An ordinary Atiyah--Patodi--Singer calculation \cite{APS}, including
real index parity and the finite connection correction, evaluates
the closed eta phase. This auxiliary filling carries the Spin and
gauge data required by the index theorem, but its anomaly four-cocycle
is not trivialized. That distinction allows the scalar to be
nontrivial.

The specified integral boundary nullhomotopy determines a relative
class $z$, recording the bulk characteristic class and its chosen
boundary trivialization. Its cohomology group and the map to absolute
cohomology are
\[
 z\in H^4(W,B;\Z),\qquad
 j:H^4(W,B;\Z)\longrightarrow H^4(W;\Z).
\]
The map $j$ forgets the boundary trivialization. Put
$c_W=c_2(\widetilde U)$, where $\widetilde U$ underlies
$\widetilde V$, and $\lambda_W=\lambda(TW)$.
The bulk integral and retained local term then combine into
\begin{equation}\label{eq:introduction:relative-response}
 \begin{aligned}
  \phi(S(V),\check s_M)&=\frac{\Krel}{3}\pmod\Z,\\
  \Krel&=\langle c_W\smile z,[W,B]\rangle\in\Z,
  \qquad jz=\lambda_W+c_W.
 \end{aligned}
\end{equation}
Here $\smile$ is the cohomological cup product and $[W,B]$ is the
relative fundamental class. In forms, the pairing is
$\int_W c_W(\lambda_W+c_W)-\int_Bh\wedge c$; the boundary subtraction
is what turns the APS expression into this integral relative product.
Changing $W$ with all boundary data fixed changes $\Krel$ by a
multiple of $24$, and $\Krel$ is even. Thus the normalized response
is a third root of unity on this sphere-bundle domain. The relative
integer is the output of the characteristic calculation; no charge
interpretation is needed to reach this conclusion.

The relative pairing connects the response to an established secondary
invariant of the base. If
$H^3(B;\Q)=H^4(B;\Q)=0$, Crowley--Goette's invariant $t_B(V)$
is defined and their extrinsic formula has the same relative pairing
with a different normalization \cite[\S1, Proposition~1.3]{CG11}:
\begin{equation}\label{eq:introduction:cg-response}
 t_B(V)=\frac{\Krel}{24},\qquad
 \phi(S(V),\check s_M)=8t_B(V)\quad\text{in }\R/\Z.
\end{equation}
These two rational-cohomology vanishings are the exact additional
domain restriction; simple connectedness is not required. They make
the rational relative lift unique, while the integral differential
data remain present. The formula for $\phi$ in
\eqref{eq:introduction:relative-response} applies more generally and
does not define an extension of $t$ outside its domain.

For the original $\Sp(2)$ background, an actual quaternionic-line
filling yields $\Krel=-2$, hence
$\phi=1/3$ and $\mathcal O=\exp(2\pi i/3)$. Recovering the
published value requires the connection and framing transport, in
addition to this number. The general sphere construction has outward
fibre period $-2$. The right-framed period $+1$ of the original
background is obtained by the special inversion and differential
transport on $\Sp(2)$; it is not a generic reidentification of the
fibre datum.

The larger domain exposes information that neither the ten-manifold
nor its fibre period specifies. At fixed metric and gauge connection,
changing $\check h$ by a degree-three differential character with
integral characteristic class $\alpha\in H^3(B;\Z)$ changes the scalar by
$-\langle\alpha\smile c,[B]\rangle/3$. Li--Sun's construction
\cite{C1Working} gives one fixed ten-manifold with choices $\check s_n$ for which
$\mathcal O(S(V),\check s_n)=\exp(4\pi i n/3)$, giving all three
values for $n=0,1,2$. These are distinct integral String data, rather
than three vacua or a phase transition. The example lies outside the
Crowley--Goette domain. It makes the role of the larger-domain
formula tangible: the response can distinguish global background
choices that share the same underlying manifold and fibre period.

The same relative formula also has a six-dimensional reading. The
virtual complex Weyl coefficient $4U-8$, where the integers denote
bundle multiplicities, has anomaly density $c(\lambda+c)/3$.
The gauge-dependent part of Lee--Tachikawa's primitive half-doublet
density is $c(\lambda+c)/24$, and their response uses backgrounds with $dH=c$
\cite[\S3.3]{LT21}. The factor of eight relates these normalizations;
the different equations $dh=\lambda+c$ and $dH=c$ retain different
background data. The resulting divisibility condition and the integral
lattice factorization are explained within the established
Green--Schwarz quantization framework of Monnier--Moore and
Monnier--Moore--Park \cite{MM18,MMP18}, with no claim to construct a
complete six-dimensional theory.

We first specify the normalized response and its background data in
Section~\ref{sec:response-data}, then follow the disk construction
and the direct APS calculation to the relative integer. The comparison,
calibration and change law explain what that integer measures.
Later sections describe the reduced six-dimensional response, the
published heterotic prescription and the separately conditional
comparison with integral topological modular forms ($\TMF$) and
vertex operator superalgebras. The latter requires actual compatible
integral lifts and real equivariant characters, and uses the selected
filling with zero local counterterm. These additional inputs are stated
in Section~\ref{sec:tmf-comparison}; the response calculation does not
construct them. Neither that conditional application nor the heterotic
identification is needed to prove the sphere-bundle evaluation.

\section{The normalized scalar and its background data}
\label{sec:response-data}

The Green--Schwarz response studied here is a phase assigned to a
ten-dimensional spin manifold with a specified differential String
datum. This datum includes an integral trivialization of the spin
characteristic class $\lambda=p_1/2$, together with its refinement at
the chosen connection. The response combines a local
eleven-dimensional integral with a closed real fermion anomaly. The
two contributions belong to the same normalization: specifying the
anomaly polynomial fixes the local density, but does not by itself
specify the real anomaly theory, its boundary line, or the reference
against which that line is evaluated.

The connections and String trivialization are background inputs. The
eleven- and twelve-dimensional manifolds used below are auxiliary
geometric fillings for evaluating their response. Their introduction
does not supply additional spacetime degrees of freedom. The issue is
which global phase remains after a local anomaly density has been
trivialized, so the integral and spectral definitions of that phase
must be specified together.

For fermions with an anomaly, the partition function is naturally a
vector in a one-dimensional anomaly line over the background. A
trivialization of the anomaly supplies a unit vector in that line; its
Hermitian pairing with the fermion vector is a number. This is the
framework used by Tachikawa--Zhang
\cite[\S4.2, equations (67)--(73)]{TZ}. The scalar below extracts the
Green--Schwarz factor by pairing with a spin reference having trivial
gauge fields. In the general invertible-field-theory formulation, an
anomaly assigns a graded line to a ten-dimensional background and a
phase to a closed eleven-manifold \cite[\S2.3]{FreedAnomalies}.
The line used here is the ordered graded tensor product of
inverse-Pfaffian lines for the real fermion coefficient specified below.
Invertibility refers to the
anomaly sector; the fermion path integral can itself vanish in the
presence of zero modes \cite[\S2.1]{FreedAnomalies}.
The scalar is an anomaly response with specified background data,
rather than a calculation of the magnitude or other sectors of a full
partition function. The real coefficient, spectral convention and
reference pairing fixed below determine its mathematical normalization
\cite[\S2]{C1Working}. The physical identification
with the published heterotic prescription is addressed separately in
Section~\ref{sec:heterotic-comparison}.

Let $N$ be an eleven-dimensional filling of $M$ carrying the integral
and differential anomaly-trivializing data, and let $W_{\rm ref}$ be a
spin filling of the same boundary tangent geometry with trivial framed
gauges. For this scalar the boundary gauge connections on $M$ are
trivial in prescribed frames. Both fillings have product collars, and
the boundary identifications
include their Spin lifts. Their glued closed spin/gauge manifold is
$Z=\overline N\cup_M W_{\rm ref}$, where the bar reverses orientation.
With the conventions developed below, define
\begin{equation}
 \mathcal O(M)=\exp\bigl(2\pi i\phi(M)\bigr),\qquad
 \phi(M)=-a_N-\mu_*(Z),\qquad
 a_N=\int_N H_3Y_8,
 \label{eq:response-data:observable}
\end{equation}
with $\phi\in\R/\Z$. Here $H_3$ is the three-form part of the
trivialization, $Y_8$ is fixed by the full fermion coefficient, and
$\mu_*$ is the real reduced eta value at the specified connection.
The spin reference need not carry an extension of the String datum.
Keeping these two sorts of filling distinct will permit the nonzero
response computed in Section~\ref{sec:aps-evaluation}.

\subsection{The common cover and the full real coefficient}

Put $\lambda=p_1/2$, and write $T$ for the tangent bundle. The
gauge backgrounds are actual principal
$\Spin(16)\times\Spin(16)$ bundles, with associated rank-sixteen
real vector bundles $E,E'$. This is the common cover on which the
comparison is made. A background for a quotient gauge group that does
not lift to this cover is outside the present domain. In particular,
the cover matters when using the bordism calculation of
Basile--Debray--Delgado--Montero
\cite[\S3.3.3, Theorem~3.78]{BDDM}.

The virtual real coefficient and its characteristic forms are
\begin{align}
 \Delta_{\R}
  &=E\otimes_{\R}E'-S^+_{\R}(E)-S^+_{\R}(E'),
  \label{eq:response-data:coefficient}\\
 n&=\lambda(E),\qquad n'=\lambda(E'),\nonumber\\
 X_4&=\lambda(T)-n-n',\qquad
 Y_8=\frac{n^2+nn'+n'^2-p_2(E)-p_2(E')}{6}.
 \label{eq:response-data:polynomial}
\end{align}
Here $S^+_{\R}$ denotes the real positive half-spin representation,
and $\Delta_{\C}$ is the complexification of the virtual real
coefficient. The two subtracted half-spin bundles each have real
rank $128$; subtraction denotes their inverse anomaly factors.
Characteristic classes below also denote their Chern--Weil forms
when connections are fixed. The real structure fixes the
half factor in the anomaly polynomial. With $\widehat A(T)$ the
Dirac index characteristic form and $[\cdot]_{12}$ the form-degree
twelve component, it reads
\begin{equation}
 I_{12}=-\frac12\bigl[\widehat A(T)\ch(\Delta_{\C})\bigr]_{12}
       =X_4Y_8.
 \label{eq:response-data:factorization}
\end{equation}
We normalize the curvature $F$ by absorbing $1/(2\pi)$ into it,
so that $\ch=\tr e^{iF}$ and $c_2=\tr F^2/2$ for the symplectic
complex representation \cite[\S2, \S4.3]{TZ}.

The factorization can be read directly from the character expansion
before specializing the gauge bundles. Set
$S=n^2+nn'+n'^2-p_2(E)-p_2(E')$. The vector and half-spin characters
give the following components, with subscripts denoting form degree:
\[
 \ch_0(\Delta_{\C})=\ch_4(\Delta_{\C})=0,\qquad
 \ch_8(\Delta_{\C})=4S,\qquad
 \ch_{12}(\Delta_{\C})=\frac{n+n'}{3}S.
\]
Since $\widehat A_4=-\lambda(T)/12$, the degree-twelve product is
$(n+n'-\lambda(T))S/3$. Multiplication by $-1/2$ produces
\eqref{eq:response-data:factorization}, including both gauge $p_2$
terms. Thus the coefficient is a statement about the representations
and their real normalization, not merely their net rank. The character
expansion and the real parity argument are completed in
Appendix~\ref{app:real-eta-conventions}.

\subsection{Integral differential trivialization}

The topological structure records an integral trivialization of the
class represented by $X_4$. Here $BG$ denotes the classifying space
of a group $G$, and $K(\Z,4)$ is the space classifying degree-four
integral cohomology. The characteristic-class map sends the tangent
and gauge bundles to $\lambda_T-\lambda_E-\lambda_{E'}$:
\begin{equation}
 B\mathcal H=\hofib\left(
 B\Spin\times B\Spin(16)\times B\Spin(16)
 \xrightarrow{\lambda_T-\lambda_E-\lambda_{E'}}K(\Z,4)
 \right).
 \label{eq:response-data:H-structure}
\end{equation}
The homotopy fibre records the Spin and gauge bundles together with an
integral nullhomotopy of their degree-four class. A differential
$\mathcal H$-structure refines this datum using the connections and a
three-form satisfying $dH_3=X_4$. The curvature equation is a necessary
part of the background, but loses the integral information if treated
as its entire definition.

This is the global counterpart of the familiar Green--Schwarz Bianchi
identity. In a geometric description of the heterotic $B$-field, a
specified homotopy joins the gravitational and gauge characteristic
cocycles, and its differential refinement supplies the three-form
field strength \cite[\S2.2.1, (2.10)--(2.11); \S3.2, (3.23)]{SSS12}.
Here the characteristic class and common cover are fixed by
\eqref{eq:response-data:H-structure}; the following cocycle equations
make their trivializing data explicit.

For example, represent the degree-four differential cocycle by
$\check X=(c_X,h_X,X_4)$, where $c_X$ is an integral four-cochain,
$h_X$ a real three-cochain, and $\mathcal I$ integrates forms over
singular chains. Write $\delta$ for the singular-cochain coboundary
and $d$ for the exterior derivative on forms. A geometric
trivialization is a triple $(b,k,H_3)$
obeying
\begin{equation}
 \delta b=c_X,\qquad
 \mathcal I(H_3)-b-\delta k=h_X,\qquad dH_3=X_4,
 \label{eq:response-data:trivialization}
\end{equation}
with $b$ integral of degree three and $k$ real of degree two. These are
the geometric trivialization equations of Redden
\cite[Propositions~3.11--3.12]{Redden}. The middle relation ties the
form to the integral nullhomotopy, with the integral cochain $b$
included in the real cochains. Such a trivialization requires the
integral characteristic class to vanish, although the curvature
form $X_4$ need not be zero. It therefore does not identify a
nonflat differential cocycle with the zero differential character.
A real primitive of $X_4$ supplies only the last equation and
cannot replace the full triple.

The isomorphism classes of geometric trivializations form a torsor
for degree-three differential cohomology $\widehat H^3(-;\Z)$;
the underlying integral trivialization classes form a torsor
for $H^3(-;\Z)$ \cite[Theorem~3.15]{Redden}. Thus two choices have
a difference in the indicated group, but no preferred zero choice.
Specified morphisms
between differential cocycles also induce morphisms of their
trivializations, and that induced identification depends on the
chosen morphism \cite[Proposition~3.17]{Redden}. This is why the
tautological boundary frame and the characteristic morphisms used in
the disk construction must be retained. They identify the background
on which the scalar is evaluated, rather than adding an optional
description of the same curvature form.

On a boundary with trivial framed gauges the structure reduces to a
differential String trivialization of $\check\lambda(TM)$. Transport
along an oriented differential String isomorphism preserves the
scalar. An independent integral shift of the trivialization is a
different input unless an actual isomorphism identifies it with the
original one. Section~\ref{sec:string-dependence} will compute the
resulting dependence on this global datum.

\subsection{Real eta values and finite connection transport}

On a closed spin/gauge eleven-manifold the spectral convention uses
the Atiyah--Patodi--Singer (APS) eta invariant, which omits zero
eigenvalues from the spectral asymmetry. Let $D$ be the ordinary
self-adjoint Levi--Civita Dirac operator, with its twisting coefficient
indicated by the subscript:
\begin{equation}
 \xi(D)=\frac{\eta_{\rm APS}(D)+\dim_{\C}\ker D}{2},\qquad
 \mu_I(Z)=-\frac12\xi(D_{Z,\Delta_{\C}})\pmod{\Z}.
 \label{eq:response-data:eta}
\end{equation}
For the virtual coefficient, the real-valued invariants of its
summands are combined before reduction modulo $\Z$. Halving an
arbitrary element of $\R/\Z$ would leave an unresolved sign in its
exponential. The half in \eqref{eq:response-data:eta} instead comes
from the real Pfaffian theory. In dimension $11\equiv3\bmod8$,
the Dirac eigenspaces for a real twist are quaternionic; this is the
spectral structure behind the half-eta function and boundary Pfaffian
line of Freed--Moore
\cite[Proposition~1.28, \S\S3.1--3.2]{FM}.

The kernel term is part of this normalization even when a chosen
metric has zero modes. If $h_D=\dim_{\C}\ker D$, orientation reversal
gives $\xi(-D)=-\xi(D)+h_D$. Quaternionic symmetry makes $h_D$ even,
so multiplication by $-1/2$ yields
$\mu_I(-Z)=-\mu_I(Z)$ in $\R/\Z$. Omitting the kernel term would
discard the reduced invariant precisely at the point where its
spectral definition requires it.

The same real structure fixes the period in the filling calculation.
For a twelve-dimensional spin manifold $Q$ with real twists and
product boundary data, APS gives
\[
 \xi(D_{\partial Q,\Delta_{\C}})
 =\int_Q\widehat A(TQ)\ch(\Delta_{\C})
  -\Ind_{\rm APS}D^+_{\Delta_{\C}}.
\]
In dimension $12\equiv4\bmod8$, the quaternionic structures
commute with the Dirac operator and preserve the APS spectral
projections. Both chiral APS kernels consequently have even
complex dimension, and hence the virtual index is even. The
half-index therefore disappears modulo $\Z$, leaving
\begin{equation}
 \mu_I(\partial Q)=\int_Q I_{12}\pmod{\Z}.
 \label{eq:response-data:APS-period}
\end{equation}
The twisted APS index theorem supplies the underlying formula
\cite[Theorem~(4.2), formula~(4.3)]{APS}. Witten--Yonekura describe
the real antilinear symmetry and its role in the half-index
normalization \cite[equations (3.8)--(3.9)]{WY}. This statement requires a spin/gauge
filling, with no requirement that the geometric trivialization extend
over $Q$.

The disk calculation uses a specified stable metric Spin connection
$A_T$, which can differ from the Levi--Civita connection. We adopt
the finite transgression convention
\begin{equation}
 dT_Q(A_0,A_1)=Q(A_1)-Q(A_0),\qquad
 \mu_*(Z;A_T)=\mu_I(Z;A_{\LC})
       +\int_Z T_{I_{12}}(A_{\LC},A_T)\pmod{\Z},
 \label{eq:response-data:finite-transport}
\end{equation}
with fixed gauge connections. In the first equality $Q$ denotes a
characteristic polynomial. The second defines the differential-index
value at $A_T$; it does not identify that value with the raw spectrum
of a Dirac operator with torsion. Finite changes of connection thus
carry a specified correction. Using the stable value in the local
term but the uncorrected Levi--Civita value in the closed term would
evaluate two different geometric prescriptions.
The APS period likewise holds with $\mu_*$ and the characteristic
forms at that stable connection. Below the closed anomaly values use
$\mu_*$, including this correction whenever it is required.

\subsection{Real bordism maps and the normalized section}

Let $L(M)$ denote the ordered graded tensor product of the inverse
Pfaffian lines for the real coefficient
\eqref{eq:response-data:coefficient}. The tensor factors are those of
the honest real summands, with duals for the subtracted summands.
Appendix~\ref{app:real-eta-conventions:real-gluing} constructs their
unit eta states using Freed--Moore's real boundary conditions and
Witten--Yonekura's real-state gluing
\cite[\S3.2]{FM} \cite[\S\S2.5, 3.1]{WY}. It retains the spectral-cut
phases and norms, zero modes, and actual Spin and gauge seam maps.
For disconnected boundaries, orientation duality reverses the order
of the dual tensor factors. If the original order is retained instead,
the corresponding sign is
$(-1)^{\sum_{r<s}p_rp_s}$, where $p_j$ is the real Pfaffian parity
of the $j$th component. Even total parity does not remove these signs.

Write $\tau_{\Delta,*}(Y)\in L(\partial Y)$ for the resulting unit
state at the specified stable tangent connection. Its transport is
\eqref{eq:real-eta-conventions:transported-state}: the ordinary
Levi--Civita state is multiplied by
$\exp[-2\pi i\int_YT_{I_{12}}(A_{\LC},A_T)]$, with the boundary
lines identified along the prescribed connection homotopy.
On a closed manifold we denote this value by
\[
 Z_-(Z)=\tau_{\Delta,*}(Z)=\exp(-2\pi i\mu_*(Z)).
\]
The minus sign in the finite factor is required by
\eqref{eq:response-data:finite-transport}; the state and the local
integral use the same connection prescription.

For a bordism $A:M_0\to M_1$, the ordered orientation-dual
identifications place its state in
$L(M_1)\widehat\otimes L(M_0)^*$, where $\widehat\otimes$ denotes
the graded tensor product. Evaluation against the second factor
defines a linear map
\begin{equation}
 Z_-(A):L(M_0)\longrightarrow L(M_1),\qquad
 Z_-(A)x=(\mathrm{id}\widehat\otimes\mathrm{ev}_x)
                   \tau_{\Delta,*}(A).
 \label{eq:response-data:bordism-map}
\end{equation}
Here $\mathrm{ev}_x$ evaluates a covector on $x$ with the prescribed
graded order. This defines the map before any composition law is
used. For a filling $N:\varnothing\to M$, the empty line is
$\mathbb C$ with unit $1$, and $Z_-(N)(1)=\tau_{\Delta,*}(N)$.

The composition needed on the component of
$\mathcal H$-boundaries follows from the actual closed-gluing
pairing. Take differential $\mathcal H$-fillings
$N_i:\varnothing\to M_i$ and a differential $\mathcal H$-bordism
$A:M_0\to M_1$. Pair both $Z_-(A)\tau_{\Delta,*}(N_0)$ and
$\tau_{\Delta,*}(A\circ N_0)$ with the same unit cap
$\tau_{\Delta,*}(N_1)$. The real contraction of
Appendix~\ref{app:real-eta-conventions:real-gluing}, whose reference
case is \eqref{eq:real-eta-conventions:state-gluing}, gives
\begin{align}
 \bigl\langle\tau_{\Delta,*}(N_1),
                Z_-(A)\tau_{\Delta,*}(N_0)\bigr\rangle
 &=Z_-(\overline N_1\cup_{M_1}A\cup_{M_0}N_0)
 \nonumber\\
 &=\bigl\langle\tau_{\Delta,*}(N_1),
                \tau_{\Delta,*}(A\circ N_0)\bigr\rangle.
 \label{eq:response-data:capped-composition}
\end{align}
For the first closed gluing, cap $A$ by
$\overline N_0\sqcup N_1$, using reversed dual tensor order on
$\overline N_0$. For the second, cap $A\circ N_0$ by $N_1$.
Both give the displayed closed manifold with the same component
order and actual seam identifications.
The first slot of the Hermitian pairing is conjugate-linear, which
reverses $N_1$. Since the cap is nonzero and $L(M_1)$ is
one-dimensional, equal pairings imply equal vectors.
The finite counterterm is additive across matched product collars,
so the equality holds for the transported states as written, not
only for their Levi--Civita values. This proves the required
composition on these states without assuming an equivalence of
complete anomaly functors.

For a differential $\mathcal H$-filling $N$ define
\begin{equation}
 v_-(M)=\exp(2\pi i a_N)Z_-(N)(1).
 \label{eq:response-data:section}
\end{equation}
This is a unit vector in $L(M)$. Its filling independence follows
from the closed cancellation identity and the real contraction.

The cancellation of the local and spectral contributions is first
a statement on closed manifolds. On a closed differential $\mathcal H$
eleven-manifold $C$, the common-cover bordism result
$\Omega_{11}^{\mathcal H}=0$
\cite[Theorem~3.78]{BDDM} extends the integral structure over a
twelve-manifold $Q_{\mathcal H}$. The metric and gauge connections
extend with product collars. The forgetful map
$\widehat H^3(Q_{\mathcal H};\Z)\to H^3(Q_{\mathcal H};\Z)$
is surjective, and the corresponding map of trivialization torsors
is equivariant. It therefore gives a geometric refinement of the
specified bulk integral trivialization
\cite[Theorem~3.15]{Redden}. Its boundary refinement can differ from
the prescribed one only by a differential character with zero
characteristic class. Such a difference is represented by a global
two-form on $C$. Extending that form into the collar and adjusting
the bulk refinement makes the boundary agree up to the specified
isomorphism; its cochain representative is retained along the
collar \cite[\S2]{C1Working}. This extends the prescribed refinement
after the integral nullhomotopy has been extended, without requiring
surjectivity of $H^3(Q_{\mathcal H};\Z)\to H^3(C;\Z)$.
Thus \eqref{eq:response-data:factorization},
\eqref{eq:response-data:APS-period}, and Stokes give
\begin{equation}
 \mu_*(C)=\int_{Q_{\mathcal H}}X_4Y_8
         =\int_C H_3Y_8\pmod{\Z}.
 \label{eq:response-data:closed-cancellation}
\end{equation}
Here $Y_8$ is closed. No assertion that $Y_8$ itself has integral
periods is needed: the real APS formula controls the full exponential.

For two such fillings $N_0,N_1$ of the same specified boundary,
let $C=\overline N_0\cup_M N_1$, including their differential
identifications. The actual real gluing and
\eqref{eq:response-data:closed-cancellation} give
\begin{align*}
 \langle v_-(M;N_0),v_-(M;N_1)\rangle
 &=\exp\bigl(2\pi i(a_{N_1}-a_{N_0})\bigr)Z_-(C)\\
 &=\exp\left(2\pi i\int_C H_3Y_8-2\pi i\mu_*(C)\right)=1.
\end{align*}
Both vectors have unit norm, so they coincide. This proves filling
independence with the ordered real contraction, rather than by
choosing a square root of a complex determinant identity.

Concatenate $A$ with $N_0$. The local integral is additive at the
matched collars, and
\eqref{eq:response-data:capped-composition} identifies its eta state
with $Z_-(A)Z_-(N_0)(1)$. Filling independence then gives
\begin{equation}
 v_-(M_1)=\exp\left(2\pi i\int_A H_3Y_8\right)
           Z_-(A)v_-(M_0).
 \label{eq:response-data:bordism-transport}
\end{equation}
Conversely, applying this law to a filling from the empty manifold
fixes the section once $v_-(\varnothing)=1$ is fixed. Its uniqueness
is on the stated component of boundaries of $\mathcal H$-fillings.

\subsection{The framed spin reference and the scalar}

Closed cancellation controls comparisons between
$\mathcal H$-fillings; it does not force the scalar formed with a
spin reference to be one. The reference has trivial framed gauges
and matching boundary tangent geometry, but need not extend the
differential $\mathcal H$ datum. Thus
$\overline N\cup_M W_{\rm ref}$ is a closed spin/gauge manifold
to which the closed eta value applies; it need not be a closed
differential $\mathcal H$ manifold satisfying
\eqref{eq:response-data:closed-cancellation}.

For fixed framed trivial principal gauge connections, the actual
parallel virtual coefficient cancels as $256-128-128=0$ and
$Y_8=0$. Its real operators and kernels cancel with their inverse
factors, fixing the canonical even reference line and unit.
The ordered evaluation and the real Hermitian gluing
\eqref{eq:real-eta-conventions:state-gluing} therefore give
\begin{align*}
 \langle v_-(M),Z_-(W_{\rm ref})(1)\rangle
 &=\exp(-2\pi i a_N)
   \langle Z_-(N)(1),Z_-(W_{\rm ref})(1)\rangle\\
 &=\exp\bigl(-2\pi i[a_N+
            \mu_*(\overline N\cup_M W_{\rm ref})]\bigr).
\end{align*}
The local factor is conjugated by the first slot, and the same slot
reverses the filling $N$ in the closed double. This is
\eqref{eq:response-data:observable}, with the local and spectral
terms at the same specified connection. It uses the normalized
section and reference framework of
Tachikawa--Zhang \cite[equations (67), (70)--(73)]{TZ}.

Changing $W_{\rm ref}$ adds a closed framed trivial-gauge background.
Its actual virtual operator, eta phase and finite tangent
transgression cancel, so the scalar is independent of that reference
choice. On an oriented differential String cylinder with trivial
gauges, the same zero coefficient and $Y_8=0$ make
\eqref{eq:response-data:bordism-transport} agree with the canonical
reference transport. The scalar is therefore preserved by such
cylinders and by oriented differential String isomorphisms, with
their Spin and differential identifications retained.

Writing $a=a_N$ and $\mu=\mu_*(Z)$, both minus signs in
$\phi=-a-\mu$ are fixed by this pairing. Changing
only the eta sign would yield $-a+\mu$, whereas complex conjugation
changes both terms to $a+\mu$. Those operations coincide only if
$2a\in\Z$. No such condition is imposed here, and the disk local
term is generally nonzero. The sign dictionary agrees with the
negative Green--Schwarz sign in the published scalar
\cite[equations (71)--(73)]{TZ}.

For the sphere bundles in this paper, let $B$ be a closed connected
spin seven-manifold with specified Spin structure and metric. A
principal right $K=\Sp(1)$ bundle $P\to B$ with connection gives
$V=P\times_K\HH$ through the left multiplication action of $K$;
right multiplication
by $i$ defines its underlying complex rank-two bundle $U$. In the
real quaternionic orientation $(1,i,j,k)$, put $c=c_2(U)$ and
choose an integral differential trivialization $\check h$ of
$\check\lambda(TB)+\check c$, with three-form $h$. Check accents
denote the corresponding differential cocycles and trivializations.
On checked cochains, $d$ denotes the full cochain differential,
whose form component is the exterior derivative.
The resulting relations are
\[
 c=c_2(U),\qquad e(V_{\R})=\lambda(V_{\R})=-c,\qquad
 d\check h=\check\lambda(TB)+\check c,\qquad dh=\lambda(TB)+c.
\]
Orient $N=D(V)$ by the base followed by this fibre and $M=S(V)$
as its outward boundary, and write $\pi:N\to B$ for the disk
projection. The gauge bundles are
$E=2\pi^*V_{\R}\oplus\R^8$ and $E'=\R^{16}$, with their actual
Spin lifts. For the connection $A$ on the pulled-back principal bundle
chosen in the disk construction, put $C=c_2(\pi^*P,A)$. Then
$n=-2C$, $p_2(E)=6C^2$, and
\eqref{eq:response-data:polynomial} specializes to
$X_4=\lambda(A_T)+2C$, $Y_8=-C^2/3$.
Section~\ref{sec:disk-construction} constructs the full induced
boundary String datum and evaluates $a_N$ at these connections.
Section~\ref{sec:aps-evaluation} then fills the actual closed
spin/gauge double to evaluate $\mu_*$. Their combination measures
the retained integral background data in
\eqref{eq:response-data:observable}.

\section{Quaternionic sphere bundles and the local counterterm}
\label{sec:disk-construction}

The response of Section~\ref{sec:response-data} requires a filling with
more structure than a spin manifold and a gauge connection. Its
three-form must refine an integral trivialization of $X_4$, and its
boundary must carry the specified String datum and gauge frames. A
quaternionic disk bundle constructs all of these data together. It also
reduces the local contribution to an integral over the seven-dimensional
base. That contribution is generally nonzero; it is needed when the
local and spectral terms are combined in
Section~\ref{sec:aps-evaluation}. We follow Li and Sun's differential disk construction
\cite[Section~3]{C1Working}, retaining the connections and boundary
identifications that determine this integral. Its evaluation uses
exact identities after integration over the disk fibre; the
corresponding pointwise curvature identities on the total space
are not needed.

\subsection{The disk geometry and its framed gauge boundary}

Let $B$ be a closed connected spin seven-manifold with specified Spin
structure and metric. Take a principal right $K=\Sp(1)$ bundle $P$ with
connection $a$, and form the quaternionic line
\[
 V=P\times_K\HH,\qquad (pk,x)\sim(p,kx).
\]
The group acts by left multiplication. Right multiplication by $i$
gives the underlying complex rank-two bundle $U$. With quaternionic
real orientation $(1,i,j,k)$, the complex orientation is
$(1,i,j,-k)$, so our characteristic conventions are
\begin{equation}
 \label{eq:disk-construction:quaternionic-classes}
 c=c_2(U),\qquad e(V_{\R})=-c,\qquad
 \lambda(V_{\R})=-c,\qquad \lambda=\frac{p_1}{2}.
\end{equation}
This orientation difference fixes the sign of the relative gauge
number below. We use the same symbols for characteristic classes and
their Chern--Weil forms when the connection is specified; statements
about differential cocycles retain their check accents.

The base input is an integral differential trivialization $\check h$:
\begin{equation}
 \label{eq:disk-construction:base-datum}
 d\check h=\check\lambda(TB)+\check c,
 \qquad dh=\lambda(TB)+c.
\end{equation}
Here $h$ is its three-form component, and $\check c$ uses the connection
$a$. The first equation includes the integral nullhomotopy and its
geometric refinement. The second records only the curvature equation.
In particular, solving the second equation over the reals does not
supply the first. The construction applies to these specified
$(B,P,a,\check h)$ data; it does not assert that every quaternionic
line on every spin base admits such a trivialization. No
rational-cohomology vanishing or restriction on the fundamental group
is imposed in this construction.

Set $N=D(V)$ and $M=S(V)=\partial N$, of dimensions eleven and ten,
and write $\pi:N\to B$ for the bundle projection.
Orient $N$ by the base followed by the quaternionic real fibre, and
$M$ by the outward-boundary convention. The horizontal distribution
of $a$ splits the tangent bundle:
\begin{equation}
 \label{eq:disk-construction:tangent}
 TN\cong\pi^*(TB\oplus V_{\R}).
\end{equation}
The specified Spin structure on $B$ and the quaternionic
representation give its product Spin structure. The actual gauge
bundles on $N$ are
\begin{equation}
 \label{eq:disk-construction:gauges}
 E=2\pi^*V_{\R}\oplus\R^8,
 \qquad E'=\R^{16}.
\end{equation}
Writing $L_u$ for left multiplication by $u$, the representation
$u\mapsto\operatorname{diag}(L_u,L_u,I_8)$ of
$K$ in $\SO(16)$ has a unique identity-normalized lift to
$\Spin(16)$ because $K$ is simply connected. Thus these are genuine
common-cover Spin gauge bundles, rather than a formal choice of
Pontryagin classes. In integral cohomology,
\begin{equation}
 \label{eq:disk-construction:integral-X}
 \lambda(TN)-\lambda(E)-\lambda(E')
 =\pi^*(\lambda(TB)+c).
\end{equation}
Equation~\eqref{eq:disk-construction:base-datum} supplies its integral
trivialization. No special splitting of $TB$ is used.

The remaining task is to choose connections that give the required
boundary identification. Write $Q=\pi^*P$. In a local base gauge,
write a nonzero disk coordinate as $x=ru$, $|u|=1$, and choose a
smooth radial function $f$ with $f(r)=r^2g(r^2)$ near the centre and
$f=1$ throughout a boundary collar. On $Q$ put
\begin{equation}
 \label{eq:disk-construction:radial-connection}
 A=(1-f)a-f\,du\,u^{-1},\qquad C=c_2(Q,A).
\end{equation}
Although $u$ is undefined at the centre,
$r^2du\,u^{-1}=\operatorname{Im}(dx\,\bar x)$ makes the connection
smooth there. Under $p\mapsto pk$, the coordinate changes by
$u\mapsto k^{-1}u$ and the expression transforms as
$A\mapsto k^{-1}Ak+k^{-1}dk$; it therefore defines a connection on
the stated bundle.

At a unit vector $v\in V_b$, the map $q\mapsto vq$ identifies
$\HH$ with the pulled-back line. These maps give the tautological
quaternionic frame $\tau$ on $M$, extended radially to the collar.
In that frame,
\begin{equation}
 \label{eq:disk-construction:framed-collar}
 A^\tau=u^{-1}Au+u^{-1}du
 =(1-f)(u^{-1}au+u^{-1}du)=0
 \quad\hbox{on the collar}.
\end{equation}
The fixed representation lifts carry this frame to the gauge bundles
in \eqref{eq:disk-construction:gauges}. Their boundary connections are
trivial in those specified Spin frames. Flatness alone would not have
specified the frame used to identify the boundary with a reference.

\subsection{Finite transgression and the induced String datum}

Choose an invariant vertical metric
$dr^2+\ell(r)^2g_{S^3}$, where $g_{S^3}$ is the unit round sphere
metric and the radial scale $\ell(r)$ makes the metric Euclidean near
the centre and product near the boundary. Its vertical natural connection
$\nabla^v_{\nat}$ is the vertical Levi--Civita connection on vertical
arguments and the associated connection under horizontal $K$-transport.
Using the radial positive isometry on the same stabilized Spin bundle,
compare
\begin{align}
 \label{eq:disk-construction:tangent-connections}
 A_*&=\nabla^v_{\nat}\oplus\pi^*A_B^{\LC},
 & A_0&=\pi^*(A_B^{\LC}\oplus a_V),\\
 \label{eq:disk-construction:form-transgressions}
 T_q&=T_{c_2}(\pi^*a,A),
 & K_3&=T_\lambda(A_0,A_*).
\end{align}
Here $A_B^{\LC}$ is the Levi--Civita connection of the base metric,
$a_V$ is the connection induced on $V_{\R}$, and for a characteristic
polynomial $\mathcal P$ the transgression convention is
$dT_{\mathcal P}(A_0,A_1)=\mathcal P(A_1)-\mathcal P(A_0)$.
The connection $A_*$ preserves the outward normal and is product on
the collar, as required for subsequent gluing. The intermediate
$A_0$ is used to calculate the finite change in characteristic forms;
it is not the gluing connection. Their identical base summands cancel
in $K_3$. What remains is the transgression of a fixed vertical
equivariant model. This matching will also control its fibre integral.

To lift these form equations to integral differential data, fix the
quaternionic and Whitney-sum characteristic morphisms. For Spin
bundles $F_1,F_2$ they read
\begin{equation}
 \label{eq:disk-construction:characteristic-morphisms}
 \check\lambda(V_{\R})\simeq-\check c_2(U),
 \qquad
 \check\lambda(F_1\oplus F_2)\simeq
 \check\lambda(F_1)+\check\lambda(F_2).
\end{equation}
The symbol $\simeq$ denotes the chosen characteristic morphism,
whose pullbacks will identify the cocycles throughout the construction.
Their integral classes and curvature polynomials agree. At the
universal level the possible flat differences lie in degree-three
$\R/\Z$ cohomology of the relevant products of the classifying spaces
$B\Spin$ and $B\Sp(1)$, which vanishes. Choose the unit morphism on framed trivial
bundles, and retain the pullbacks and compositions in this construction.
This specifies the morphisms being used; it does not make arbitrary
morphisms on $B$ unique. In the geometric model of differential
trivializations, the induced identification depends on the actual
morphism, as Redden explains \cite[Proposition~3.17]{Redden}.

Let $\check q=\check c_2(Q,A)$,
$\check q_0=\pi^*\check c$, and
$\check\lambda_B=\check\lambda(TB,A_B^{\LC})$.
Interpolating between the connections along an interval, with the
specified endpoint morphisms, gives geometric transgression cochains
$\check T,\check K$ satisfying
\begin{equation}
 \label{eq:disk-construction:cochain-transgressions}
 d\check T=\check q-\check q_0,
 \qquad
 d\check K=\check\lambda(A_*)
             -\pi^*(\check\lambda_B-\check c).
\end{equation}
They include integral and real cochain components as well as the forms
$T_q,K_3$. These equations are not isomorphisms between differential
characters of unequal curvature. Appendix~\ref{app:differential-data}
describes the cochain information needed here and in the relative
pairing.

The full disk datum is
\begin{align}
 \label{eq:disk-construction:H}
 \check H_N&=\pi^*\check h+2\check T+\check K,
 &H_N&=\pi^*h+2T_q+K_3,\\
 \label{eq:disk-construction:H-equation}
 d\check H_N&=\check\lambda(A_*)+2\check q,
 &dH_N&=\lambda(A_*)+2C.
\end{align}
Indeed, the full cochain differential is
\begin{align}
 \label{eq:disk-construction:cochain-cancellation}
 d\check H_N
 &=\pi^*(\check\lambda_B+\check c)
   +2(\check q-\pi^*\check c)
   +\check\lambda(A_*)-\pi^*(\check\lambda_B-\check c)\nonumber\\
 &=\check\lambda(A_*)+2\check q.
\end{align}
Both base contributions cancel with their indicated coefficients.
This is the differential refinement
of \eqref{eq:disk-construction:integral-X} at the selected connections.
The finite transgressions carry the base trivialization to the
connections and boundary geometry used in the normalized response.
Although their form contributions to the local integral will vanish
below, the integral datum still retains their cochains and endpoint
morphisms.

The tautological boundary frame supplies a specified cochain morphism
$\check\gamma_\tau$ with
$d\check\gamma_\tau=\check q|_M$. Its three-form component is zero
in the normalized framed flat model. Its remaining components still
specify the identification with the trivial gauge datum. Splitting
off the outward normal by the product Spin identification gives
\begin{equation}
 \label{eq:disk-construction:induced-String}
 \check s_M=\check H_N|_M-2\check\gamma_\tau,
 \qquad d\check s_M=\check\lambda(TM).
\end{equation}
The factor two removes the two quaternionic gauge summands in their
specified frames. What remains is a geometric trivialization of the
integral tangent class $\lambda(TM)$, namely the differential String
datum on the sphere bundle. An independent integral shift of
$\check s_M$ changes this input, even when the curvature equation
has the same appearance; an actual differential isomorphism is needed
to identify two such data.

\subsection{The exact fibre integral of the local term}

We can now evaluate the local term at these connections. The full
gauge polynomial of Section~\ref{sec:response-data} uses
$n=\lambda(E)$ and $n'=\lambda(E')$:
\[
 Y_8=\frac{n^2+nn'+n'^2-p_2(E)-p_2(E')}{6}.
\]
For one quaternionic line, the total Pontryagin form is
$1-2C+C^2$. Two copies consequently give
$n=-2C$ and $p_2(E)=6C^2$, whereas the framed trivial second gauge
factor has $n'=p_2(E')=0$. Substitution yields
\begin{equation}
 \label{eq:disk-construction:XY}
 X_4=\lambda(A_*)+2C,
 \qquad Y_8=\frac{4C^2-6C^2}{6}=-\frac13C^2.
\end{equation}
The second Pontryagin term is essential to this coefficient. The
local integral is therefore
\begin{equation}
 \label{eq:disk-construction:local-before-pushforward}
 a_N=\int_N H_NY_8
 =-\frac13\int_N(\pi^*h+2T_q+K_3)C^2.
\end{equation}
Its evaluation requires identities for the selected forms, not merely
their cohomology classes: $h$ need not be closed, and the transgression
terms are part of the specified differential background.
There is a related distinction between $C$ and $\pi^*c$. They
represent the same ordinary characteristic class on the disk, but
$C$ is the curvature form of the radial connection, with vertical
components and a fixed trivialization on the boundary. A pullback
form $\pi^*c$ would have no vertical components; moreover, $c^2$ is
an eight-form on a seven-dimensional base. Substituting it for $C$
inside \eqref{eq:disk-construction:local-before-pushforward} would
erase the relative fibre information that the calculation needs.
The connection and frame specify precisely how that information enters.

The universal $K$-equivariant quaternionic disk supplies those
identities. Its radial connection is $-f\,du\,u^{-1}$, with moment
correction $(1-f)$ times the Lie algebra variable. Cartan association
with the base connection $a$ gives exactly
\eqref{eq:disk-construction:radial-connection}. In relative equivariant
cohomology, its second Chern class is $C_K=-U_{\Th}$, where
$U_{\Th}$ is the oriented Thom class. The tautological section has
positive Euler number, while \eqref{eq:disk-construction:quaternionic-classes}
gives $e=-c$. In this universal calculation $c$ denotes the class in $H^4(BK)$
whose associated Chern--Weil form is $c$ on $B$. The Thom relation
and fibre normalization are
\[
 U_{\Th}^2=-c\,U_{\Th},\qquad
 \int_{D^4}U_{\Th}=1,
 \qquad \int_{D^4}C_K^2=-c.
\]
After fibre integration the universal expression lies in the Cartan
complex of a point: the $K$-invariant polynomials on its Lie algebra,
with each polynomial variable assigned degree two. Association
replaces these variables by the curvature of $a$. The differential
of this target complex is zero, so there are no nonzero exact
polynomials; the cohomology calculation determines the integrated
polynomial itself. Cartan association commutes with oriented integration. Writing
$\pi_*$ for integration over the oriented disk fibre, we obtain the
exact form identity
\begin{equation}
 \label{eq:disk-construction:C-pushforward}
 \pi_*C^2=-c.
\end{equation}
Exactness here refers to the fibre-integrated expressions. The relative
cohomology calculation alone would not establish a corresponding
pointwise form identity on $N$; the zero differential in the universal
target is what makes the integrated identity exact.

The same target eliminates the other two contributions.
$T_q$ and $K_3$ arise from degree-three equivariant forms on the fixed
vertical model; for $K_3$, this uses the cancellation of the base
connections in \eqref{eq:disk-construction:tangent-connections}.
Multiplying by $C_K^2$ and integrating over a four-dimensional disk
leaves total degree seven. The Cartan complex of a point has no
odd-degree part, so after association
\begin{equation}
 \label{eq:disk-construction:odd-pushforwards}
 \pi_*(T_qC^2)=0,
 \qquad \pi_*(K_3C^2)=0
\end{equation}
as forms. The gauge curvature is zero throughout the framed collar;
the relative calculation has no additional boundary contribution.
The projection formula now gives
\[
 \pi_*\bigl((\pi^*h+2T_q+K_3)C^2\bigr)=-h\wedge c.
\]
Here the two degree-seven transgression pushforwards vanish, while
the pullback of $h$ multiplies $\pi_*C^2=-c$. Substitution into
\eqref{eq:disk-construction:local-before-pushforward} gives
\begin{equation}
 \label{eq:disk-construction:local-term}
 a_N=\frac13\int_Bh\wedge c.
\end{equation}
The eleven-dimensional integral has reduced to a seven-dimensional
one, with degree-three and degree-four forms. This is a local
counterterm in the specified anomaly response, entering the definition
of $\phi$ with sign $-a_N$. It is not by itself the closed response. A special
representative such as the $h=0$ datum used in the $\Sp(2)$
calibration makes it vanish; a general base datum does not.

\subsection{The outward fibre period}

The disk also fixes a useful boundary period. Restrict the chosen
tangent connection $A_*$ to a quaternionic disk fibre. Its base
summand has no fibre curvature. For the warped vertical metric, each
curvature entry is proportional to either $dr\wedge\theta_a$ or
$\theta_a\wedge\theta_b$, where the $\theta_a$ are angular coframe
forms. The self-wedges of these entries vanish, and hence
$\tr(F\wedge F)=0$. The tangent $\lambda$ form restricts to zero
on the fibre. The relative gauge number in the tautological frame is
$\int_{D^4}C=-1$. Stokes' theorem applied to
\eqref{eq:disk-construction:H-equation} therefore gives
\begin{equation}
 \label{eq:disk-construction:outward-period}
 \int_{\partial D^4}H_N
 =\int_{D^4}\bigl(\lambda(A_*)+2C\bigr)=-2.
\end{equation}
Here the fibre sphere is oriented as the boundary of the quaternionic
disk. This is a real period at the specified connection. The original
right-framed positive-unit datum on $\Sp(2)$ requires its separate
inversion and differential transport; its period $+1$ is not the
generic period of this construction.

Section~\ref{sec:string-dependence} will use the distinction between
fibre period and full String datum when changing backgrounds. For the
immediate calculation, the disk has now provided the actual boundary
geometry and the local contribution
\eqref{eq:disk-construction:local-term}. What remains is the spectral
term of the closed manifold $\overline N\cup_M W_{\rm ref}$, formed
by gluing the orientation-reversed disk to the reference filling of
Section~\ref{sec:response-data}. The next section computes it with an
auxiliary twelve-dimensional spin and gauge filling. That filling
serves a different purpose from the differential $\mathcal H$-filling
constructed here.

\section{The closed eta phase from a twelve-dimensional filling}
\label{sec:aps-evaluation}

The disk calculation in Section~\ref{sec:disk-construction} determines
the local contribution $a_N$, but the normalized response also contains
the closed eta invariant of $Z=\overline N\cup_M W_{\rm ref}$.
The gauge connection is trivial in the specified frame on the common
boundary $M=S(V)$; it is not trivial throughout the closed manifold.
Its extension across the two halves carries information that the local
boundary curvature cannot record. We evaluate this closed term by
constructing a compact spin twelve-manifold with boundary $Z$ and
extending its actual gauge bundles and connections. Ordinary
Atiyah--Patodi--Singer (APS) theory then relates the eta invariant to a
bulk integral, with the real parity fixing the factor of one half.

The filling is auxiliary geometry for this calculation. Its existence
does not assert an additional spacetime or an extension of the
differential String trivialization. This distinction allows the closed
term to be nonzero. The following construction realizes the mixed
double as an associated four-sphere bundle and extends it over a spin
filling of $(B,V)$ \cite[\S4]{C1Working}. The resulting APS integral
requires no shrinking-fibre limit.

\subsection{Doubling the disk with its framed reference}

Choose $W_{\rm ref}$ to be a second spin disk bundle with the same
product tangent geometry on $M$, equipped with trivial principal gauge
bundles and their constant Spin frames. On the other half, the
tautological frame of the pulled-back quaternionic line supplies the
boundary identification. Gluing these specified frames produces
\begin{equation}
 Z=\overline N\cup_M W_{\rm ref}
   =P\times_K F,
 \qquad F=S(\HH\oplus\R),\qquad K=\Sp(1).
 \label{eq:aps-evaluation:double}
\end{equation}
The term ``mixed double'' refers to the different gauge data on the two
halves. Although their underlying disk geometry agrees, one half has
the quaternionic gauge bundle and the other has the framed trivial
reference. An ordinary double with identical gauges would compute a
different closed background.

Orient $F$ as the outward sphere of $\HH\oplus\R$, with $\HH$
listed first. In the two disk charts, the lower chart has orientation
$-\HH$ and the upper chart has orientation $+\HH$. The gauge line
on the lower chart transforms by left multiplication by $K$; on the
upper chart its chosen coefficient frame has trivial $K$-action. The
transition on the equator is the unit quaternion $u$. Denote the resulting
$K$-equivariant quaternionic line by $L_F$. Its fibre number is
\begin{equation}
 \int_F c_2(L_F)=1.
 \label{eq:aps-evaluation:fibre-number}
\end{equation}
Indeed, the relative disk number in the tautological frame was $-1$.
That disk appears with reversed orientation in $\overline N$, giving
$+1$ after gluing to the trivial upper half. This is the orientation
needed in the subsequent pushforwards.

The coefficient is the actual equivariant tautological line on
$\HH P^1$, rather than a line chosen solely to have this characteristic
number. The projective charts exhibit the identification. The action
of $\operatorname{diag}(u,1)$ on $[1:z]$ sends $z$ to $zu^{-1}$
and acts by $u$ on the coefficient frame. With the coordinate
$x=\bar z$, the base action becomes $x\mapsto ux$. On $[w:1]$,
the action is $w\mapsto uw$, while the coefficient frame is invariant.
These are precisely the lower and upper chart actions of the glued
line. Retaining them fixes how the fibre construction will associate
to the bundle $P$ on $B$.

The disk connection also glues as a connection, including the collar.
In a local base gauge its form is
$A=(1-f)a-f\,du\,u^{-1}$, where $f=1$ throughout a boundary
collar. In the tautological frame,
$A^\tau=(1-f)(u^{-1}au+u^{-1}du)$, so it vanishes on that collar.
It therefore extends by zero in the specified upper frame, smoothly
across the equator. No change of gauge connection at the seam is
hidden in \eqref{eq:aps-evaluation:double}.

The Spin identification is equally concrete. The equivariant stable
tangent isomorphism
\begin{equation}
 TF\oplus\R\cong F\times(\HH\oplus\R)
 \label{eq:aps-evaluation:vertical-spin}
\end{equation}
uses the action $u\mapsto\operatorname{diag}(L_u,1)$ on the
right-hand side, where $L_u$ denotes left multiplication by $u$ on
$\HH$. Since $K$ is simply connected, this representation
lifts to $\Spin(5)$. Removing the outward normal recovers the Spin
lifts used on the disk charts. Combining the vertical lift with the
specified Spin structure of $B$ gives the Spin structure already
present on the mixed double. Thus the associated description fixes
the manifold, the twisting bundle and the Spin data of the operator
whose eta invariant enters the response.

\subsection{A spin and gauge filling of the base pair}

To fill $Z$, first fill the pair $(B,V)$ by an eight-dimensional
spin manifold with an actual quaternionic line. The relevant bordism
group is ordinary spin bordism with an $\SU(2)$ bundle. Write
$B\SU(2)$ for the classifying space for that bundle and
$\Omega_q^{\Spin}$ for the bordism group of closed spin $q$-manifolds.
The degree-seven vanishing follows directly from the
Atiyah--Hirzebruch spectral sequence, whose second page is
\begin{equation}
 E^2_{p,q}=H_p(B\SU(2);\Omega_q^{\Spin}),
 \qquad B\SU(2)=\HH P^\infty.
 \label{eq:aps-evaluation:bordism-page}
\end{equation}
The cells of $\HH P^\infty$ occur only in dimensions
divisible by four. Its cellular chain complex therefore has no adjacent
nonzero groups, even with the coefficient group $\Omega_q^{\Spin}$.
In total degree seven, the only possibly nonzero positions
are $(p,q)=(0,7)$ and $(4,3)$, but
$\Omega_7^{\Spin}=\Omega_3^{\Spin}=0$. Positions with negative
$q$ do not contribute. The entire degree-seven diagonal therefore
vanishes already on the second page, giving
\begin{equation}
 \Omega_7^{\Spin}(B\SU(2))=0.
 \label{eq:aps-evaluation:bordism-zero}
\end{equation}
This is the ordinary spin/$\SU(2)$ result discussed by
Davighi--Lohitsiri \cite[\S4.1, Figure~4]{DL21}. It extends the
classifying map of the actual bundle, not merely its Chern class.
Neither this argument nor its application imposes a condition on
$\pi_1(B)$.

Choose such a compact spin eight-manifold $W$ with $\partial W=B$
and quaternionic line $\widetilde V$ restricting to $V$. Fix the
boundary Spin and bundle isomorphisms. Extend the metric and connection
with product collars, and write
\begin{equation}
 c_W=c_2(\widetilde U),\qquad \lambda_W=\lambda(TW),
 \qquad \lambda=p_1/2.
 \label{eq:aps-evaluation:base-forms}
\end{equation}
Here $\widetilde U$ is the underlying complex rank-two bundle of
$\widetilde V$, with the same right-complex convention as $U$ on $B$.
These symbols denote the characteristic forms when used in integrals;
their integral classes have the same normalization. The boundary
trivialization $\check h$ of $\check\lambda(TB)+\check c$
is still fixed. We do not demand that it extend over $W$. The
bordism result concerns spin and gauge data, whereas extending that
trivialization would be a stronger requirement. Its boundary form $h$
will remain in the answer.

Let $\widetilde P$ be the principal $K$ bundle of $\widetilde V$.
Associate the same equivariant fibre and coefficient as above:
\begin{equation}
 \begin{aligned}
 \mathcal Q&=\widetilde P\times_K F\xrightarrow{p}W,
 &\mathcal L&=\widetilde P\times_K L_F,\\
 \mathcal E&=2\mathcal L_{\R}\oplus\R^8,
 &\mathcal E'&=\R^{16}.
 \end{aligned}
 \label{eq:aps-evaluation:twelve-filling}
\end{equation}
The vertical Spin lift combines with that on $W$. The homomorphism
$K\to\SO(16)$ acting on the two real quaternionic summands has
the specified lift to $\Spin(16)$, so these are actual principal
gauge bundles on the $\Spin(16)\times\Spin(16)$ cover. On the base
collar every datum is pulled back from $B$. Consequently
$\partial\mathcal Q=Z$ with the required Spin, gauge and connection
identifications, rather than only with isomorphic characteristic
classes. Orient $\mathcal Q$ by $W$ followed by $F$. Its boundary
is then oriented by $B$ followed by $F$; the even fibre dimension
introduces no additional sign.

For the characteristic-form calculation, use the invariant fibre
metric, its natural vertical connection, and the split tangent
connection
\begin{equation}
 A_{\mathcal Q,*}=\nabla^v_{\nat}\oplus p^*A_W^{\LC}.
 \label{eq:aps-evaluation:split-connection}
\end{equation}
It restricts to the specified stable tangent connection on $Z$.
The analytic APS operator will first be taken with a product-collar
Levi--Civita connection; we account below for the finite difference
between that connection and \eqref{eq:aps-evaluation:split-connection}.

\subsection{Exact fibre integration of the anomaly polynomial}

Set $C=c_2(\mathcal L)$ and $\ell=\lambda(T^v\mathcal Q)$,
where $T^v\mathcal Q$ is the tangent bundle along the fibres of $p$.
The exact integrations needed for the density can be determined on
the universal $K$-equivariant four-sphere. Write $C_K$ and $\ell_K$
for its equivariant gauge and vertical tangent characteristic classes,
respectively, and $c$ for the universal base class. Then
\begin{equation}
 C_K^2=cC_K,\qquad \ell_K=-c,
 \qquad \int_F C_K=1
 \label{eq:aps-evaluation:equivariant-classes}
\end{equation}
in equivariant cohomology. The first relation retains the chart
orientation: the lower chart has orientation $-\HH$ and Euler
class $+c$. Its compactly supported Chern class is the positive
Thom class, normalized to integrate to one along the fibre; its square
is $c$ times that class. The gauge class
restricts to $c$ at the lower pole and to zero at the upper pole.
The second relation follows from the stable tangent identification
\eqref{eq:aps-evaluation:vertical-spin} and
$\lambda(V_{\R})=-c$.

These cohomology relations alone would not justify pointwise
substitution in an index density. What makes the present calculation
exact at the level of forms is the target of the fibre integration.
The equivariant integrals lie in the Cartan complex of a point:
the invariant polynomials on the Lie algebra of $K$, with polynomial
degree counted twice in the total degree. Its differential is zero.
Thus the classes of the integrated expressions determine the invariant
polynomials themselves, with no exact-form ambiguity in this target.
The Cartan association map substitutes the curvature of the extended
connection for the Lie algebra variable and commutes with fibre integration.
Associating this universal model with the extended connection on
$\widetilde P$ gives exact integrated forms. With $p_*$ denoting
oriented integration over the four-dimensional fibre, they are
\begin{equation}
 p_*C=1,\qquad p_*C^2=c_W,\qquad
 p_*C^3=c_W^2,\qquad p_*(\ell C^2)=-c_W^2.
 \label{eq:aps-evaluation:pushforwards}
\end{equation}
The first identity fixes the normalization, the next two integrate the
gauge terms, and the last keeps the vertical tangent contribution.
They are equalities of forms at the associated connections. We have
not asserted the pointwise curvature equalities
$C^2=p^*c_W C$ or $\ell=-p^*c_W$ on $\mathcal Q$.

For the actual gauge bundles in
\eqref{eq:aps-evaluation:twelve-filling}, the same full anomaly
polynomial used on the disk gives
\begin{equation}
 I_{12,*}=-\frac13(p^*\lambda_W+\ell+2C)C^2.
 \label{eq:aps-evaluation:full-density}
\end{equation}
The tangent term splits as $p^*\lambda_W+\ell$. For one real
quaternionic line, $p_1=-2C$ and $p_2=C^2$. The Whitney product
for its two copies is $(1-2C+C^2)^2$, so
$\lambda(\mathcal E)=-2C$ and $p_2(\mathcal E)=6C^2$;
the second gauge is trivial.
Hence $X_4=p^*\lambda_W+\ell+2C$ and $Y_8=-C^2/3$,
with the gauge $p_2$ contribution included. The projection formula
and \eqref{eq:aps-evaluation:pushforwards} now yield
\begin{align}
 p_*I_{12,*}
 &=-\frac13\bigl(\lambda_Wc_W-c_W^2+2c_W^2\bigr)
 \nonumber\\
 &=-\frac13c_W(\lambda_W+c_W).
 \label{eq:aps-evaluation:integrated-density}
\end{align}
The vertical tangent term cancels one of the two gauge contributions
to $c_W^2$. A remaining gauge term combines with $\lambda_Wc_W$;
the density does not vanish. The derivation is polynomial, so it
uses no division by $c_W$ and remains applicable when that form
vanishes or its cohomology class is torsion.

This also displays why $\mathcal Q$ is a spin/gauge filling rather
than an $\mathcal H$-filling. On a fibre, the pullback base term
has zero integral, and the vertical class $\ell_K=-c$ contributes
zero after restriction to the ordinary fibre. Therefore
\begin{equation}
 \int_F X_4=2\int_F C=2.
 \label{eq:aps-evaluation:not-h-filling}
\end{equation}
A global trivialization of $X_4$ cannot exist there. In particular,
one cannot replace the bulk polynomial by $d(H_3Y_8)$ on
$\mathcal Q$ and invoke closed differential $\mathcal H$
cancellation. Ordinary APS needs the Spin structure and twisting
bundles that we have constructed, and does not require that stronger
trivialization.

\subsection{Real parity and the finite tangent correction}

Use the real virtual coefficient
\begin{equation}
 \Delta_{\R}=\mathcal E\otimes_{\R}\mathcal E'
       -S^+_{\R}(\mathcal E)-S^+_{\R}(\mathcal E'),
 \qquad \Delta_{\C}=\Delta_{\R}\otimes_{\R}\C.
 \label{eq:aps-evaluation:real-coefficient}
\end{equation}
The half-spin representations are real. With the ordinary
Levi--Civita Dirac operator on $Z$, combine the eta representatives
of these actual bundles before reduction, and use
\begin{equation}
 \xi(D)=\frac{\eta_{\mathrm{APS}}(D)+\dim_{\C}\ker D}{2},
 \qquad \mu_{\LC}(Z)=-\frac12\xi(D_{Z,\Delta_{\C}})
       \pmod\Z.
 \label{eq:aps-evaluation:real-eta}
\end{equation}
The kernel term is part of the reduced invariant. No invertibility
assumption is needed, and the real half is taken on the representatives
in this formula.

Apply the twisted APS index theorem on the compact manifold
$\mathcal Q$ with a product-collar Levi--Civita metric and the
extended gauge connection \cite[Theorem~\textup{(4.2)},
formula~\textup{(4.3)}]{APS}. It gives
\begin{equation}
 \xi(D_{Z,\Delta_{\C}})
 =\int_{\mathcal Q}[\widehat A(T\mathcal Q)
             \ch(\Delta_{\C})]_{12}
       -\Ind_{\mathrm{APS}}D^+_{\Delta_{\C}}.
 \label{eq:aps-evaluation:complex-aps}
\end{equation}
Because $12\equiv4\pmod8$, the chiral Spin representations
with real twists have parallel quaternionic structures. They are
compatible with the Dirac operator and the APS projections,
including the zero eigenspaces. They preserve the kernel of the
chiral APS problem and the kernel of its adjoint, whose boundary
condition is complementary. Both kernels therefore have even
complex dimension, so their difference, the complex APS index, is even.
The statement persists under subtraction for the real virtual
coefficient. This is the parity that removes the possible half-index
ambiguity; the corresponding antilinear symmetry and APS
normalization are described in \cite[(2.47)--(2.52),
(3.8)--(3.9)]{WY} and in Appendix~\ref{app:real-eta-conventions}.

Multiplying \eqref{eq:aps-evaluation:complex-aps} by $-1/2$,
with $I_{12}=-[\widehat A\ch(\Delta_{\C})]_{12}/2$, gives
\begin{equation}
 \mu_{\LC}(Z)
 =\int_{\mathcal Q}I_{12,\LC}
      +\frac12\Ind_{\mathrm{APS}}D^+_{\Delta_{\C}}
 =\int_{\mathcal Q}I_{12,\LC}\pmod\Z.
 \label{eq:aps-evaluation:real-aps}
\end{equation}
Without the evenness of the actual index, this congruence would not
follow from complex APS with this normalization. In particular, an
arbitrary square root of a complex exponentiated identity cannot
replace the real-coefficient argument.

The response uses $\mu_*$ at the stable connection
\eqref{eq:aps-evaluation:split-connection}. Keep the gauge connection
fixed, and let $T_I(A_{\LC},A_{\mathcal Q,*})$ be the tangent
Chern--Simons transgression with
\begin{equation}
 dT_I(A_{\LC},A_{\mathcal Q,*})
       =I_{12,*}-I_{12,\LC}.
 \label{eq:aps-evaluation:transgression}
\end{equation}
Stokes' theorem changes the bulk integral by the boundary integral
of this form. By the differential-index convention of
Section~\ref{sec:response-data}, that same finite boundary integral
is added to $\mu_{\LC}$ to define $\mu_*$. The two changes
therefore agree:
\begin{align}
 \mu_*(Z)
 &=\mu_{\LC}(Z)+\int_ZT_I(A_{\LC},A_{\mathcal Q,*})
 \nonumber\\
 &=\int_{\mathcal Q}I_{12,*}
 =-\frac13\int_Wc_W(\lambda_W+c_W)\pmod\Z.
 \label{eq:aps-evaluation:closed-term}
\end{align}
All spectral operators in this derivation are ordinary Levi--Civita
operators. The stable connection enters through its finite
transgression, not through a proposed spectrum for a torsionful
operator. Thus the form calculation and the analytic boundary value
refer to the same specified data.

\subsection{Combining the disk and closed contributions}

The disk result was $a_N=\int_Bhc/3$. With the normalized scalar
$\phi=-a_N-\mu_*(Z)$, equation
\eqref{eq:aps-evaluation:closed-term} gives
\begin{equation}
 \phi(S(V),\check s_M)
 =\frac13\left(
       \int_Wc_W(\lambda_W+c_W)-\int_Bhc\right)
       \pmod\Z.
 \label{eq:aps-evaluation:response}
\end{equation}
The positive bulk term comes from the minus sign on the closed eta
term; the boundary correction retains the negative sign on the disk
term. Dropping $a_N$ would change this response for general boundary
data. Neither the polynomial cancellation above nor the existence
of a spin/gauge filling sets $\int_Bhc$ to zero.

The evaluation \eqref{eq:aps-evaluation:response}
\cite[\S4]{C1Working} applies to the sphere-bundle background already
specified: a closed spin seven-manifold with the actual quaternionic
line and the integral
differential trivialization of $\lambda+c$. No rational-cohomology
vanishing was used. The existence of the actual filling is essential:
it permits APS to evaluate the closed eta phase, including information
that a match of local polynomials alone would not fix. The separate
family calculation in Appendix~\ref{app:family-reduction} provides
another route; it is not an assumption of this evaluation.

The parentheses in \eqref{eq:aps-evaluation:response} have the form
of a relative integral: the bulk eight-form has boundary primitive
$hc$ because $dh=\lambda+c$ and $dc=0$. Its integrality, however,
requires the retained integral nullhomotopy in $\check h$, not just
that curvature equation. This nullhomotopy supplies a relative class
$z\in H^4(W,B;\Z)$ whose absolute image is $\lambda(TW)+c_2(\widetilde U)$.
Section~\ref{sec:relative-invariant} identifies the integral combination
with $\Krel=\langle c_W\smile z,[W,B]\rangle$, the pairing with the
relative fundamental class of the oriented filling. That identification converts
the geometric evaluation into $\phi=\Krel/3$ and specifies its
filling ambiguity. It is also the starting point for the
Crowley--Goette comparison on its stated rational-cohomology domain.

\section{The relative characteristic number and the factor of eight}
\label{sec:relative-invariant}

The APS calculation in Section~\ref{sec:aps-evaluation} has reduced the
ten-dimensional response to a bulk integral on an eight-manifold and a
boundary integral on its seven-dimensional base. Their combination is
\begin{equation}
 \phi(S(V),\check s_M)
 =\frac13\left\{\int_W c_W(\lambda_W+c_W)-\int_B h\wedge c\right\}
 \quad\text{in }\R/\Z.
 \label{eq:relative-invariant:form-response}
\end{equation}
Here $\partial W=B$, the actual quaternionic line $\widetilde V$ extends
$V$, and $c_W=c_2(\widetilde U)$, where $\widetilde U$ is its underlying
complex rank-two bundle. The convention is $\lambda_W=p_1(TW)/2$;
the extension includes the specified boundary identifications of the
Spin structures, bundles, and connections. The three-form $h$ is the form component of the
integral differential trivialization $\check h$ on $B$; it satisfies
$dh=\lambda(TB)+c$. The orientation of $B$ is the boundary orientation
of $W$, compatible with the base-then-fibre and outward-boundary
orientations used to construct $S(V)$.

Neither integral in \eqref{eq:relative-invariant:form-response} is by
itself the retained topological datum. The bulk integral uses an
auxiliary extension, and the boundary term is generally nonzero. The
full integral nullhomotopy in $\check h$ ties these two pieces together:
their difference is an integral relative characteristic number. This
explains why the response is quantized even though its calculation
passes through differential forms. Li and Sun identify this difference
with an integral relative pairing and use it to compare the response
with the Crowley--Goette invariant \cite[\S5]{C1Working}.

\subsection{The boundary trivialization fixes the relative class}

Let $j:H^4(W,B;\Z)\to H^4(W;\Z)$ forget the boundary
trivialization. The specified integral nullhomotopy determines
\begin{equation}
 z\in H^4(W,B;\Z),\qquad jz=\lambda(TW)+c_2(\widetilde U).
 \label{eq:relative-invariant:integral-lift}
\end{equation}
This is a relative class: it contains the bulk characteristic class
together with its chosen trivialization at the boundary. Choosing only
an absolute class whose boundary restriction vanishes would omit the
information specifying that trivialization. Likewise, solving the
real equation for $h$ would not determine $z$.

One can see the relation between the integral and differential data
directly. Write the bulk differential cocycle for $\lambda+c$ as
$(x_{\Z},x_{\R},X)$, where $x_{\Z}$ is an integral four-cochain,
$x_{\R}$ is a real three-cochain, and the curvature four-form is
$X=\lambda_W+c_W$. Its boundary geometric trivialization is
$(b,k,h)$, with integral three-cochain $b$, real two-cochain $k$,
and three-form $h$. These components are the differential-cocycle
model described in Appendix~\ref{app:differential-data}. Let
$i:B\hookrightarrow W$ be the inclusion and $\delta$ the singular
cochain coboundary. With $\mathcal I$ denoting integration of forms
on singular chains, the trivialization equations are
\begin{equation}
 \delta b=x_{\Z}|_B,\qquad
 \mathcal I h-b-\delta k=x_{\R}|_B,\qquad dh=X|_B.
 \label{eq:relative-invariant:boundary-cochains}
\end{equation}
The integral cocycle $(x_{\Z},b)$ represents $z$. In the relative
complex, whose differential is
$d_{\mathrm{rel}}(u,v)=(du,i^*u-dv)$, these equations give
\begin{equation}
 (\mathcal I X,\mathcal I h)-(x_{\Z},b)
 =d_{\mathrm{rel}}(x_{\R},-k).
 \label{eq:relative-invariant:real-image}
\end{equation}
Thus the pair of forms $(X,h)$ represents the real image of this
particular integral class. The equation identifies the real
representative without replacing the integral cocycle or its
boundary morphism by form data.

Multiplying by the absolute class $c_2(\widetilde U)$ and evaluating
on $[W,B]$ yields
\begin{equation}
 \Krel:=\left\langle c_2(\widetilde U)\smile z,[W,B]\right\rangle
 =\int_W c_W(\lambda_W+c_W)-\int_B h\wedge c\in\Z.
 \label{eq:relative-invariant:K}
\end{equation}
The minus sign is the relative-integration convention: a pair
$(\alpha,\beta)$ integrates as $\int_W\alpha-\int_B\beta$.
At the differential-cochain level the product includes the homotopy
between cup and wedge products. Redden's product of a geometric
trivialization with a differential cocycle has integral trivializing
component $b\smile c_2(\widetilde U)$ and form component
$h\wedge c$ \cite[\S3.4, Proposition~3.23]{Redden}. These are
precisely the two components used in
\eqref{eq:relative-invariant:K}; Appendix~\ref{app:differential-data}
gives the relative-complex conventions and product details. A freely
chosen real primitive need not have this integral lift, so the
curvature equation alone would not prove the displayed integrality.

Substitution into the evaluated response gives
\begin{equation}
 \phi(S(V),\check s_M)=\frac{\Krel}{3},\qquad
 \mathcal O(S(V),\check s_M)=\exp\left(2\pi i\frac{\Krel}{3}\right).
 \label{eq:relative-invariant:response}
\end{equation}
The number $\Krel$ measures the relative characteristic pairing of
the specified background; the construction does not identify it
with a new physical charge. Its integrality implies that this response
takes values among the third roots of unity. The exponentiation
forgets all but the residue of $\Krel$ modulo three, whereas the
relative construction retains more information about the background.
No rational-cohomology vanishing has entered this calculation.

\subsection{Closed periods and relative evenness}

A choice of filling gives an integer $\Krel$, but the complete fixed
boundary data determine its residue modulo $24$. To compare two
fillings, glue one to the orientation reversal of the other, using
the same boundary Spin and bundle identifications and the same
integral nullhomotopy. The boundary cochains cancel. The resulting
closed spin eight-manifold $X$ carries an actual quaternionic line
$L$, and the difference of the two relative pairings is
\begin{equation}
 \Delta\Krel
 =\left\langle c_2(L)(\lambda(TX)+c_2(L)),[X]\right\rangle
 =12\,\Ind D_{L-2}\in24\Z.
 \label{eq:relative-invariant:closed-periods}
\end{equation}
Here $L$ is regarded as a complex rank-two coefficient and $2$ means
two trivial complex coefficients. Since
$\ch(L-2)=-c_2(L)+c_2(L)^2/12+\cdots$ and
$\widehat A_4(TX)=-\lambda(TX)/12$, the degree-eight density is
$c_2(L)(\lambda(TX)+c_2(L))/12$. The last inclusion uses
quaternionic index parity
\cite[(1.1)--(1.4)]{CG11}. In eight dimensions the chiral Spin
representations are real. Tensoring with a quaternionic line supplies
a quaternionic structure commuting with the twisted Dirac operator,
so its index is even. Subtracting twice the untwisted index preserves
evenness. Consequently the closed characteristic number is a
multiple of $24$, rather than merely a multiple of $12$.

This gluing compares spin/gauge extensions of the same boundary
datum. It does not require $\check h$ to extend as a trivialization
over either eight-manifold. Such an extension would restrict the
relative class and is not a premise of the response formula. The
$24\Z$ ambiguity changes $\Krel/3$ by an integer, and therefore
leaves the normalized scalar and its exponentiated response unchanged.

There is a separate parity statement for each relative pairing,
\begin{equation}
 \Krel
 =\left\langle z\smile jz-\lambda(TW)\smile z,[W,B]\right\rangle
 \in2\Z.
 \label{eq:relative-invariant:evenness}
\end{equation}
It follows from \eqref{eq:relative-invariant:integral-lift}, since
$c_2(\widetilde U)=jz-\lambda(TW)$. To see the parity without
assuming an extension of the trivialization, extend the relative
class $z$ by zero to the spin double of $W$. For a degree-four
mod-two class its square is the fourth Steenrod square
$\mathrm{Sq}^4$; the Wu formula on the double pairs this square with
the fourth Wu class. On a spin manifold this class is
$w_4=\lambda\bmod2$. The zero extension contributes nothing from
the other half of the double, so the two terms in
\eqref{eq:relative-invariant:evenness} agree modulo two. Their
difference is even.

The two statements have different uses. Closed index parity controls
the change under a different auxiliary filling; the relative Wu
argument constrains the allowed integer for one filling with its
specified boundary trivialization. In particular, relative evenness
does not force the phase to vanish: an even integer can have a
nonzero residue modulo three. This distinction will be useful for
the six-dimensional normalization in
Section~\ref{sec:six-dimensional}.

\subsection{An intrinsic seven-dimensional expression}

The same answer can be obtained from an ordinary Dirac operator on
$B$, with no explicit choice of $W$ in the final expression. Use the
actual complex bundle $U$ underlying $V$, and the virtual complex
coefficient $4U-8$. The numbers $4$ and $-8$ are integer
multiplicities of actual bundles; the virtual coefficient has rank
zero. Its degree-eight index density is
\begin{align}
 \ch(\widetilde U)&=2-c_W+\frac{c_W^2}{12}+\cdots,
 &\widehat A_4(TW)&=-\frac{\lambda_W}{12},\nonumber\\
 [\widehat A(TW)\ch(4\widetilde U-8)]_8
 &=\frac13c_W(\lambda_W+c_W).
 \label{eq:relative-invariant:intrinsic-density}
\end{align}
The cancellation of the rank term removes the pure gravitational
degree-eight contribution. The remaining product contains exactly
the characteristic density in \eqref{eq:relative-invariant:K},
with its positive sign and factor $1/3$.

For product-collar geometry, complex APS for $\widetilde U$ and
the trivial bundle, with these integer multiplicities, gives
\cite{APS}
\[
 \Ind D_{4\widetilde U-8}
 =\frac13\int_W c_W(\lambda_W+c_W)-\xi(B,4U-8).
\]
The index is an integer, so reduction modulo $\Z$ removes it.
Subtracting the boundary term therefore gives
\begin{equation}
 \phi(S(V),\check s_M)
 =\xi(B,4U-8)-\frac13\int_B h\wedge c
 =\frac{\Krel}{3}\quad\text{in }\R/\Z.
 \label{eq:relative-invariant:intrinsic}
\end{equation}
The reduced invariant $\xi(D)=(\eta(D)+\dim_{\C}\ker D)/2$ is
extended additively to the virtual coefficient, including its zero
modes. This is an intrinsic expression for the response
already evaluated by the twelve-dimensional real APS calculation,
not a replacement of its real eta normalization by an arbitrary
half of a class in $\R/\Z$. The boundary correction still carries
the differential trivialization. Omitting it would retain the
spectral term while changing the scalar being computed.

\subsection{The Crowley--Goette normalization}

The comparison with the established secondary invariant requires an
additional, precise restriction on the base:
\begin{equation}
 H^3(B;\Q)=H^4(B;\Q)=0.
 \label{eq:relative-invariant:CG-domain}
\end{equation}
In the cited arXiv version, Crowley and Goette define their
quaternionic-line $t$-invariant on
closed spin manifolds with these rational-cohomology vanishings
\cite[Definition~1.4]{CG11}. There is no simple-connectedness
condition in this definition or in the filling comparison used here.
The restriction has a concrete algebraic role. The sequence
\begin{equation}
 H^3(B;\Q)\longrightarrow H^4(W,B;\Q)
 \xrightarrow{j}H^4(W;\Q)\longrightarrow H^4(B;\Q)
 \label{eq:relative-invariant:rational-sequence}
\end{equation}
makes $j$ an isomorphism. Hence $c_2(\widetilde U)$ has a unique
rational relative lift, denoted $\bar c$. The vanishings remove
both the obstruction to this lift and its rational ambiguity;
they do not set integral torsion to zero.

For an actual spin/quaternionic filling, the dimension-seven
extrinsic formula of Crowley--Goette is
\begin{equation}
 t_B(V)=-\frac12
 \left\langle[\widehat A(TW)\ch'(\widetilde U)]_4
              \smile\bar c,[W,B]\right\rangle\pmod\Z,
 \qquad 2-\ch(\widetilde U)=c_W\ch'(\widetilde U).
 \label{eq:relative-invariant:CG-extrinsic}
\end{equation}
Their general dimension $4k-1$ formula contains $1/a_{k+1}$,
with $a_j=2$ for odd $j$ and $a_j=1$ for even $j$. Here $k=2$
and $a_3=2$, giving the factor $1/2$ in
\eqref{eq:relative-invariant:CG-extrinsic}
\cite[(1.3)--(1.6)]{CG11}. Their Proposition~1.3 identifies this
filling expression with the intrinsic invariant. The actual
spin/quaternionic filling constructed in
Section~\ref{sec:aps-evaluation} supplies its geometric premise;
the cohomology restriction supplies the unique relative lift.

The degree-four calculation is short but fixes both sign and
coefficient:
\begin{equation}
 \ch'(\widetilde U)=1-\frac{c_W}{12}+\cdots,
 \qquad
 [\widehat A(TW)\ch'(\widetilde U)]_4
 =-\frac{\lambda_W+c_W}{12}.
 \label{eq:relative-invariant:CG-density}
\end{equation}
The minus sign cancels the one in the extrinsic definition.
If $z_{\Q}$ denotes the rational image of the specified integral
class $z$, uniqueness implies that it is the relative lift of
$\lambda_W+c_W$. Compatibility and graded commutativity of
relative and absolute products then give
\begin{align}
 t_B(V)
 &=\frac1{24}\left\langle(\lambda_W+c_W)\smile\bar c,[W,B]\right\rangle
 \nonumber\\
 &=\frac1{24}\left\langle c_2(\widetilde U)\smile z_{\Q},[W,B]\right\rangle
 =\frac{\Krel}{24}\pmod\Z.
 \label{eq:relative-invariant:CG-relative}
\end{align}
Thus the rational calculation evaluates the same pairing as the
integral construction. It does not introduce a different choice of
boundary data. At the level of forms, choose a primitive
$d\widehat c=c$ on $B$. The identity
$d(h\wedge\widehat c)=\widehat c\wedge(\lambda+c)-h\wedge c$
then gives
$\int_B\widehat c\wedge(\lambda+c)=\int_Bh\wedge c$ by Stokes
on closed $B$. Thus the two relative representatives have the
same boundary correction, including its sign.

Combining \eqref{eq:relative-invariant:response} and
\eqref{eq:relative-invariant:CG-relative} gives the comparison
\begin{equation}
 \phi(S(V),\check s_M)=8t_B(V),\qquad
 \mathcal O(S(V),\check s_M)=\exp(2\pi i\,8t_B(V)).
 \label{eq:relative-invariant:factor-eight}
\end{equation}
The factor eight compares two normalizations of one relative
characteristic pairing: $1/3$ for the Green--Schwarz scalar and
$1/24$ for $t$. Its denominator includes the quaternionic factor
$a_3=2$; omitting that factor would change the answer. This is a
comparison with Crowley--Goette's existing invariant, rather than a
new definition of $t$ or a count of microscopic degrees of freedom.

The integral and comparison formulas retain different domains.
Equation~\eqref{eq:relative-invariant:response} applies to the full
specified sphere-bundle backgrounds. Equation~\eqref{eq:relative-invariant:factor-eight}
additionally requires exactly \eqref{eq:relative-invariant:CG-domain}.
Outside that domain the relative integer still computes the response,
but it is not justified to call $\Krel/24$ the Crowley--Goette
invariant. Section~\ref{sec:sp2-calibration} evaluates the actual
line on the original $\Sp(2)$ example, which lies in the comparison
domain. Section~\ref{sec:string-dependence} then uses the larger
domain to show how changing the integral String datum can change
the response on a fixed ten-manifold.

\section{The calibrated response on \texorpdfstring{$\Sp(2)$}{Sp(2)}}
\label{sec:sp2-calibration}

The relative formula of Section~\ref{sec:relative-invariant} becomes a
concrete response on the group manifold $\Sp(2)$. An actual
quaternionic-line filling fixes the sign of the relative pairing and
gives $\Krel=-2$, hence $\mathcal O=e^{2\pi i/3}$. This is the
known third-root phase obtained by Tachikawa--Zhang in their published
calculation \cite[\S3.2, equation~(63)]{TZ}. Li and Sun's actual-line
evaluation and Spin frame identifications \cite[\S6]{C1Working}
explain how the sphere-bundle calculation reaches the specified
Lie-framed background. The pairing evaluates the induced left datum;
finite differential transport then identifies its response with that
of the original right-framed positive-unit datum. The connections,
frames and integral differential trivialization are essential to this
comparison.

Let $e_1,e_2$ be the standard coordinate vectors of $\HH^2$.
Take $B=S(\HH^2)$ with its outward orientation and let
$P=\Sp(2)\to B$ send $p$ to $pe_2$. Its right stabilizer is
$K=\Sp(1)$ embedded as $\operatorname{diag}(u,1)$. The associated
quaternionic line is $V_y=y^{\perp_{\HH}}$, and hence
\begin{equation}
\label{eq:sp2-calibration:sphere-bundle}
 K\longrightarrow P\longrightarrow S^7,
 \qquad N=D(V),\qquad M=S(V)\cong\Sp(2).
\end{equation}
The identification of the unit-vector bundle with $P$ sends
$[p,u]$ to $p\operatorname{diag}(u,1)$. Thus the same principal
bundle supplies the base gauge data and the total space on which the
response is evaluated. Write $\pi=\pi_N:N\to B$ for the disk
projection. We orient $N$ by the base followed by the real
quaternionic fibre, and $M$ by the outward boundary convention. The
integer $\Krel$ remains distinct from the stabilizer group $K$.

\subsection{The actual line and the Thom pairing}

To evaluate $\Krel$, the eight-dimensional filling must extend the
actual line $V$ with its boundary identification. A stable extension
with a matching characteristic number would not specify the bundle
needed by the associated spin and gauge filling. The useful filling
here is the disk of the tautological line over $\HH P^1$, rather than
a disk $D^8$ in which the gauge group has been enlarged.

Let $L$ be that tautological line over $\HH P^1\cong S^4$,
with complement $L^\perp$ in
$\HH P^1\times\HH^2$, and let $\rho$ be the disk projection.
Set
\begin{equation}
\label{eq:sp2-calibration:actual-line-filling}
 W_0=D(L_{\R}),\qquad \widetilde V=\rho^*L^\perp,
 \qquad \partial W_0=S(L_{\R})\cong S(\HH^2).
\end{equation}
The boundary map $(\ell,y)\mapsto y$ identifies the fibre of
$\widetilde V$ literally with $y^{\perp_{\HH}}$. This is the
eight-dimensional base filling used to calculate the relative integer;
it is different from the eleven-dimensional disk $N$ that defines the
boundary response.

Orient the chart $w\mapsto[1:w]$ by $(1,i,j,k)$, and normalize
$x\in H^4(\HH P^1;\Z)$ by $\int_{\HH P^1}x=1$.
The signs can be seen directly by projecting $e_1$ into $L$. The
projection has one zero at $[0:1]$; in the chart $[w:1]$ its
quaternionic coefficient is $\bar w/(1+|w|^2)$. Quaternionic
conjugation reverses real orientation, while the transition between
the charts is quaternionic inversion and preserves orientation in
dimension four. Consequently
\begin{equation}
\label{eq:sp2-calibration:line-classes}
 e(L_{\R})=-x,\qquad c_2(L)=x,\qquad c_2(L^\perp)=-x.
\end{equation}
These signs also retain the convention $e(V_{\R})=-c_2(U)$:
the complex structure is right multiplication by $i$, whose complex
orientation is opposite to $(1,i,j,k)$.

Orient $W_0$ by base then fibre. Near $[1:0]$, the blowdown to
$\HH^2$ is $(w,q)\mapsto(q,wq)$. At $(0,1)$ its differential
interchanges two four-dimensional blocks with sign $(-1)^{16}=+1$
and sends the outward radial normal to the outward radial normal.
It therefore induces precisely the outward $S^7$ orientation used
above. The boundary Spin structure is the unique one on $S^7$.
The tangent bundle of the disk splits as
$TW_0\cong\rho^*(TS^4\oplus L_{\R})$. Since
$\lambda(TS^4)=0$ and $\lambda(L_{\R})=-c_2(L)$, this gives
\begin{equation}
\label{eq:sp2-calibration:bulk-classes}
 \lambda(TW_0)=-\rho^*x,\qquad
 c_W=c_2(\widetilde V)=-\rho^*x.
\end{equation}

Let $u\in H^4(W_0,S^7;\Z)$ be the positive Thom class of
$L_{\R}$, and let $j$ denote the map from relative to absolute
cohomology. The Thom identity gives $ju=\rho^*e(L_{\R})=-\rho^*x$.
Both $H^3(S^7;\Z)$ and $H^4(S^7;\Z)$ vanish, so the exact
sequence of the pair makes $j$ an isomorphism here. Consequently the
class specified by the boundary nullhomotopy is $z=2u$, since
$jz=\lambda(TW_0)+c_W=-2\rho^*x$. Thom integration sends
$\rho^*x\smile u$ to $x$ on $\HH P^1$, whose integral is one.
The signed relative product therefore evaluates to
\begin{equation}
\label{eq:sp2-calibration:relative-value}
 \Krel=\langle(-\rho^*x)\smile2u,[W_0,S^7]\rangle=-2,
 \qquad t_{S^7}(V)=-\frac1{12},\qquad
 \phi(S(V))=\frac13\pmod\Z.
\end{equation}
The last two values are in $\R/\Z$: the Crowley--Goette comparison
gives $t_{S^7}(V)=\Krel/24$, and the response formula gives
$\phi=\Krel/3=-2/3=1/3$ modulo $\Z$. The comparison applies
because $S^7$ has vanishing rational cohomology in degrees three and
four. Product-collar extensions of the specified boundary geometry and
connections preserve this relative
evaluation; an extension of the base String trivialization over $W_0$
is not required.

\subsection{The selected zero local term and finite transport}

The relative number computes the response of the induced sphere-bundle
datum. To identify it with the left Lie-framed datum $s_L$, one must
also identify the actual Spin bundles, boundary frames and differential
trivialization. Over $S^7$ there are special tangent splittings
$TB=V_{\R}\oplus\R^3$ and
$TN\cong\pi^*(2V_{\R}\oplus\R^3)$. They identify the actual
gauge bundle $E_L=2Q_{\HH}\oplus\R^8$ with
$TN\oplus\R^5$, where $Q=\pi^*P$ and
$Q_{\HH}=\pi^*V_{\R}$ is its associated real quaternionic line.
In the tautological
boundary frame taken twice, the induced stable frame is the left
Lie frame, with its fixed Spin lift. The diagonal Spin identification
supplies the integral nullhomotopy. Appendix~\ref{app:sp2-framings}
gives the frame comparison and its smooth product-collar connections.

This geometry admits a particularly useful representative. If $a$ is
the base principal connection and $a_V$ its associated connection,
choose
\begin{equation}
\label{eq:sp2-calibration:base-connection}
 A_B^*=a_V\oplus d_{\R^3},\qquad
 \lambda(A_B^*)+c=0.
\end{equation}
The second equality is an equality of forms at this connection. The
corresponding differential character is flat, and
$H^3(S^7;\R/\Z)=0$ makes it trivial. The integral base
trivialization class is unique, so its geometric representative can
be chosen with $h=0$, retaining the characteristic morphisms and
integral cochains.

For the smooth gauge connection trivial in the tautological boundary
frame, put $C=c_2(Q,A)$. The product tangent connection $A_*$
uses the natural vertical connection of
Section~\ref{sec:disk-construction} and base summand $\pi^*A_B^*$.
For two copies of the quaternionic line,
$\lambda(E_L)=-2C$ and $p_2(E_L)=6C^2$. Thus the full
polynomial of Section~\ref{sec:response-data} reduces to
$X_4=\lambda(A_*)+2C$ and
$Y_8=(4C^2-6C^2)/6=-C^2/3$. The three-form
transgressions are $T_q=T_{c_2}(\pi^*a,A)$ and
$K_3=T_\lambda(A_0,A_*)$, with
$A_0=\pi^*(A_B^*\oplus a_V)$. Thus both tangent connections
have the same base summand. As in the disk calculation, the
transgressions in $H_L=2T_q+K_3$ have vanishing integrated
odd-degree terms:
\begin{equation}
\label{eq:sp2-calibration:zero-local-term}
 \pi_{N*}C^2=-c,\qquad
 \pi_{N*}(T_qC^2)=\pi_{N*}(K_3C^2)=0,
 \qquad a_{N,*}=\int_N H_LY_8=0.
\end{equation}
These are exact fibre-integrated form identities for the specified
connections, obtained from the equivariant disk calculation. A
cohomology-level vanishing alone would not justify dropping this local
integral. The selected representative makes it zero; the general
formula remains $a_N=\frac13\int_Bhc$.

The relation to the Levi--Civita base representative is a finite
connection transport, not a replacement of the String datum. Let
$\alpha=T_\lambda(A_B^{\LC},A_B^*)$ and transport the same
trivialization by $h_*=h_{\LC}+\alpha$, up to an exact form
with its retained morphism. Write $\pi_Z:Z\to B$ for the projection
of the mixed double. The corresponding tangent transgression is
$-\pi_Z^*\alpha C^2/3$, and its closed-fibre pushforward uses
$\pi_{Z*}C^2=c$. In the following formula, the subscripts
$*,\LC$ distinguish the two base connections; both closed terms
include the finite tangent correction, using the real eta convention
that includes kernels. Thus
\begin{equation}
\label{eq:sp2-calibration:base-transport}
 a_{N,*}-a_{N,\LC}=\frac13\int_B\alpha c,
 \qquad \mu_* -\mu_{\LC}=-\frac13\int_B\alpha c.
\end{equation}
The two finite changes cancel in $\phi=-a_N-\mu$. Choosing the
$h=0$ representative therefore puts the entire response in the
corrected closed term. Changing back to the Levi--Civita base
representative redistributes the value between the local and spectral
terms, while preserving their combination. For the selected
zero-local-term filling, $\mu_*=2/3$ modulo $\Z$, and hence
the left Lie-framed response is
\begin{equation}
\label{eq:sp2-calibration:phase}
 \phi(P,s_L)=\frac13\pmod\Z,\qquad
 \mathcal O(P,s_L)=e^{2\pi i/3}.
\end{equation}
Section~\ref{sec:tmf-comparison} uses this zero-counterterm filling.

\subsection{The right-framed positive-unit datum}

The outward stabilizer-fibre period of the sphere-bundle construction
is $-2$. The original right-framed positive-unit datum is reached
by additional geometry specific to $\Sp(2)$. In particular, the
right and left integral String structures differ over the identity
map: $s_R-s_L=3\nu$, with $\nu\in H^3(P;\Z)$ having positive unit period on
the stabilizer $S^3$. Agreement of a local polynomial or a bare flux
number would not identify these data.

The right filling uses the actual bundle
$E_R=TN\oplus\R^5$, with a full $\Spin(16)$ gauge
connection extending the flat connection in the right boundary
frame. It is not constrained to the two-quaternionic-block reduction
used on the left. A product tangent connection $A_T$ can be chosen
with vanishing $p_1$ form on each disk fibre. Its diagonal
transgression with the right gauge connection $A_G$ supplies
$H_R=T_\lambda(A_G,A_T\oplus d_{\R^5})$, including the
integral refinement. The relative gauge number in that boundary
frame is $-1$, so Stokes gives the real period
\begin{equation}
\label{eq:sp2-calibration:right-period}
 \int_{S^3}H_R
 =\int_{D^4}\bigl(\lambda(A_T)-\lambda(A_G)\bigr)
 =0-(-1)=1.
\end{equation}
The full Spin polynomial and the specified boundary connections are
essential here; a connection change does not change the relative
class, and its boundary Chern--Simons term accounts for the change
in the bulk integral.

Inversion $\iota(p)=p^{-1}$ preserves orientation on the
ten-dimensional group and satisfies $\iota^*s_R=s_L$.
Its differential sends a left-invariant vector $pX$ to
$-Xp^{-1}$, a right-invariant vector with a constant minus sign.
The constant transformation $-I_{10}$ has positive determinant
and can be joined to $I_{10}$ while retaining the outward normal
and the specified Spin lift. This identifies the framings after
inversion; it does not identify them over the identity map of $P$.
Pulling back the complete right metric, connections and refinement,
then transporting them by differential String cylinders with trivial
gauges, identifies its scalar with
\eqref{eq:sp2-calibration:phase}. Appendix~\ref{app:sp2-framings}
describes the Spin lift and the required interval term in the differential
cylinder; interpolation of endpoint three-forms alone would not
supply this identification. On these cylinders the trivial gauges
give $Y_8=\Delta=0$, so the transport preserves the normalized
response. Bare three-form periods need not remain fixed as the
tangent connection changes. The result recovers the known third root
on both specified data without asserting a positive-unit transport
for an arbitrary sphere bundle.

Finally, reversing orientation together with the associated boundary,
differential and reference data changes $\phi$ to $-\phi$ and
$\mathcal O$ to its complex conjugate. For this datum the reversed
response is $e^{-2\pi i/3}$. Independently resetting a reversed
period to $+1$ would specify a different input. The dependence on
integral String data, which this calibration makes concrete, is the
subject of Section~\ref{sec:string-dependence}; its reduced
six-dimensional consequence is discussed in
Section~\ref{sec:six-dimensional}.

\section{String dependence on a fixed ten-manifold}
\label{sec:string-dependence}

The calibrated value on $\Sp(2)$ concerns a specified String datum.
To see how much that specification matters, we now hold the sphere
bundle fixed and vary its integral background data. The scalar can
change while the metric, gauge connections and outward fibre period
remain unchanged. This dependence is already contained in the relative
integer of section~\ref{sec:relative-invariant}; it does not require a
different ten-dimensional geometry. We explain the change law and then
two examples of Li and Sun~\cite[section~7]{C1Working}. The first lies in the
Crowley--Goette domain, whereas the second isolates the additional
information carried by the String datum outside that domain.

\subsection{The pairing that detects a change of trivialization}
\label{sec:string-dependence:variation}

Fix a closed spin seven-manifold $B$ and its actual quaternionic line
$V$, with underlying complex rank-two bundle $U$ and $c=c_2(U)$.
We also write $c$ for its Chern--Weil form at the fixed connection.
Keep the Spin structure, metric, gauge connection, and all
the characteristic and tautological morphisms of the disk
construction fixed. Let $\check h$ be the integral differential
trivialization of $\check\lambda(TB)+\check c$. Another such
trivialization is obtained by adding a degree-three differential
character $\check\alpha$: the isomorphism classes of these
trivializations form a torsor for $\widehat H^3(B;\Z)$, as explained
in Appendix~\ref{app:differential-data}. Write
$\alpha\in H^3(B;\Z)$ for its
characteristic class and $\omega$ for its curvature. Then
\begin{equation}
 \check h\longmapsto\check h+\check\alpha,
 \qquad h\longmapsto h+\omega,
 \qquad d\omega=0.
 \label{eq:string-dependence:shift}
\end{equation}
The equation $dh=\lambda+c$ therefore remains the same. It does
not by itself identify the two trivializations: the shift specifies
their difference as integral differential data, including any part
invisible to curvature.

The intrinsic expression of section~\ref{sec:relative-invariant}
makes the effect especially transparent. Here $\xi(B,4U-8)$ is the reduced
Dirac invariant with the indicated virtual coefficient:
\begin{equation}
 \phi(S(V),\check s_M)
 =\xi(B,4U-8)-\frac13\int_Bh\wedge c
 \quad\text{in }\R/\Z.
 \label{eq:string-dependence:intrinsic}
\end{equation}
The reduced Dirac invariant, including its kernel term, depends on the
geometry and connection that we have held fixed. Only the second term
changes. For the same spin/quaternionic filling $W_8$, the bulk
term in \eqref{eq:relative-invariant:K} is likewise fixed; the
boundary subtraction changes by $-\int_B\omega\wedge c$.
Since $\omega$ represents the real image of $\alpha$,
$\int_B\omega\wedge c=\langle\alpha\smile c,[B]\rangle$.
Thus
\begin{align}
 \Delta\Krel
 &=-\langle\alpha\smile c,[B]\rangle,\nonumber\\
 \Delta\phi
 &=-\frac13\langle\alpha\smile c,[B]\rangle
 \quad\text{in }\R/\Z,\label{eq:string-dependence:change-law}\\
 \frac{\mathcal O(S(V),\check s'_M)}
      {\mathcal O(S(V),\check s_M)}
 &=\exp\!\left(-\frac{2\pi i}{3}
                 \langle\alpha\smile c,[B]\rangle\right).
 \label{eq:string-dependence:phase-ratio}
\end{align}
The sign also follows directly from the relative
cochain convention: the change of the boundary component of $z$ is
represented by $(0,\alpha)$, and relative integration subtracts the
boundary integral. Both expressions therefore measure the same
change of the input rather than a change of filling.

The disk cochain changes by $\pi^*\check\alpha$, while the
tautological boundary morphism stays fixed. Consequently the induced
differential String datum on $M=S(V)$ changes by the same pullback,
and its underlying integral class obeys
\begin{equation}
 \check s'_M-\check s_M=\pi^*\check\alpha,
 \qquad
 [s'_M]-[s_M]=\pi^*\alpha.
 \label{eq:string-dependence:string-class}
\end{equation}
An integral String shift is therefore not generally a gauge
transformation or an isomorphism of the original differential object.
Equation~\eqref{eq:string-dependence:phase-ratio} specifies which
part of that shift this scalar detects.

It also specifies the information the scalar misses. A flat
$\check\alpha$ has zero curvature, so it leaves
\eqref{eq:string-dependence:intrinsic} unchanged. If $\alpha$ is
torsion, its top-dimensional pairing with $c$ is zero: a torsion
product cannot evaluate nontrivially on the integral fundamental class
of the closed oriented seven-manifold. These are statements about
response insensitivity, not equality of differential String objects.
More generally, even a nonzero integral pairing leaves the
exponentiated response unchanged when it is divisible by three. An
equality of scalar values is thus a substantially weaker statement
than an identification of the background data.

On the exact domain $H^3(B;\Q)=H^4(B;\Q)=0$ used in
section~\ref{sec:relative-invariant}, every integral degree-three class
has zero real image and the pairing vanishes. This agrees with
$\phi=8t_B(V)$: the scalar there is read from the quaternionic-line
invariant, although the integral differential data remain part of its
construction. The larger-domain formula
\eqref{eq:string-dependence:intrinsic} is what permits the following
comparison between bases with and without those vanishings.

\subsection{An evaluation on the unit tangent bundle of \texorpdfstring{$S^4$}{the four-sphere}}
\label{sec:string-dependence:sts4}

Let $x$ generate $H^4(S^4;\Z)$ with $\int_{S^4}x=1$, and take
$W_8=D(TS^4)$ with boundary the unit tangent sphere bundle
$B=STS^4$. Orient its tangent disk
fibres positively, so $e(TS^4)=2x$. The Gysin sequence relates the
cohomology of a sphere bundle to pullback, fibre integration and cup
product with its Euler class. Its relevant portion here is
\begin{equation}
 0\longrightarrow H^3(B;\Z)\longrightarrow\Z
 \xrightarrow{\,\times2\,}\Z
 \longrightarrow H^4(B;\Z)\longrightarrow0.
 \label{eq:string-dependence:sts4-gysin}
\end{equation}
It gives $H^3(B;\Z)=0$ and $H^4(B;\Z)=\Z/2$, hence precisely
the two rational vanishings needed for the $t$ comparison. The
integral torsion in degree four has not been discarded.

Choose an actual quaternionic line on $S^4$ with $c_2=2x$ and
pull it back to $W_8$. We write $x$ also for its pullback. The tangent
bundle of $W_8$ splits as the pullback of $TS^4\oplus TS^4$;
since $TS^4\oplus\R$ is trivial, $\lambda(TW_8)=0$.
For the positive Thom class $u\in H^4(W_8,B;\Z)$, normalized to
integrate to one on each disk fibre, the map $j$ to absolute
cohomology satisfies $ju=2x$. Hence $z=u$ is the relative
class lifting $\lambda_W+c_W$. The absolute class $2x$ restricts
to zero integrally on $B$, furnishing the required trivialization;
the absence of
$H^3(B;\Z)$ fixes its integral class. Thom integration then yields
\begin{equation}
 \Krel=\langle2x\smile u,[W_8,B]\rangle=2,
 \qquad t_B(V)=\frac1{12},
 \qquad\phi(S(V),\check s_M)=\frac23.
 \label{eq:string-dependence:sts4-value}
\end{equation}
The last two values are understood modulo $\Z$. This calculation
uses an actual line and an integral boundary nullhomotopy, so both
the relative formula and the Crowley--Goette comparison apply. It
provides a second evaluation of the response, with the orientation
accounted for by the positive tangent Thom class.

\subsection{Three response values with the geometry held fixed}
\label{sec:string-dependence:fixed-manifold}

For a base with a detectable degree-three class, let $Q_{-1}$ be
an actual quaternionic line on $S^4$ with $c_2(Q_{-1})=-x$, and set
\begin{equation}
 F_0=\R\oplus\Ad_{\R}(Q_{-1}),\qquad
 W_8=D(F_0),\qquad B=S(F_0).
 \label{eq:string-dependence:first-sphere}
\end{equation}
Here $\Ad_{\R}(Q_{-1})$ is the real rank-three adjoint bundle.
Use $\rho$ for the projection to $S^4$ and its extension over the
disk bundle. The adjoint action lifts to $\Spin(3)$, making $F_0$
an oriented Spin four-plane bundle. Its real-line section forces
$e(F_0)=0$. The adjoint characteristic identity
$p_1(\Ad_{\R}Q)=-4c_2(Q)$ gives $\lambda(F_0)=2x$.
Choose a second actual quaternionic line $Q_{-2}$ with
$c_2(Q_{-2})=-2x$ and put $\widetilde V=\rho^*Q_{-2}$ on
$W_8$, with boundary restriction $V=\widetilde V|_B$.
The disk tangent splitting is the pullback of $TS^4\oplus F_0$.
The $TS^4$ summand has zero $\lambda$, so it gives
\begin{equation}
 \lambda(TW_8)=2\rho^*x,
 \qquad c_W=-2\rho^*x,
 \qquad \lambda(TW_8)+c_W=0\quad\text{integrally}.
 \label{eq:string-dependence:zero-bulk-class}
\end{equation}
The integral vanishing permits a geometric trivialization on
$W_8$; choose one and restrict it to a boundary datum
$\check h_0$. Its form component need not vanish, because the
chosen characteristic forms need not vanish pointwise. What matters
for the relative pairing is that this particular trivialization
extends over the filling: its relative class is zero, so
$\Krel_0=0$.

The first Gysin sequence, for $\rho:B\to S^4$, now reads
\begin{equation}
 0\longrightarrow H^3(B;\Z)
 \xrightarrow{\,\rho_*\,}\Z
 \xrightarrow{\,\smile e(F_0)=0\,}\Z
 \xrightarrow{\,\rho^*\,}H^4(B;\Z)\longrightarrow0.
 \label{eq:string-dependence:first-gysin}
\end{equation}
Thus $H^3(B;\Z)\cong\Z$ and $H^4(B;\Z)\cong\Z$.
The map $\rho_*$ integrates over the oriented sphere fibre and
is an isomorphism here. Choose its primitive generator $\alpha$
so that $\rho_*\alpha=1$; the pullback $\rho^*x$ is likewise
a primitive generator in degree four. The projection formula
then computes the pairing needed by the change law:
\begin{equation}
 \langle\alpha\smile c,[B]\rangle
 =-2\langle(\rho_*\alpha)\smile x,[S^4]\rangle=-2.
 \label{eq:string-dependence:pairing}
\end{equation}
Choose a differential refinement $\check\alpha$ and use
$\check h_n=\check h_0+n\check\alpha$. Equations
\eqref{eq:string-dependence:change-law} and
\eqref{eq:string-dependence:pairing} give
\begin{equation}
 \Krel_n=2n,\qquad\phi_n=\frac{2n}{3}\pmod\Z,
 \qquad\mathcal O(M,\check s_n)=\exp(4\pi i n/3).
 \label{eq:string-dependence:three-values}
\end{equation}
For $n=0,1,2$ these are the three distinct values
$1,e^{4\pi i/3},e^{2\pi i/3}$.

The phase calculation alone does not prove that the induced integral
String classes on $M$ differ: a pullback can kill a base class.
This is the role of the second Gysin sequence, for
$\pi:M=S(V)\to B$. Our quaternionic orientation gives
$e(V_{\R})=-c=2\rho^*x$, so its relevant terms are
\begin{equation}
 0\longrightarrow H^3(B;\Z)\xrightarrow{\,\pi^*\,}
 H^3(M;\Z)\longrightarrow H^0(B;\Z)
 \xrightarrow{\,\smile 2\rho^*x\,}H^4(B;\Z).
 \label{eq:string-dependence:second-gysin}
\end{equation}
The final map is multiplication by two between copies of $\Z$,
and is injective. Exactness therefore makes $\pi^*$ an
isomorphism in degree three. Equation
\eqref{eq:string-dependence:string-class} now shows that all the
classes $s_n$ and $s_m$ differ by the nonzero class
$(n-m)\pi^*\alpha$ whenever $n\ne m$.
In particular, $s_n$ and $s_{n+3}$ remain distinct integral classes
even though their scalar values agree. The response detects the
integer parameter only modulo three; the integral String class
retains the whole parameter.

Throughout this family the ten-manifold $M$, its sphere projection,
Spin data, metric, gauge bundles, connections and tautological
frames are held fixed. The changed data are the base integral
differential trivialization and the String datum it induces on $M$.
The change of $\check H_N$ is pulled back from $B$ and has zero
restriction to every disk fibre. All backgrounds consequently retain
the outward fibre period $-2$ of
section~\ref{sec:disk-construction}, at the specified connection.
There is no use here of the special right-framed $+1$ transport on
$\Sp(2)$.

Both $H^3(B;\Q)$ and $H^4(B;\Q)$ are nonzero in this example.
Equation~\eqref{eq:string-dependence:three-values} uses the general
relative scalar formula, not three different values of $t_B(V)$ in
its original domain. The three values are responses to distinct
background data; they do not count vacua or establish a phase
transition. What the calculation establishes is that the underlying
ten-manifold and a bare fibre period do not together determine this
response. The degree-three pairing makes that missing information
explicit, and leads to the six-dimensional anomaly interpretation
of the same corrected base expression in
section~\ref{sec:six-dimensional}.

\section{Six-dimensional response and integral lattice factorization}
\label{sec:six-dimensional}

The intrinsic expression in Section~\ref{sec:relative-invariant} gives
the ten-dimensional scalar a second interpretation: it is a closed
seven-dimensional anomaly response for a virtual chiral system in six
dimensions. This interpretation uses the same quaternionic line and the
same integral differential trivialization that entered the sphere-bundle
calculation. In particular, the sensitivity to the String datum found in
Section~\ref{sec:string-dependence} remains part of the background on
which the reduced response is evaluated.

Let $U$ be the complex rank-two bundle underlying the actual
quaternionic line $V$, and retain $c=c_2(U)$ and
$\lambda=p_1(TB)/2$. Define
\begin{align}
 R_+&=4U-8, & I_8^+&=\frac13c(\lambda+c),
 \label{eq:six-dimensional:weyl-density}\\
 \mathcal A_7(B,V,\check h)
 &=\exp\left(2\pi i\left[
     \xi(B,R_+)-\frac13\int_Bh\wedge c\right]\right).
 \label{eq:six-dimensional:reduced-response}
\end{align}
Here the integer $8$ denotes eight trivial complex summands. Without an
additional reality condition, $R_+$ describes four complex Weyl
doublets of one chirality and eight complex Weyl singlets of the
opposite chirality. Subtraction records their opposite contributions
to the anomaly. It is a virtual anomaly system, rather than a specified
interacting matter theory. The quaternionic origin of $U$ does not
impose an additional reality condition on these six-dimensional Weyl
fields.

The reduced invariant in
\eqref{eq:six-dimensional:reduced-response} is the complex invariant
$\xi=(\eta_{\mathrm{APS}}+\dim_{\C}\ker D)/2$, evaluated additively on
the virtual coefficient. Its kernel term is retained. There is no
additional real Pfaffian half multiplying this seven-dimensional
expression. The real normalization of the original eleven-dimensional
operator has already entered the direct APS reduction. The
eight-form $I_8^+$ is consequently the index density appropriate to
this complex coefficient, with the chirality sign fixed by
\eqref{eq:six-dimensional:reduced-response}.

One can see its normalization directly. Through degree eight, the
rank-two $\SU(2)$ bundle has
\[
 \ch(U)=2-c+\frac{c^2}{12}+\cdots,
 \qquad \widehat A_4=-\frac{\lambda}{12}.
\]
The rank of $4U-8$ vanishes, so the purely gravitational
degree-eight term proportional to that rank cancels. Its remaining
degree-four and degree-eight Chern character terms are $-4c$ and
$c^2/3$. Thus
\begin{equation}
 [\widehat A\ch(R_+)]_8
 =\left(-\frac{\lambda}{12}\right)(-4c)+\frac{c^2}{3}
 =\frac13c(\lambda+c).
 \label{eq:six-dimensional:density-expansion}
\end{equation}
This calculation fixes both the mixed gravitational--gauge coefficient
and the pure gauge coefficient; agreement of only the latter would
not specify the response.

The local correction uses the geometric component $h$ of the full
datum
\begin{equation}
 d\check h=\check\lambda+\check c,
 \qquad dh=\lambda+c.
 \label{eq:six-dimensional:background-trivialization}
\end{equation}
At the form level, $d(h\wedge c)=c(\lambda+c)$ because $dc=0$.
The subtraction in \eqref{eq:six-dimensional:reduced-response}
therefore cancels the local variation of the eta response on these
backgrounds. The full differential equation also fixes the integral
nullhomotopy and its compatibility with the connection data; a freely
chosen real primitive would not supply them. Local cancellation leaves
a closed response whose value can still depend on that integral datum.
The eight-dimensional spin/gauge filling used to evaluate it extends
the actual line with the prescribed boundary identification, without
requiring $\check h$ to extend as a trivialization over the filling.

For these data the direct calculation and the relative pairing give
\begin{equation}
 \mathcal A_7(B,V,\check h)
 =\mathcal O(S(V),\check s_M)
 =\exp\left(2\pi i\frac{\Krel}{3}\right).
 \label{eq:six-dimensional:response-identification}
\end{equation}
The sphere-bundle formula~\cite[Theorem~1.1; Sections~4--5 and~8]{C1Working},
derived in Section~\ref{sec:relative-invariant}, identifies the closed
response on the permitted backgrounds, with their orientation and
differential data. It is stronger than matching local polynomials,
since the actual filling evaluates the eta term, but does not assert
an equivalence of six- and ten-dimensional theories. No rational
$H^3$ or $H^4$ vanishing is needed here. Those restrictions belong to
the comparison with the Crowley--Goette invariant, rather than to
\eqref{eq:six-dimensional:response-identification}.

\subsection{The cubic character and the half-doublet normalization}
\label{sec:six-dimensional:cubic-character}

Integrality of $\Krel$ immediately implies $\mathcal A_7^3=1$
on every permitted background. To decide whether a smaller number of
copies has trivial response everywhere, one must also know that a
primitive third root occurs. Section~\ref{sec:sp2-calibration}
provides that test: the actual quaternionic line in the
$\Sp(2)\to S^7$ fibration, with the specified orientation and
transported datum, has $\Krel=-2$. Hence its response is
$\exp(2\pi i/3)$. The integrality argument and this example together
prove, for a nonnegative integer $r$, that the stacked response is
identically one on the permitted background category precisely when
three divides $r$. Writing $\mathcal A_7^r\equiv1$ for this universal
statement, we have
\begin{equation}
 \mathcal A_7^r\equiv1
 \quad\Longleftrightarrow\quad 3\mid r.
 \label{eq:six-dimensional:cubic-copies}
\end{equation}
For the forward implication, the response of $r$ copies on the
primitive background is one only when $3\mid r$. For the reverse implication,
$r=3k$ makes the exponent $k\Krel$ integral on every background.
Thus the sufficiency is universal within the stated domain, while
the necessity is witnessed by the calibrated example. Stacking copies
means multiplying these closed anomaly responses; this calculation
does not classify full quantum field theories or their states.

The coefficient has a known six-dimensional normalization precedent.
Lee--Tachikawa~\cite[\S2.2, (2.14); \S3.3, (3.14)--(3.17)]{LT21}
give the gauge-dependent part of the primitive half-doublet density as
\begin{equation}
 I_8^{\mathrm{half}}=\frac1{24}c(\lambda+c),
 \qquad I_8^+=8I_8^{\mathrm{half}}.
 \label{eq:six-dimensional:eightfold-density}
\end{equation}
The half-doublet includes a reality condition; our virtual complex
coefficient has no such condition. Its eight neutral singlet
subtractions cancel the purely gravitational term, so the comparison
in \eqref{eq:six-dimensional:eightfold-density} concerns the remaining
gauge-dependent density. Their modified anomaly response detects a
generator with value
$\exp(2\pi i/12)$, giving an order-twelve condition for copies of
that primitive system. Eight copies per $R_+$ system give the same
arithmetic condition as \eqref{eq:six-dimensional:cubic-copies}:
\begin{equation}
 12\mid 8r\quad\Longleftrightarrow\quad 3\mid r.
 \label{eq:six-dimensional:normalization-divisibility}
\end{equation}
The divisibility in \eqref{eq:six-dimensional:cubic-copies} is thus a
specialization of the known six-dimensional normalization condition.

The backgrounds in the two calculations differ. Lee--Tachikawa add a
three-form satisfying $dH=c$; their local correction is
$-\int H\wedge(c+\lambda)/24$. In this companion the prescribed
integral differential datum trivializes $\lambda+c$, and the local
correction is $-\int h\wedge c/3$. Their seven-form integrands have
proportional exterior derivatives, but the objects trivialized and their global
data are different. The density comparison therefore explains the
coefficient and the resulting divisibility without identifying the
two anomaly theories on all backgrounds or families. In particular,
the primitive $\Sp(2)$ evaluation used above is the one already
computed with this companion's background datum.

This distinction also keeps local cancellation separate from a
dynamical tensor construction. Lee--Tachikawa's discussion following
(3.17) explains why cancellation for twelve copies with their classical
three-form does not alone define the path integral over gauge and
two-form fields. The closed character is one part of the anomaly
problem. Its triviality does not supply the missing quantum tensor
data.

\subsection{Universal coefficient matching in an integral lattice}
\label{sec:six-dimensional:lattice-factorization}

In six-dimensional theories with chiral two-form fields, the lattice
records charges of strings coupled to those fields. Locality of
probe-string surface operators gives an integral pairing, and
consistency under dimensional reduction requires a self-dual, or
unimodular, charge lattice \cite[\S2.3, (2.9)--(2.11)]{SeibergTaylor11}.
Seiberg--Taylor's even signature-$(1,1)$ example has the hyperbolic
pairing used below \cite[\S2.4, (2.16)]{SeibergTaylor11}.
This motivates the arithmetic setting for our coefficient-matching
question; it does not introduce tensor fields into the original
background.

The same divisibility has a precise arithmetic formulation in the
usual tensor Green--Schwarz normalization. Monnier--Moore write the
eight-form as $\tfrac12(Y,Y)$ and, for a single simple gauge factor,
use $Y=a\,p_1/4-bc$~\cite[(2.5), (2.10)--(2.12)]{MM18}.
Here $Y$ is a lattice-valued degree-four polynomial, distinct from the
degree-eight $Y_8$ of the ten-dimensional calculation. The vectors
$a$ and $b$ are respectively the gravitational and gauge coefficients
in this normalization. Monnier--Moore--Park relate their integrality
to string-charge quantization and distinguish conditions for the
covering gauge group from those for the full gauge global
form~\cite[\S2.3, constraints~9, $9^*$ and~10]{MMP18}. Those
integrality constraints use a generalized completeness hypothesis:
the theory is assumed to admit arbitrary spin spacetimes with arbitrary
smooth gauge-field configurations. These background assumptions enter
their physical derivation of charge quantization.
In that normalization, we ask whether $rI_8^+$ admits an integral
lattice factorization. The scalar calculation does not establish the
broader background assumptions used in the supergravity analysis.

Take $c$ to be the primitive integral second Chern class of an actual
$\SU(2)=\Sp(1)$ bundle. An integral lattice $\Lambda$ means a free
abelian group with an integer-valued symmetric nondegenerate pairing.
Require $a,b\in\Lambda$ and require $a$ to be characteristic:
\begin{equation}
 (a,v)\equiv(v,v)\pmod2\qquad\text{for every }v\in\Lambda.
 \label{eq:six-dimensional:characteristic-vector}
\end{equation}
The desired identity is
\begin{equation}
 \frac12(Y,Y)=rI_8^+,
 \qquad Y=\frac a2\lambda-bc.
 \label{eq:six-dimensional:lattice-factorization}
\end{equation}
It is an identity of universal polynomials in $\lambda$ and $c$.
One must compare its coefficients before restricting to a background
on which $\lambda+c$ is trivialized. Substituting that relation first
would erase the information the factorization is meant to test.

Expanding the left-hand side gives
\begin{equation}
 \frac12(Y,Y)
 =\frac{(a,a)}8\lambda^2
  -\frac{(a,b)}2\lambda c
  +\frac{(b,b)}2c^2.
 \label{eq:six-dimensional:coefficient-expansion}
\end{equation}
Comparison with $r c(\lambda+c)/3$ therefore requires
\begin{equation}
 (a,a)=0,\qquad (a,b)=-\frac{2r}{3},\qquad
 (b,b)=\frac{2r}{3}.
 \label{eq:six-dimensional:coefficient-conditions}
\end{equation}
The last equation already proves necessity: $(b,b)$ is an integer,
so $3\mid r$. This implication uses neither unimodularity nor the
characteristic condition. Those conditions matter for constructing
the converse with the specified gravitational coefficient.

For $r=3k$, take the rank-two hyperbolic lattice with basis $e,f$ and
pairing
\begin{equation}
 (e,e)=(f,f)=0,\qquad (e,f)=1,
 \qquad a=2kf,\quad b=-e-kf.
 \label{eq:six-dimensional:hyperbolic-data}
\end{equation}
Its Gram matrix has determinant $-1$, so the lattice is unimodular.
For $v=me+nf$, $(v,v)=2mn$ is even, so it is also even. Both
coefficients are integral, and
\begin{equation}
 (a,v)-(v,v)=2m(k-n)\in2\Z.
 \label{eq:six-dimensional:characteristic-check}
\end{equation}
Thus $a$ is characteristic. With these choices the degree-four
polynomial becomes
\begin{equation}
 Y=ec+kf(\lambda+c),\qquad
 \frac12(Y,Y)=kc(\lambda+c)=rI_8^+.
 \label{eq:six-dimensional:hyperbolic-factorization}
\end{equation}
Only the cross-pairing contributes, which makes both the sign of $b$
and the absence of a $\lambda^2$ term transparent. This construction
also covers $r=0$ with $k=0$. Consequently, the universal factorization
with integral gauge coefficient and characteristic gravitational
coefficient exists precisely for multiples of three, and an even
unimodular hyperbolic lattice suffices~\cite[Section~8,
Proposition~8.1]{C1Working}.

The conclusion concerns this polynomial and these integral
coefficients. The primitive class and actual gauge global form are
essential to its meaning; a change of matter content, gauge group or
anomaly sector changes the question. Moreover, the tensor formulation
uses a lattice-valued differential trivialization $d\check H=\check Y$,
whose curvature equation is $dH=Y$. Monnier--Moore's construction
shifts $\check Y$ by $\tfrac12\nu\otimes a$, where $\nu$ represents
the degree-four Wu class. On spin manifolds this class equals the
fourth Stiefel--Whitney class $w_4$. The
cochain $\check H$ has a compatible shift~\cite[\S3.2, (3.9)--(3.12)]{MM18}.
These data differ from the integral trivialization used by our
original scalar in
\eqref{eq:six-dimensional:background-trivialization}. The explicit
lattice identity does not construct quantum self-dual tensors or
establish the remaining matter and gravitational conditions of a
complete supergravity theory. It supplies an arithmetic consequence
of the reduced response, with the known tensor normalization kept
visible.

The identification with a particular published heterotic prescription
requires a different comparison: the actual real operator, its line
and its differential transport must agree. Section~\ref{sec:heterotic-comparison}
addresses those inputs. The lattice calculation alone supplies none
of them.

\section{Comparison with the heterotic fermion prescription}
\label{sec:heterotic-comparison}

The response computed above can be compared with the ordered heterotic
Green--Schwarz phase of Tachikawa--Zhang \cite{TZ}. The relevant object is
the $\Spin(32)/\Z_2$ phase divided by the $E_8\times E_8$ phase, evaluated
on the same boundary geometry and differential String datum. Identifying
this object requires more than recovering its known third-root value:
the closed fermion systems, their chirality, their lines and their
reference normalization must agree. The comparison holds on the full
filling data, before either the local integral or closed eta invariant
is specialized \cite[\S9]{C1Working}. Its operator and connection
content is separate from the conditional integral topological modular
forms ($\TMF$) comparison in Section~\ref{sec:tmf-comparison}.

\subsection{The ordered phase and its reference}

The physical input is the fermion prescription used in the published
analysis of Tachikawa--Zhang \cite[\S2.1, \S\S4.2--4.4]{TZ}. The two supersymmetric
theories have adjoint gauginos of the same chirality and identical
gravitational fermion and ghost sectors. Their anomaly is the real
fermion eta anomaly of these systems, with no additional independent
invertible sector. For each theory the differential $H_3$-dependent
section obeys the consistency law of \cite[equation (67)]{TZ}, takes
the empty background to the unit, and is paired with that same theory's
trivial-gauge spin reference as in \cite[equation (70)]{TZ}. The Hermitian
pairing is conjugate-linear in its first slot. All of these clauses
specify the observable; they are not an assertion that every interacting
theory is determined by its massless spectrum. An independently added
flat anomaly sector would change this prescription while leaving its
local polynomial unchanged.

For a bordism $A:M_0\to M_1$, the section law for either theory has
the form
\[
 v_j(M_1)=\exp\left(2\pi i\int_A H_3Y_{8,j}\right)
             Z_j(A)v_j(M_0),\qquad v_j(\varnothing)=1,
\]
where $j$ denotes that theory and $Z_j(A)$ is its anomaly evolution
map. The law relates vectors in anomaly lines, rather than assigning
an absolute complex number to the fermion integral. The spin reference
supplies the comparison vector in the same line. Applied to a filling
from the empty manifold, the law fixes the normalized section; its
first-slot pairing then conjugates the local factor. Taking the ratio
of the two theories uses their own sections and references, and the
common $X_4$ turns the difference of $Y_{8,j}$ into $Y_8$. This explains
why the reference and the order of the quotient are essential to the
sign of the scalar.

Let $M$ be a connected closed spin ten-manifold with framed trivial
gauge connections, differential String data and specified tangent
geometry. Use a product-collar differential $\mathcal H$-filling $N$
and a trivial-gauge spin reference $W_{\rm ref}$ with their specified
boundary Spin and differential identifications. Tachikawa--Zhang's
consistency law and reference pairing, followed by its prescribed
gluing, give \cite[equations (71)--(73)]{TZ}
\begin{equation}
 \mathcal O_{\rm TZ}(M)
 =\exp\left(-2\pi i\int_N H_3Y_8\right)I_{\rm diff}(C),
 \qquad C=\overline N\cup_M W_{\rm ref}.
 \label{eq:heterotic-comparison:source-closed}
\end{equation}
Here $I_{\rm diff}(C)$ is the quotient of the closed anomaly partition
functions in the stated order. The negative local sign and the
orientation reversal of $N$ both come from the first-slot conjugation.
On the common cover the two $X_4$ polynomials agree, so $Y_8$ is the
full difference in \eqref{eq:response-data:polynomial}; the reference
does not replace this difference by a first-root specialization.

The reference $W_{\rm ref}$ is a spin filling with trivial gauges;
it is not required to fill the differential $\mathcal H$ structure.
Thus $C$ is a closed spin/gauge manifold, rather than necessarily a
closed differential $\mathcal H$ manifold to which closed cancellation
would apply. A nontrivial scalar is possible even though the section
on $\mathcal H$ boundaries is independent of its $\mathcal H$ filling.
Replacing the reference by an $H$-filling in this argument would
change the comparison being made. The actual seam identifications
also matter: trivial characteristic classes on the boundary alone
do not specify the framed Spin connection or the differential morphism.

The real Pfaffian contraction in
\eqref{eq:heterotic-comparison:source-closed} is constructed in
Appendix~\ref{app:real-eta-conventions:real-gluing}. It matches the
Witten--Yonekura and Freed--Moore allowed boundary domains and spectral-cut
Pfaffian frames, retaining kernels and the actual Spin and gauge seam maps.
The transported state in
\eqref{eq:real-eta-conventions:transported-state} includes the finite
stable tangent correction; its Hermitian contraction in
\eqref{eq:real-eta-conventions:state-gluing} reverses the first filling.

The Dai--Freed erratum \cite{DFerr} corrects the odd-dimensional
collar operator by a $J^{-1}$ factor, with $J$ the outward normal
Clifford action, and amends orientation reversal and graded gluing.
These complex determinant-line formulas do not replace the real
component parities. The real construction retains ordered duality
of its component lines and the specified seam identifications, rather
than taking a square root of a complex gluing identity. It supplies
the contraction for the stated
fermion prescription; the closed-operator comparison below does not
assert an equivalence of complete anomaly functors.

\subsection{Parallel real branching and the fermion line}

The gauge backgrounds used in the calculation lift to the actual
common cover $G=\Spin(16)\times\Spin(16)$. With its induced
connections, the two real adjoint bundles decompose as
\begin{align}
 \operatorname{ad}(\Spin(32))|_G
 &=\Lambda^2E\oplus(E\otimes_{\R}E')\oplus\Lambda^2E',
 \label{eq:heterotic-comparison:branching}\\
 \operatorname{ad}(E_8\times E_8)|_G
 &=\Lambda^2E\oplus S^+_{\R}(E)
      \oplus\Lambda^2E'\oplus S^+_{\R}(E').\nonumber
\end{align}
The first equality is the skew-matrix block decomposition. For the
second, the $112$ roots $\pm e_i\pm e_j$ and eight Cartan directions
give the $120$-dimensional $D_8$ adjoint; the remaining $128$ roots
$\frac12(\pm e_1\pm\cdots\pm e_8)$ with even sign parity give the
real half-spin module. The embedding fixes which half-spin module
occurs. These are equivariant real-module isomorphisms. Invariant
inner products make them orthogonal, and equivariance makes them
parallel for every connection induced from the common cover.

Tensoring these maps with the spacetime spinors therefore intertwines
the actual closed Dirac operators. It preserves their kernels and
flat holonomy, as well as the characteristic forms. The common
adjoint blocks cancel spectrally, and the identical gravitational
fermion and ghost factors cancel against their inverses. The remaining
coefficient is the real virtual bundle
\[
 \Delta_{\R}=E\otimes_{\R}E'-S^+_{\R}(E)-S^+_{\R}(E'),
\]
up to the common gaugino chirality. This stronger statement is what
allows a comparison of global phases: equality of Chern characters
alone would not control the spectrum or its flat dependence.

The sign of the phase also depends on the line in which the fermion
integral takes values. For a real twisting coefficient $R$, fix
positive-index chirality by
$I(R)=\frac12[\widehat A\ch(R_{\C})]_{12}$ and denote the associated
boundary Pfaffian line by $\mathcal P_R$. Freed--Moore place their real
half-eta state $\exp(+\pi i\xi_R)$ in $\mathcal P_R^*$
\cite[equation (3.11)]{FM}; the positive fermion integral itself belongs
to $\mathcal P_R$ \cite[equations (4.33)--(4.34)]{FM}. Only this line
assignment is used, rather than their later conjectural M-theory
cancellation. Tachikawa--Zhang put the fermion integral and anomaly
state in the same line. Their positive-index closed fermion phase is
accordingly the dual state,
\begin{equation}
 \chi_R(C)=\exp(-\pi i\xi_R(C)),\qquad
 \xi_R=\frac{\eta_{\rm APS}(D_R)+\dim_{\C}\ker D_R}{2}.
 \label{eq:heterotic-comparison:real-phase}
\end{equation}

Witten--Yonekura's negative eta exponent describes inflow in the line
dual to the boundary fermion \cite[equations (2.10), (2.52)]{WY}.
To compare it with \eqref{eq:heterotic-comparison:real-phase}, the
closed operator must be translated along with the line. With
Freed--Moore's skew Clifford action $c(v)$ and normalized odd volume
action $\omega$, the translation is
$\gamma_{\rm WY}(v)=i\omega c(v)$. With outward unit normal $\nu$
and boundary volume action $\omega_\partial$, the collar relation is
$\gamma_{\rm WY}(\nu)=-i\omega_\partial$, so the displayed boundary
chiralities agree, but the closed odd operators have opposite signs:
\begin{equation}
 \begin{gathered}
 D_{\rm WY}=-d_{\rm FM},\qquad
 \eta_{\rm WY}=-2\xi_R+2h_R,\\
 e^{-i\pi\eta_{\rm WY}/2}=e^{+\pi i\xi_R},
 \quad h_R=\dim_{\C}\ker d_{\rm FM}\in2\Z.
 \end{gathered}
 \label{eq:heterotic-comparison:operator-dictionary}
\end{equation}
Here $D$ is the Clifford-module Dirac operator and
$d_{\rm FM}=\omega D|_{\mathscr M^0}$ is its odd self-adjoint
operator on the even Clifford-module summand. The volume
action commutes with the Clifford action. The induced boundary
grading intertwines the oppositely signed tangential operators without
reversing their graded index. The dictionary therefore preserves the
boundary fermion while changing the sign of the closed odd operator,
which is precisely the distinction needed for the eta comparison.
The kernel multiplicity is even because closed real-twisted
eleven-dimensional Dirac eigenspaces are quaternionic. Thus the inflow
state agrees with Freed--Moore's inverse-line state, whose dual gives
\eqref{eq:heterotic-comparison:real-phase}. The convention
$\eta=2\xi$ discussed after \cite[equation (2.19)]{WY} applies to
Witten--Yonekura's own odd operator; it cannot identify differently
signed operators merely because their boundary chirality is written
the same way.

Direct sums add reduced eta invariants including their kernels, and
virtual subtraction inverts phases. Quaternionic spectral crossings
introduce no jump in the real phase, whereas
$\xi(-D)=-\xi(D)+h_R$ gives its complex conjugate under operator
reversal. The real-valued
representatives are combined before reduction; no half of an arbitrary
$\R/\Z$ class is selected. We can now determine the remaining binary
chirality choice. The already identified coefficient has
$I(\Delta_{\R})=-X_4Y_8$, whereas the ordered source polynomial is
$+X_4Y_8$ \cite[equations (74)--(76)]{TZ}. This nonzero universal
polynomial fixes the source coefficient as $-\Delta_{\R}$ in the
positive-index convention, yielding
\begin{equation}
 I_{\rm diff}(C)=\chi_{-\Delta_{\R}}(C)
 =\exp(+\pi i\xi_\Delta(C))
 =\exp(-2\pi i\mu_I(C)).
 \label{eq:heterotic-comparison:eta-comparison}
\end{equation}
The polynomial fixes the chirality of this actual system; it does not
classify additional flat anomaly sectors. As a separate sign check,
the even real APS index in twelve dimensions gives
$\chi_R(\partial Q)=\exp(-2\pi i\int_Q I(R))$, the inverse bordism
phase required by the source's positively signed consistency law.

For framed trivial gauge connections the reference coefficient is the
actual zero virtual real module $256-128-128$, so its difference phase
is the unit. On the connected boundary the remaining trivial gauge
summands have even real ranks $120$, $128$ and $256$, and hence even
chiral kernel multiplicities. These facts are useful parity data for
the specified contraction, but they do not prove that all
component-ordering or orientation signs in global real gluing vanish.

\subsection{Finite connection transport of the complete scalar}

The closed operator comparison initially uses product-collar
Levi--Civita conventions. Write $\mu_{\LC}(C)=\mu_I(C;A_{\LC})$
for this uncorrected spectral value. The disk calculation uses the
stable tangent connection $A_*$. Let
\begin{equation}
 \alpha=\CS_\lambda(A_{\LC},A_*),\qquad
 d\alpha=\lambda(A_*)-\lambda(A_{\LC}),\qquad
 H_*=H_{\LC}+\alpha.
 \label{eq:heterotic-comparison:connection-path}
\end{equation}
The last equality transports the existing differential characteristic
trivialization along the specified connection homotopy. It retains its
integral class, its endpoint isomorphisms and the boundary identification;
it is not the selection of an unrelated primitive for the new curvature.

Since $Y_8$ depends only on the gauge connections, the tangent
transgression of $I_{12}$ is $\alpha Y_8$. The paths match on the
collars, and the reference gauges are trivial, so $Y_8|_{W_{\rm ref}}=0$.
The differential-index convention and real APS variation then give
\begin{align}
 \mu_*(C)-\mu_{\LC}(C)
 &=\int_C\alpha Y_8=-\int_N\alpha Y_8\pmod\Z,
 \nonumber\\
 a_*-a_{\LC}&=\int_N\alpha Y_8,\qquad
 (a+\mu)_*=(a+\mu)_{\LC}\pmod\Z.
 \label{eq:heterotic-comparison:finite-transport}
\end{align}
The orientation of $\overline N$ fixes the first minus sign. Neither
term is separately invariant: their corrections cancel in the complete
scalar. On differential String cylinders with trivial gauges, both
$Y_8$ and $\Delta_{\R}$ vanish, so the source law also preserves the
scalar with its transported reference.

The transported $\mu_*$ is the differential-index value defined in
Section~\ref{sec:response-data}, with its finite Chern--Simons correction.
It is not being identified with the raw spectrum of a torsionful Dirac
operator. This matters even for a topological final answer: using
$a_*$ alongside $\mu_{\LC}$ would omit one of the two compensating
terms in \eqref{eq:heterotic-comparison:finite-transport}. Conversely,
setting $a_*=0$ in a chosen geometric representative does not establish
that the local term vanishes in every connection convention or on
every permitted background.

Substituting \eqref{eq:heterotic-comparison:eta-comparison} into
\eqref{eq:heterotic-comparison:source-closed}, then applying
\eqref{eq:heterotic-comparison:finite-transport}, gives the identification
on the full permitted filling/reference data \cite[\S9]{C1Working}:
\begin{equation}
 \mathcal O_{\rm TZ}(M)
 =\exp\bigl[-2\pi i(a_*+\mu_*(C))\bigr]
 =\mathcal O(M),\qquad \phi=-a_*-\mu_*\pmod\Z.
 \label{eq:heterotic-comparison:scalar}
\end{equation}
The generally nonzero local integral has not been suppressed. Only
after this equality do we use the selected $\Sp(2)$ calculation in
Section~\ref{sec:sp2-calibration}, with $a_*=0$ and $\mu_*=2/3$ and
the specified right-frame transport. It gives the known value
$\mathcal O_{\rm TZ}=\exp(2\pi i/3)$
\cite[\S3.2, equation (63)]{TZ}.

Under the same prescription, Tachikawa--Zhang's published analysis
\cite[\S4.4]{TZ} identifies the non-tachyonic fermion spectrum with the
first system plus the chirality-flipped second system. It consequently
has the same closed eta expression and scalar. This equality concerns
the response under the stated spectrum prescription; it
does not construct an ultraviolet string theory or identify complete
anomaly functors on arbitrary families or nonlifting quotient
backgrounds.

\subsection{The first-root bundle, the effective system and the published slant}

The preceding identification uses the full gauge polynomial. It is
compatible with distinguishing, rather than interchanging, three
specific systems appearing in the published discussion. For
the first-root bundle $\HH_{\R}\oplus\R^{12}$ with
$c=c_2(\HH)$, the total Pontryagin class is $1-2c+c^2$.
Thus $n=-c$ and $p_2=c^2$. With the other factor trivial,
\begin{equation}
 Y_8=\frac{n^2-p_2}{6}=\frac{c^2-c^2}{6}=0.
 \label{eq:heterotic-comparison:first-root}
\end{equation}
This does not give the $c^2/6$ printed in \cite[equation (77)]{TZ}.
The $p_2$ term in their full equation (76) is indispensable. The
two-summand disk bundle used here has $E=2Q_{\R}\oplus\R^8$,
$n=-2C$, $p_2=6C^2$ and $Y_8=-C^2/3$, where $C=c_2(Q,A)$
is the disk curvature form. Its actual connections, boundary frames
and integral differential data have been constructed above. The
different substitution cannot by itself supply those data.

The seven-dimensional representation printed in
\cite[equations (57)--(62)]{TZ} differs from the effective coefficient
of the associated-family calculation. Let $\mathbf4$ and $\mathbf{10}$
denote the complex defining and adjoint representations of $\Sp(2)$.
For defining roots $x,-x,y,-y$, write $T_4=2(x^4+y^4)$. The adjoint
roots $\pm2x,\pm2y,\pm x\pm y$, together with two zero roots, give
\[
 \tr_{\mathbf{10}}F^4
 =12\tr_{\mathbf4}F^4+3(\tr_{\mathbf4}F^2)^2.
\]
In Chern-character normalization the primitive quartic coefficients
are $1/24$ and $12/24$. The printed system therefore gives
\begin{equation}
 R_{\rm TZ}=-10\cdot\mathbf4+\mathbf{10}+30,
 \qquad \frac{-10+12}{24}T_4=\frac4{48}T_4.
 \label{eq:heterotic-comparison:old-coefficient}
\end{equation}
The representation-valued fibre index of the present actual-bundle
construction gives
$R_{\eff}=-4\cdot\mathbf4+16$, whose primitive coefficient is
$-8T_4/48$ \cite[Appendix B]{C1Working}.
The qualifier ``printed'' is important here. Wada--Yonekura instead use
$\widetilde R_{\rm aux}=10\cdot\mathbf4-\mathbf{10}-30\cdot\mathbf1$
\cite[Section~9]{WadaYonekura2026}. With the trace normalization above,
its primitive quartic coefficient is
\[
\frac{10-12}{24}T_4=-\frac4{48}T_4,
\]
which is half the primitive coefficient of $R_{\eff}$. This comparison
of coefficients is not an identity of virtual representations.
More precisely, in their specified $S^8$ construction Wada--Yonekura
establish an equality of index densities and hence the index relation
\cite[equations (9.13)--(9.14)]{WadaYonekura2026}
\[
\Ind_{\C}\!\left(D^+_{S^8}\otimes\widetilde R_{\rm aux}\right)
=\frac12\Ind_{\C}\!\left(D^+_{S^8}\otimes\mathcal V_{g'}\right),
\]
where the bundles and the notation $\mathcal V_{g'}$ are those of
\cite{WadaYonekura2026}. We do not identify $\mathcal V_{g'}$ with
$R_{\eff}$ as a virtual bundle or assert an exact dictionary between
the intermediate geometric and spectral data of the two approaches.
In particular, the index relation does not identify their eta lifts,
connections, framings or differential fillings. Comparisons of signed
scalar values also require matching orientation and phase conventions.

The full effective polynomial also keeps more information than that
primitive coefficient. In the source normalization it is
\[
 \widetilde I_{8,\eff}=-\frac16T_4-\frac16p_1c_2.
\]
The representation has rank zero, so the pure gravitational term
cancels. On the stabilizer $K=\Sp(1)$, the restriction
$\mathbf4|_K=U+2$ gives $R_{\eff}|_K=-4U+8$ and $T_4=2c^2$,
where $U$ is the defining complex representation of $K$ and
$c=c_2(U)$. With $p_1=2\lambda$, this yields
$I_{8,\eff}=-(\lambda+c)c/3$. Thus the auxiliary factor used below
comes from a specified representation and its associated connection,
not from selecting a factorization that resembles the published brane
polynomial.

The published slant construction provides another way to relate a
seven-dimensional base to an eleven-dimensional relative anomaly
evaluation \cite[equations (A.16)--(A.23)]{TZ}. Comparing its domain
with the present disk construction is necessary before using such an
evaluation for the same response: a shared dimension or third-root
value does not identify the gauge twist or the differential filling.

Here $\MSpin$ and $\MString$ are the spectra representing spin and
String bordism: stable objects packaging bordism in all dimensions.
Their homotopy groups record manifolds with the
corresponding structures up to bordism. The map
$\MString\to\MSpin$ forgets the String trivialization. Its cofiber
$C_0=\cofib(\MString\to\MSpin)$ represents relative bordism: a
spin filling with a specified String structure on its boundary defines
a relative class. In particular, $\pi_{11}C_0$ contains classes of
eleven-dimensional spin fillings of ten-dimensional String manifolds.
The space $B\SU(2)$ classifies the principal $\SU(2)$ gauge bundle;
the added twist $+2c$ specifies a trivialization of $\lambda-2c$.

The notation $\IZ$ denotes integral Anderson duality. For these
bordism spectra it supplies anomaly classes with an evaluation on
bordism cycles, including the torsion pairing valued in $\Q/\Z$
that curvature forms alone cannot detect. A torsion class $\zeta$ in
degree twelve of $\IZ C_0$ pairs with a torsion relative cycle in
degree eleven, as in \cite[equations (A.16)--(A.17)]{TZ}. The published
slant class extends the four-dimensional relative disk class
$[D^4,S^3]$ equivariantly, with the Lie-group-framed String structure
on its boundary. Its cohomological degree $-4$ records this
four-dimensional shift: acting on a seven-dimensional twisted String
cycle $b$ produces an eleven-dimensional relative cycle. The induced
dual map $\IZ a_{\rm slant}$ precomposes an anomaly evaluation with
that slant map. In these conventions,
\begin{equation}
 \begin{gathered}
 a_{\rm slant}\in C_0^{-4+2c}(B\SU(2)),\qquad
 (a_{\rm slant})_*:\MString_{7+2c}(B\SU(2))\longrightarrow\pi_{11}C_0,
 \\
 \langle\zeta,(a_{\rm slant})_*b\rangle
 =\langle(\IZ a_{\rm slant})(\zeta),b\rangle.
 \end{gathered}
 \label{eq:heterotic-comparison:old-slant}
\end{equation}
The last equality is Anderson evaluation naturality for compatible
classes and cycles, with the torsion hypotheses when its values lie
in $\Q/\Z$. It is a formal adjunction, rather than an identification
of the actual-bundle gauge and differential data with the published
instanton datum. The published $+2c$ domain imposes $\lambda=2c$.
Substitution into the auxiliary
seven-dimensional polynomial would instead give
\[
 -\frac13(\lambda+c)c\big|_{\lambda=2c}=-c^2,
\]
which does not vanish. The auxiliary factor $\lambda+c$, the actual
eleven-dimensional twist $\tTw=-2c$, and the published $+2c$ slant thus
have different roles. The integral $\TMF$ and vertex operator
superalgebra (VOSA) discussion that follows requires its own compatible
lifts and selected differential filling;
neither the ordered scalar identification nor a matching third-root
number constructs those additional inputs.

\section{Conditional integral TMF and VOSA consequences}
\label{sec:tmf-comparison}

The calibrated response on $\Sp(2)$ can test a supplied integral class in
topological modular forms ($\TMF$). Here $\TMF$ is a generalized
cohomology theory whose
integral groups retain torsion information beyond the modular forms
seen rationally \cite[\S1.3]{TY}.
This is a different use of the anomaly calculation from the published
fermion comparison in Section~\ref{sec:heterotic-comparison}. The
normalized scalar is defined without a vertex operator superalgebra
(VOSA) assignment. An assignment to $\TMF$ would instead let us ask
which torsion class is detected by that scalar. The answer below is
conditional on compatible integral lifts with their real equivariant
characters; the character calculation does not construct the lifts.
We explain the selected comparison of Li--Sun
\cite[\S10]{C1Working}, using the relative Anderson construction of
Tachikawa--Yamashita in its fixed version~3 \cite{TY}.

The broader motivation is the Stolz--Teichner proposal that
concordance classes of local $2|1$-dimensional Euclidean field
theories realize $\TMF$ \cite[Conjecture~1.17]{StolzTeichner11}.
These are two-dimensional supersymmetric theories in their geometric
formulation. A modular-form-valued elliptic index does not retain all
of the proposed integral information. Gaiotto--Johnson-Freyd's
secondary elliptic-genus invariant detects some torsion, but vanishes
for the proposed supersymmetric quantum field theory associated with
the Lie-framed $\Sp(2)$ class \cite[\S1, \S2.5]{GJF19}.
The relative Anderson pairing below tests a supplied integral class
against this degree-ten background. It supplies a torsion detector
under the stated lift hypotheses, without establishing the proposed
field-theory description of $\TMF$.

The relevant test pairs a degree-$-32$ class with the String-oriented
image of the ten-dimensional, left Lie-framed $\Sp(2)$. Two ingredients
make this test computable. A constant-coefficient functional selects
exactly the virtual real fermion coefficient already used to define
the response. The selected differential disk filling of
Section~\ref{sec:sp2-calibration} has zero local term, so its relative
Anderson evaluation reduces to a closed spin/gauge evaluation. Both
ingredients are needed: the representation identity determines the
closed coefficient, while the differential filling determines whether
an additional relative contribution remains.

\subsection{Integral lifts and the eleven-dimensional twist}

The bordism notation records which geometric data may extend across
a filling. The spectra $\MSpin$ and $\MString$ represent spin and
String bordism: their groups in dimension $n$ consist of closed
$n$-manifolds with the indicated structure, identified when joined
by a bordism preserving that structure. A String structure includes
an integral trivialization of $\lambda$. A spectrum is the stable
object that collects these groups and their operations across
dimensions; $\pi_n$ denotes its degree-$n$ group. The map
$\MString\to\MSpin$ forgets the integral String trivialization.
Its cofiber $C_0$ represents relative spin/String bordism, in which
a spin filling carries a specified String structure on its boundary
\cite[\S2.3, Appendix~E]{TY}. This relative object retains the
boundary datum even when the boundary already bounds as a spin
manifold. It is therefore suited to evaluating the globally
normalized response rather than only its local polynomial.

The other coefficient theory used below is real $K$-theory, $\KO$,
whose degree-zero classes are virtual real vector bundles.
$\KO((q))$ allows formal Laurent series in $q$ with those
$\KO$ coefficients. The Tate-character map from $\TMF$ to
$\KO((q))$ extracts the real bundle-valued character series,
retaining a gauge action when working equivariantly
\cite[\S1.3, \S4.5]{TY}. Forgetting that action gives a numerical
character with less information. We use these theories only through
their stated character map, orientations and relative pairings.

Keep $K=\Sp(1)$, with $BK$ the space classifying principal
$K$-bundles, and let $c=c_2\in H^4(BK;\Z)$ denote its universal
class. The notation $BK_+$ adds a disjoint basepoint. Smashing a
bordism spectrum with it adjoins a gauge background, while the
subscript $\tTw$ imposes the twisted String trivialization.
We denote spin/gauge bordism by $B_{\tTw}$, its twisted String
refinement by $S_{\tTw}$, and their relative bordism theory by
$C_{\tTw}$ \cite[Appendix~E]{TY}. In this section the twist is
\begin{align}
 \tTw&=-2c,\qquad
 S_{\tTw}=\MString\wedge_{\tTw}BK_+,\notag\\
 B_{\tTw}&=\MSpin\wedge BK_+,\qquad
 C_{\tTw}=\cofib(S_{\tTw}\longrightarrow B_{\tTw}).
\label{eq:tmf-comparison:twist}
\end{align}
A $\tTw$-twisted String structure trivializes
$\lambda-\tTw=\lambda+2c$. This is the eleven-dimensional gauge
twist supplied by $\lambda(E_L)+\lambda(E'_L)=-2c$. It differs from
the seven-dimensional base trivialization of $\lambda+c$ used to
construct the disk datum. It also differs from the Crowley--Goette
invariant $t_B(V)$: $\tTw$ names a degree-four integral twist,
whereas $t_B(V)$ is a secondary invariant with its separate domain.

Write $C_0=\cofib(\MString\to\MSpin)$. The basepoint of $BK$
induces $i:C_0\to C_{\tTw}$, and the cofiber map is
$j:B_{\tTw}\to C_{\tTw}$. Geometrically, $i$ equips a relative
cycle with the trivial gauge background, while $j$ regards a closed
spin/gauge cycle as a relative cycle with empty boundary. With
$M=\Sp(2)$ carrying $s_L$ and the
same trivial-gauge spin reference $W_{\mathrm{ref}}$, put
\begin{equation}
\label{eq:tmf-comparison:relative-cycle}
 z=[W_{\mathrm{ref}},M]\in\pi_{11}C_0,\qquad
 Z=\overline N\cup_M W_{\mathrm{ref}}.
\end{equation}
Here $N$ and $W_{\mathrm{ref}}$ are eleven-dimensional fillings.
The base filling in the relative characteristic calculation is
eight-dimensional, and the spin/gauge filling in the APS calculation
is twelve-dimensional. Their dimensions and roles differ from those
of the relative cycle above. The orientation of the closed double
reverses the disk side, as in the earlier evaluation of $\mu$.

Let $\Phi$ denote the Tate-character map. For any degree $n$, the group
\[
 A_n=\ker\bigl(\Phi:\pi_n\TMF\longrightarrow
                              \pi_n\KO((q))\bigr)
\]
consists of classes invisible to the character. In particular
$A_{10}$, used below, is the degree-ten group containing the
String-oriented test manifold's image. A superscript $32+\tTw$
denotes cohomological degree~32 with the integral twist, and
restriction to a point has homotopy degree~$-32$. The proposed
assignments must carry the gauge-equivariant data needed to relate
these two degrees through the relative evaluation.

\begin{hypothesis}[Compatible integral lifts]
\label{hyp:tmf-comparison:lifts}
There are classes $u_D,u_E\in\TMF^{32+\tTw}(BK)$ with compatible
real equivariant Tate characters. Their restrictions along
$\mathrm{pt}\to BK$ are $x_D,x_E$, and
$x=x_D-x_E$ belongs to $A_{-32}$.
The subscripts denote the rank-sixteen lattice inputs associated to
$\Spin(32)/\Z_2$ and $E_8\times E_8$, in that order. Their characters
have the real parity normalization of published
Tachikawa--Zhang~(A.13) \cite[(A.13)]{TZ}. If a further VOSA is
compared, its assignment supplies an actual
$u_V\in\TMF^{32+\tTw}(BK)$ with the same equivariant
$\KO((q))$ image as $u_D-u_E$.
\end{hypothesis}

These are hypotheses about integral classes, restriction maps and
real representation-valued characters. Equality of numerical
$q$-series after forgetting the gauge action would not provide them.
In the published comparison, the VOSA assignments are formulated as
conjectures \cite[Conjectures~A.7 and~A.12]{TZ}. We use an assignment
only when it supplies the stated class, and do not infer it from the
existence of a lattice algebra or from agreement of local anomaly
polynomials.

\subsection{The real coefficient selected by the character series}

Under Hypothesis~\ref{hyp:tmf-comparison:lifts}, the common leading
$q^{-2}$ terms of the two lattice inputs cancel. The next term is
the difference of their gauge adjoints. On the common
$\Spin(16)\times\Spin(16)$ cover, the actual representation
decompositions are
\begin{align}
 \Lambda^2(E\oplus E')
   &=\Lambda^2E\oplus(E\otimes E')\oplus\Lambda^2E',\notag\\
 \mathfrak e_8|_{\Spin(16)}
   &=\Lambda^2E\oplus S^+_{\R}(E).
 \label{eq:tmf-comparison:adjoints}
\end{align}
Subtracting the two $E_8$ adjoints cancels both exterior squares.
The remaining virtual real coefficient is therefore
\begin{equation}
\label{eq:tmf-comparison:real-coefficient}
 \Delta_{\R}=E\otimes_{\R}E'-S^+_{\R}(E)-S^+_{\R}(E'),\qquad
 \Phi(u_D-u_E)=q^{-1}\Delta_{\R}+O(q^0).
\end{equation}
The real structures and Ramond parity in the hypothesis fix this
sign and normalization. This identity retains the representations,
rather than replacing them by their ranks. It is the coefficient
identity used by Li--Sun \cite[Lemma~10.2]{C1Working}, and in the
published character comparison \cite[(A.13)]{TZ}.

Distinguish the modular discriminant
$\Delta_{\mathrm{mod}}(q)=q+O(q^2)$ from the virtual coefficient
$\Delta_{\R}$. The Witten bundle series satisfies
$\Wit(q)=1+O(q)$. Thus
\begin{equation}
\label{eq:tmf-comparison:constant-coefficient}
 [\Delta_{\mathrm{mod}}(q)\,\Wit(q)\,
                           \Phi(u_D-u_E)]_{q^0}=\Delta_{\R}.
\end{equation}
This extraction is exact. After the $q^{-2}$ cancellation, the
character series has lowest power $q^{-1}$, while the product of
the discriminant and Witten series has lowest power $q$. Their
leading terms are the only terms that can produce $q^0$. Higher
$q$-coefficients consequently do not enter this particular
closed evaluation; this says nothing about equality of the full
theories or about other character-dependent observables.

\subsection{The selected differential term and the positive pairing}

Anderson duality supplies the evaluation that detects integral
torsion. For a spectrum $F$, its Anderson dual $\IZ F$ records
integer-valued homomorphisms on its coefficient groups together with
extension data from torsion. In degree $r$, that torsion part can
evaluate a torsion class of degree $r-1$ in $\Q/\Z$; exponentiation
turns the result into a finite response phase. This is why the
relative evaluation can detect a class that has zero Tate character,
rather than merely reproduce a curvature integral
\cite[\S1.7, Definition~3.48]{TY}.

The untwisted relative Anderson map of Tachikawa--Yamashita is
$\alpha_{\mathrm{spin/string}}:\TMF\to\Sigma^{-20}\IZ C_0$.
Its degree-$-32$ input determines an Anderson class in degree
twelve, which pairs with the torsion relative cycle $z$ in degree
eleven. The suspension $\Sigma^{-20}$ shifts the degree of the
dual by twenty; cohomological degree twelve is consequently the
degree that evaluates this eleven-dimensional relative cycle.
The String orientation $\sigma:\MString\to\TMF$ sends a
String bordism class to its associated $\TMF$ class. Applied to
$[M,s_L]$, it gives the ten-dimensional class against which $x$
is tested. The relative torsion evaluation is precisely
$((x,\sigma[M,s_L]))$ by Tachikawa--Yamashita
\cite[Definition~3.48 and Proposition~3.55]{TY}.
The general differential construction can use the underlying integral
class when $H^{11}(C_{\tTw};\R)=0$ \cite[\S3.3.1]{TY}.
The rational calculation below establishes this vanishing, so no
additional choice of differential $\TMF$ lift is needed here.

Naturality of the twisted construction separates the evaluation
into the closed double and the extending differential filling:
\begin{equation}
\label{eq:tmf-comparison:relative-decomposition}
 i_*z=j_*[Z]+[N,M]_{\mathrm{diff}}.
\end{equation}
The last term retains the extending $\tTw$-twisted differential
String structure, gauge connection, product collars and boundary
identifications. Its underlying relative class is nullbordant, but
this alone would not set its differential evaluation to zero.
Differential information is precisely what can distinguish
refinements of a topologically null class.

For the selected filling, that evaluation can be computed using
the cylinder $N\times[0,1]$ with the intervals over $M$ collapsed
to the relative corner and smoothly rounded. One face carries the
spin data and the other its extending twisted String structure;
the fixed collars identify their common boundary. Gauge data are
pulled back and tangent data are stably product, including the
connection transgressions at the corner. In the mapping-cone
characteristic formula, the degree-twelve bulk form is pulled
back from the eleven-dimensional $N$ and vanishes by dimension.
Write $Y_{\mathrm{rel}}$ for the factor in the relative
characteristic form of Tachikawa--Yamashita. The remaining
String-face integral uses
\begin{equation}
\label{eq:tmf-comparison:zero-term}
 Y_{\mathrm{rel}}=-Y_8=\frac13C^2,\qquad
 \int_N H_LY_{\mathrm{rel}}=-\int_NH_LY_8=-a_{N,*}=0.
\end{equation}
The last equality is the exact local calculation at the connection
of Section~\ref{sec:sp2-calibration}. There the three-form is
$H_L=2T_q+K_3$ because the base curving is $h=0$, while its
integral refinement is retained. Both fibre integrals
$\pi_*(T_qC^2)$ and $\pi_*(K_3C^2)$ vanish as forms, so the
String-face integral is zero at these actual connections.
The mapping-cone construction of
Tachikawa--Yamashita is described in
\cite[Appendix~E, (E.12)--(E.14) and~(E.18)]{TY}; Li--Sun
\cite[Lemma~10.3]{C1Working} supply this selected application.
It proves zero evaluation of this term,
without asserting that every differential refinement of the
underlying null class is zero. Additivity in
\eqref{eq:tmf-comparison:relative-decomposition} now reduces the
pairing to the closed evaluation on $Z$.

For degree~32 and a closed eleven-dimensional spin manifold, the
real differential $\KO$ pushforward has period
$\kappa_{12}\Z((q))=2\Z((q))$. The positive functional of
Tachikawa--Yamashita is
\begin{equation}
\label{eq:tmf-comparison:positive-functional}
 f(q)\longmapsto
       \frac12[\Delta_{\mathrm{mod}}(q)f(q)]_{q^0}\pmod\Z,
\end{equation}
with the real-period interpretation of their equation~(3.95), and
the functional in~(3.99) and Proposition~3.100 \cite{TY}. Its
differential relative target is the quotient
\[
 \R((q))/\bigl((\mathrm{MF}_{-10})_{\R}+2\Z((q))\bigr),
\]
where $(\mathrm{MF}_{-10})_{\R}$ denotes the real modular-form
ambiguity. Multiplication by the weight-twelve discriminant sends
that ambiguity to weight two, whose constant coefficient vanishes
\cite[Fact~2.8]{TY}; the functional therefore descends to the quotient.
The factor $1/2$ acts on this real coefficient quotient: a change by an even
APS period changes the output by an integer. It is not division
of an arbitrary $\R/\Z$ eta class by two. In particular, the
reduced eta invariant retains its kernel contribution,
$\xi(D)=(\eta_{\mathrm{APS}}(D)+\dim_{\C}\ker D)/2$.

Applying \eqref{eq:tmf-comparison:constant-coefficient} to this
functional at the Levi--Civita tangent connection $A_{\LC}$
yields the raw spectral value
$+\frac12\xi(D^{\LC}_{Z,\Delta_{\C}})=-\mu_I(Z;A_{\LC})$,
where $\Delta_{\C}=\Delta_{\R}\otimes_{\R}\C$. To evaluate the
selected stable tangent connection $A_T$, we then apply the finite
transgression. The polynomial of the positive term is $-I_{12}$,
so its Chern--Simons correction is the negative of the correction
to $\mu_I$. With the convention of
\eqref{eq:response-data:finite-transport}, the corrected value is
\begin{equation}
\label{eq:tmf-comparison:finite-correction}
 +\frac12\xi(D^{\LC}_{Z,\Delta_{\C}})
   -\int_Z T_{I_{12}}(A_{\LC},A_T)
   =-\mu_*(Z;A_T)\pmod\Z.
\end{equation}
The raw eta value and this single finite correction are separate
steps; the stable connection is represented by $\mu_*$.
The selected calibration then gives
\begin{equation}
\label{eq:tmf-comparison:pairing-value}
 ((x,\sigma[M,s_L]))=-\mu_*(Z)
     =-\frac23=\frac13\quad\text{in }\R/\Z,
 \qquad e^{2\pi i((x,\sigma[M,s_L]))}=e^{2\pi i/3}.
\end{equation}
For these data it agrees with the normalized scalar
$\phi=-a_{N,*}-\mu_*$ because $a_{N,*}=0$. This equality uses
the selected filling and its finite corrections, as proved by
Li--Sun \cite[Proposition~10.4]{C1Working}. The generally nonzero
local term of a different differential filling has not been
discarded, and this calculation does not supply a general
comparison formula with that term present.

Two source qualifications matter here. Tachikawa--Yamashita
version~3, equation~(3.95), prints reduction
modulo $\Z((q))$ while displaying a target with period
$\kappa_{n+1}\Z((q))$. In the present real degree,
$\kappa_{12}=2$ by their equation~(3.75); the interpretation above
retains that real period and the corresponding APS parity
\cite[(3.75), (3.95)]{TY}. Also, their Proposition~4.59 assumes
cohomological degree
$d\equiv4\pmod8$, which excludes degree~32. The present
comparison uses the underlying differential relative construction,
the applicable closed formula and the explicit zero-term
calculation, rather than extending that proposition's range.
The suspension $-20$ fixes the relative-cycle dimension as
$32-20-1=11$ and the curvature and bordism degree as twelve.
The general boundary Witten-eta assertion, Conjecture~3.108 in
that version, is unused \cite[Proposition~4.59 and
Conjecture~3.108]{TY}. Li--Sun make these delimitations explicit
\cite[Remark~10.5]{C1Working}.

\subsection{Restricted Tate detection and the order-three class}

The result can identify a further assigned VOSA class because
equality of equivariant Tate images determines the evaluation on
this particular relative cycle. The selected twisted String
nullbordism is essential. It permits the underlying $i_*z$ to
come from a closed spin/gauge class through $j$; an arbitrary
relative cycle need not have this property.

Let $\alpha_{\tTw}$ denote the twisted counterpart of the relative
Anderson map defined above. For two integral lifts with the same
equivariant Tate image, let $v$ be their difference and
$\gamma=\alpha_{\tTw}(v)\in\IZ^{12}(C_{\tTw})$. Compatibility
with the Tate map gives $j^*\gamma=0$. To see why $\gamma$ is
torsion, first consider the rational characteristic ring
$\Q[p_1,p_2,\ldots,c]$ of $B_{\tTw}$. The twisted String
condition gives its quotient by $p_1/2+2c$ for $S_{\tTw}$.
In the rational model of the homotopy fibre, the added
degree-three generator has differential $p_1/2+2c$.
Since this polynomial is a non-zero-divisor, the quotient has
only even-degree cohomology, and the map from the spin/gauge
ring onto it is surjective. Finite-type rational duality makes
the map on even homology injective. The cofiber sequence then
implies that $\pi_{11}B_{\tTw}$ and $\pi_{11}C_{\tTw}$ are
finite torsion groups. It also gives an injection
\[
 H^{12}(C_{\tTw};\Q)\longrightarrow
 H^{12}(B_{\tTw};\Q),
\]
because $H^{11}(S_{\tTw};\Q)=0$.
Thus $j^*\gamma=0$ forces the rational image of $\gamma$ to
vanish. The Anderson universal coefficient sequence
\begin{equation}
\label{eq:tmf-comparison:universal-coefficients}
 0\longrightarrow\operatorname{Ext}^1(\pi_{11}C_{\tTw},\Z)
 \longrightarrow\IZ^{12}(C_{\tTw})
 \longrightarrow\operatorname{Hom}(\pi_{12}C_{\tTw},\Z)
 \longrightarrow0
\end{equation}
places $\gamma$ in the finite torsion term on the left, while
retaining the integer-valued homomorphisms on the right. This
identifies the part of Anderson duality that can contribute to
the next evaluation. The full
rational and universal coefficient proof is given by Li--Sun
\cite[Lemma~10.6]{C1Working}.

Since the selected nullbordism gives $i_*z=j_*w$ for
$w\in\pi_{11}B_{\tTw}$, naturality of the torsion pairing gives
\begin{equation}
\label{eq:tmf-comparison:restricted-detection}
 \langle\gamma,i_*z\rangle
  =\langle\gamma,j_*w\rangle
  =\langle j^*\gamma,w\rangle=0.
\end{equation}
Thus the two lifts have equal relative evaluations on the
selected class. This is a statement about one pairing with its
nullbordism hypothesis, not a claim that the Tate character is
injective or determines every relative anomaly evaluation.

Tachikawa--Zhang \cite[(A.3)--(A.5)]{TZ}, and
Tachikawa--Yamashita \cite[Proposition~3.49 and Remark~3.56]{TY},
identify
\begin{equation}
\label{eq:tmf-comparison:torsion-groups}
 A_{10}=A_{-32}=\Z/3,
 \qquad ((-,-)):A_{-32}\times A_{10}\longrightarrow\Q/\Z
\end{equation}
as a perfect pairing, with the Lie-framed $\Sp(2)$ image
generating $A_{10}$ \cite{TZ,TY}. The nonzero value
\eqref{eq:tmf-comparison:pairing-value} therefore implies that
$x$ has exact order three. If the additional $u_V$ in
Hypothesis~\ref{hyp:tmf-comparison:lifts} is supplied, restriction
to the basepoint places $i^*u_V$ in $A_{-32}$. Restricted
Tate detection gives it the same pairing with the generator;
perfectness then implies
\begin{equation}
\label{eq:tmf-comparison:vosa-restriction}
 i^*u_V=x_D-x_E=x.
\end{equation}
Here $i^*$ denotes restriction along $\mathrm{pt}\to BK$;
the map $i$ in \eqref{eq:tmf-comparison:relative-decomposition}
is the induced cofiber map. This is the conditional conclusion
of Li--Sun \cite[Corollary~10.7]{C1Working}.

The conditional branch thus gives the calibrated third-root
response a precise role as a torsion detector. It tests supplied
integral assignments using a known perfect pairing, without
constructing those assignments or proving an equivalence of
supersymmetric quantum field theories with $\TMF$. Its output
is compatible with the published heterotic scalar comparison,
while its premises remain the separate integral lifts, real
characters and selected differential filling specified here.

\section{Consequences for global background dependence}
\label{sec:conclusion}

On the ten-manifold $M=S(V)$, the normalized Green--Schwarz response
depends on the complete background, including its integral differential
String datum.
The local anomaly polynomial fixes the curvature dependence, while
the closed value $\mathcal O(M)=\exp(2\pi i\phi(M))$ also uses
the real fermion normalization, Spin lifts, boundary frames and
differential identifications described in Section~\ref{sec:response-data}.
The sphere-bundle formula derived by Li--Sun~\cite{C1Working}
expresses this dependence through a relative characteristic pairing
on the base filling. It makes the integral datum visible in the
calculation and determines which changes of that datum the response
can distinguish.

In the base-filling notation of Section~\ref{sec:relative-invariant},
$W$ is a spin eight-manifold filling the seven-dimensional base $B$.
The forms $c_W$ and $\lambda_W$ are the extended gauge and tangent
characteristic forms, and $h$ is the three-form component
of the boundary trivialization. The disk construction and the
twelve-dimensional Atiyah--Patodi--Singer (APS) calculation give
\begin{equation}
 \label{eq:conclusion:relative-response}
 \phi(S(V))
 =\frac13\left[\int_Wc_W(\lambda_W+c_W)-\int_Bhc\right]
 =\frac{\Krel}{3}\pmod{\Z}.
\end{equation}
The local contribution $a_N=\int_Bhc/3$ is generally nonzero.
Let $z$ be the integral relative class fixed by the boundary
nullhomotopy. Combining the disk and closed eta contributions with
the sign $-a_N-\mu$ evaluates the expression in brackets as the integral pairing
$\langle c_W\smile z,[W,B]\rangle$. Finite tangent-connection
transgressions cancel between the two contributions when the same
differential datum and its boundary identifications are transported.
The even real APS index removes the half-index ambiguity, with the
kernel terms retained. Changing the base filling changes $\Krel$
by a multiple of $24$, so the response is unchanged. Thus the auxiliary
spin/gauge twelve-manifold evaluates the spectral contribution while
the boundary nullhomotopy supplies the relative correction. The APS
filling itself admits no $\mathcal H$-trivialization, since $X_4$
has integral $2$ on its four-sphere fibre. The phase retains only
$\Krel$ modulo three; it does not recover the complete background
from that residue.

The relation to the Crowley--Goette invariant identifies this retained
information with an established secondary invariant on exactly the
domain $H^3(B;\Q)=H^4(B;\Q)=0$. There the rational relative lift is
unique, and the same product gives $t_B(V)=\Krel/24$, hence
$\phi=8t_B(V)$~\cite{CG11,C1Working}. Simple connectedness plays
no role in this comparison. Outside that domain, the relative formula
continues to evaluate the scalar and makes its dependence on the
integral trivialization explicit. It therefore describes more
backgrounds than the comparison with $t_B(V)$ alone.

The examples distinguish calibration from background dependence.
For the $\Sp(2)$ example, the filling built from the actual quaternionic
line gives $\Krel=-2$ and recovers the known response $e^{2\pi i/3}$ after the
specified Lie-frame and connection transport~\cite{TZ,C1Working}.
Inversion on $\Sp(2)$ pulls the right Lie framing back to the left
one; the subsequent differential String cylinders retain the endpoint
identifications and trivial boundary gauges. These identifications
equate the scalar with that of the special right-framed datum of
period $+1$; the general outward fibre period remains $-2$.
In the fixed-ten-manifold example of
Section~\ref{sec:string-dependence}, changes pulled back from the
base leave this latter period unchanged and give $\phi_n=2n/3$
modulo $\Z$. For $n=0,1,2$ the phases are distinct, while the
integral String classes remain distinct even when $n$ changes by
three and the phase repeats. These response values concern different
background data on one geometry, rather than a count of vacua.
The example lies outside the Crowley--Goette domain and exhibits the
information a fibre period fails to specify.

The reduced six-dimensional virtual fermion system returns the same
closed response on backgrounds trivializing $\lambda+c$. Its primitive
order-three value makes $3\mid r$ the exact condition for $r$ copies,
with $r$ a nonnegative integer, to have trivial closed response on
every permitted background. The
Lee--Tachikawa primitive normalization explains this divisibility;
the matching integral lattice factorization uses the
Monnier--Moore and Monnier--Moore--Park conventions~\cite{LT21,MM18,MMP18}.
With primitive $c$, integral gauge coefficient and characteristic
gravitational coefficient, that factorization exists precisely for
multiples of three. The density comparison preserves the different
background trivializations: the reduced response uses $\lambda+c$,
whereas Lee--Tachikawa trivialize $c$. The lattice identity answers the
universal coefficient-factorization problem in
Section~\ref{sec:six-dimensional}; it supplies neither the quantum
tensor fields nor the remaining constraints of an interacting
supergravity theory.

The published ordered heterotic comparison matches actual closed real
fermion operators within its fermion and Green--Schwarz section
prescription, including the prescribed real gluing and exclusion of an
independent invertible sector~\cite{TZ}. Section~\ref{sec:heterotic-comparison}
explains how those inputs relate its scalar to the response computed
here. Equality of the response is obtained with that prescription;
the scalar alone does not identify complete theories.

The integral application to topological modular forms and vertex
operator superalgebras has the separate premises of
Hypothesis~\ref{hyp:tmf-comparison:lifts}: compatible integral lifts
and real equivariant characters with twist $\tTw=-2c$.
For the selected left Lie-framed $\Sp(2)$ datum and its
zero-counterterm filling, the relative pairing in
Section~\ref{sec:tmf-comparison} is $1/3$. The known perfect torsion
pairing then makes the difference of the restricted classes of the
two rank-sixteen lattice inputs an element of exact order three.
This detects the class under
the stated lift hypotheses; it does not construct those lifts from
their characters. The role of the calibrated response in this branch
is consequently a test of supplied integral assignments, with the
selected differential filling retained as part of the test.

\section*{Acknowledgments}
\label{sec:acknowledgments}

We thank Yuji Tachikawa for helpful correspondence on the surrounding TMF program and generously providing his lecture notes. The initial exploration, source synthesis, and drafting of this manuscript were assisted by OpenAI's GPT-5.6, GPT-6 and GPT-6.1 under the authors' direction. The authors take responsibility for the mathematical claims and the physical exposition. 
\appendix
\section{Differential trivializations and relative products}
\label{app:differential-data}

The boundary term $\int_B h\wedge c$ in the response is tied to an
integral nullhomotopy, rather than an arbitrary solution of
$dh=\lambda+c$. This appendix explains the cochain information that
makes that distinction precise. It supports the background definition
in Section~\ref{sec:response-data}, the induced String datum in
Section~\ref{sec:disk-construction}, and the integer pairing in
Section~\ref{sec:relative-invariant}. We use the differential-cocycle
model and product of Redden \cite[\S\S3.1--3.2, 3.4]{Redden}.
For a quaternionic line $V$ over a closed spin seven-manifold $B$,
with underlying complex rank-two bundle $U$, these constructions
give the differential and relative data of Li and Sun's
sphere-bundle response \cite[\S\S2--3, 5]{C1Working}. The conventions
are $\lambda=p_1/2$, $c=c_2(U)$, and the quaternionic real
orientation $(1,i,j,k)$, for which $\lambda(V_{\R})=-c$.
No rational-cohomology vanishing is needed below.

\subsection{Cocycles, trivializations, and their differences}

Let $\mathcal I$ send a differential form to the real singular cochain
obtained by integration. Write $\delta$ for the singular-cochain
coboundary and $d$ for the exterior derivative on forms; a check
accent denotes a differential cocycle or its trivialization.
A degree-four differential cocycle on a
manifold $L$ has components
\[
 \check X=(x_{\Z},x_{\R},X),\qquad
 \delta x_{\Z}=0,\qquad
 \delta x_{\R}=\mathcal I X-x_{\Z},\qquad dX=0.
\]
Here $x_{\Z}$ is an integral four-cochain, $x_{\R}$ is a real
three-cochain, and $X$ is its curvature four-form. Integral cochains
are included in real cochains where these expressions are added.
The characteristic class is $[x_{\Z}]$; the real-cochain relation
records its agreement with the de Rham class of $X$. It does not
identify the whole differential cocycle with either component alone.

A geometric trivialization is a triple $\check s=(b,k,H)$ satisfying
\begin{equation}
 \delta b=x_{\Z},\qquad
 \mathcal I H-b-\delta k=x_{\R},\qquad dH=X.
 \label{eq:differential-data:trivialization}
\end{equation}
The degrees of $b,k,H$ are respectively three, two, and three, with
$b$ integral and $k$ real. These are Redden's trivialization equations
\cite[Propositions~3.11--3.12]{Redden}. The triple belongs to the
differential cochain complex whose form components begin in degree
three; the complex defining degree-four differential characters would
have no three-form component in degree three. This distinction is
what permits nonzero curvature
$X$: a geometric trivialization here refines the integral nullhomotopy
and has a connection form $H$ whose derivative is $X$. It is not a
connection-preserving identification of a nonflat cocycle with the
zero differential character.

Existence requires the integral characteristic class to vanish.
Vanishing only in real cohomology is weaker, and can leave an integral
obstruction. Once a trivialization exists, its isomorphism classes
form a torsor for the degree-three differential cohomology group
$\widehat H^3(L;\Z)$, while integral
trivializations form a torsor for $H^3(L;\Z)$
\cite[Theorem~3.15]{Redden}. A torsor means that any two choices have
a difference in the indicated group, but there is no preferred zero
choice. Forgetting geometric refinement is equivariant for
$\widehat H^3(L;\Z)\to H^3(L;\Z)$. Thus a shift of the integral
trivialization is part of the background data; it is not made into an
isomorphism by solving the same curvature equation again. Torsion
shifts are also retained. A torsion element of $H^3(L;\Z)$ lifts to
a flat degree-three differential character through the Bockstein
map for the coefficient sequence $\Z\hookrightarrow\R\to\R/\Z$.
It can therefore change the integral choice without changing $H$
\cite[Proposition~2.2, equations (2.4)--(2.5)]{Redden}.

Even within a fixed integral choice, the form is constrained by the
middle equation of \eqref{eq:differential-data:trivialization}. On a
closed three-cycle $\sigma$ it gives
\[
 \int_\sigma H=x_{\R}(\sigma)+b(\sigma),
 \qquad b(\sigma)\in\Z.
\]
The real two-cochain drops out because $\delta k(\sigma)=0$.
Consequently the form periods have to refine the specified holonomy
and integral nullhomotopy. Merely prescribing $dH$ does not impose
this compatibility.

Morphisms distinguish a change of representative from a change of
datum. In this model a morphism represented by an integral
two-cochain $a$ and real one-cochain $\ell$ changes a trivialization
by
\[
 (b,k,H)\longmapsto(b+\delta a,\ k-a-\delta\ell,\ H).
\]
This preserves all three equations. Likewise a specified morphism
between representing degree-four cocycles induces an identification
of their trivialization categories. That identification depends on
the morphism, not just on the equality of differential-cohomology
classes \cite[Proposition~3.17]{Redden}. Boundary gluing therefore
requires the actual identifications used in the construction.

\subsection{Extending a prescribed boundary refinement}

The torsor structure also explains the geometric extension used in
closed cancellation in Section~\ref{sec:response-data}. Here an
$\mathcal H$-structure consists of the Spin and gauge bundles with
an integral nullhomotopy of $\lambda_T-\lambda_E-\lambda_{E'}$,
where the subscripts denote the tangent bundle and the two gauge
bundles of Section~\ref{sec:response-data}. Suppose the
integral $\mathcal H$-structure of a closed eleven-manifold $C$ has
already been extended over a twelve-dimensional bordism
$Q_{\mathcal H}$, including its boundary identification. Extend the
metric and gauge connections with product collars. Surjectivity of
the geometric-to-integral torsor map gives a geometric refinement of
that prescribed bulk integral trivialization
\cite[Theorem~3.15]{Redden}.

Its boundary value need not initially equal the prescribed geometric
refinement on $C$. Their difference has zero characteristic class,
because the integral boundary identification was already fixed.
The exact sequence
\begin{equation}
 \Omega^2(C)/\Omega^2_{\Z}(C)\longrightarrow
 \widehat H^3(C;\Z)\longrightarrow H^3(C;\Z)\longrightarrow0
 \label{eq:differential-data:zero-characteristic}
\end{equation}
represents this difference by a global two-form $\beta$;
$\Omega^2_{\Z}$ denotes closed two-forms with integral periods.
The character represented by $(0,\mathcal I\beta,d\beta)$ has
zero integral component and curvature $d\beta$. Extend $\beta$
using a collar cutoff, and act on the bulk refinement by this
character. Its form component adds $d\beta$ on the boundary,
while its real-cochain component supplies the matching change in
the middle trivialization equation. This corrects the boundary
value without changing the integral class. The specified boundary
isomorphism and its cochain representative extend along the collar,
so the trivialization equations hold with the prescribed boundary
identification
\cite[\S2, Lemma ``Extension of the prescribed refinement'']{C1Working}.

The integral extension comes first. No surjectivity of
$H^3(Q_{\mathcal H};\Z)\to H^3(C;\Z)$ is asserted or required.
Nor does this argument turn the spin/gauge twelve-filling
$\mathcal Q$ of Section~\ref{sec:aps-evaluation} into an
$\mathcal H$-filling: those are distinct constructions with distinct
premises.

\subsection{Characteristic morphisms and the tautological boundary frame}

For the disk $N=D(V)$ and Spin bundles $F_1,F_2$, the identities
\begin{equation}
 \check\lambda(V_{\R})\simeq-\check c_2(U),\qquad
 \check\lambda(F_1\oplus F_2)\simeq
 \check\lambda(F_1)+\check\lambda(F_2)
 \label{eq:differential-data:characteristic-morphisms}
\end{equation}
are used with specified characteristic morphisms. Their integral
classes and curvature polynomials agree. At the universal level,
possible flat differences lie in degree-three $\R/\Z$ cohomology
of the relevant products of $B\Spin$ and $B\Sp(1)$, which
vanishes. Choose the unit morphism on framed trivial bundles and
use its pullbacks and compositions on $N$
\cite[\S3.2]{C1Working}. This normalization specifies the morphisms
in the disk construction; it does not assert uniqueness of arbitrary
morphisms on the base $B$.

Write $\pi:N\to B$ for the disk projection and $P\to B$ for the
principal right $\Sp(1)$ bundle defining $V$. The radial gauge
connection $A$ is trivial in the tautological frame on a boundary
collar, while $A_*$ is the stable tangent connection with product
collar geometry. Their connection changes in
Section~\ref{sec:disk-construction} have full transgression cochains
$\check T,\check K$. Write $\check\lambda_B=\check\lambda(TB)$
and let $\check c$ use the base gauge connection. With
$\check q=\check c_2(\pi^*P,A)$ and
$\check q_0=\pi^*\check c$, their equations are
\begin{equation}
 d\check T=\check q-\check q_0,\qquad
 d\check K=\check\lambda(A_*)-
                 \pi^*(\check\lambda_B-\check c).
 \label{eq:differential-data:transgressions}
\end{equation}
The differential here acts on integral, real, and form components.
For a quadratic invariant polynomial $\mathcal P$, normalized so
that its characteristic form is $\mathcal P(F_A,F_A)$, a smooth
path $A_t$ on the same actual bundle gives the form transgression
\begin{align*}
 T_{\mathcal P}(A_0,A_1)
   &=2\int_0^1\mathcal P(\dot A_t,F_{A_t})\,dt,\\
 dT_{\mathcal P}&=\mathcal P(F_{A_1},F_{A_1})
                 -\mathcal P(F_{A_0},F_{A_0}).
\end{align*}
Here $F_{A_t}$ is the curvature of $A_t$.
Invariance and the Bianchi identity turn the derivative of the
characteristic form into the exterior derivative of the integrand;
integration fixes the endpoint sign. This is the finite form change
underlying the Chern--Simons variation formula
\cite[Proposition~3.8]{CS}. A connection cylinder, with the chosen
endpoint cocycles and their characteristic morphisms, lifts it to a
cochain homotopy with the sign in
\eqref{eq:differential-data:transgressions}; the differential Stokes
formula retains its two endpoints \cite[equation (3.27)]{Redden}.
The form transgression alone does not specify its integral and real
components. Those components and the endpoint morphisms are retained
in $\check T$ and $\check K$. These homotopies are not isomorphisms
between differential characters with unequal curvature.

The base trivialization obeys
$d\check h=\check\lambda_B+\check c$.
Adding these equations to its pullback gives
the disk refinement already used in the local calculation,
\[
 \check H_N=\pi^*\check h+2\check T+\check K,
 \qquad d\check H_N=\check\lambda(A_*)+2\check q.
\]
The unit vector on $M=S(V)$ gives the tautological gauge frame
$\tau$. It supplies a specified morphism $\check\gamma_\tau$
with $d\check\gamma_\tau=\check q|_M$. Its three-form is zero
in the normalized framed flat model, but its cochain components
still identify that gauge datum with the trivial one. Using the
product Spin identification to split off the outward normal, the
induced String trivialization is
\begin{equation}
 \check s_M=\check H_N|_M-2\check\gamma_\tau,
 \qquad d\check s_M=\check\lambda(TM).
 \label{eq:differential-data:induced-String}
\end{equation}
The factor two removes the two quaternionic gauge summands with
their chosen frames. The resulting equation includes their
identifications, even though the boundary gauge curvature vanishes.
Discarding $\check\gamma_\tau$ because its three-form vanishes
would discard part of the induced background.

\subsection{The relative cone and the integral product}

Let $W$ be the spin eight-dimensional filling of
Section~\ref{sec:relative-invariant}, with $\partial W=B$ and
actual quaternionic extension $\widetilde V$. Give $B$ the outward
boundary orientation. Put
$\lambda_W=\lambda(TW)$ and $c_W=c_2(\widetilde U)$, where
$\widetilde U$ is the underlying complex rank-two bundle, and
$X=\lambda_W+c_W$, using the specified extending connections.
Write its bulk differential cocycle as
$\check X=(x_{\Z},x_{\R},X)$. For $i:B\hookrightarrow W$,
the relative cochain complex consists of pairs with differential
\begin{equation}
 d_{\rm rel}(u,v)=(du,i^*u-dv).
 \label{eq:differential-data:cone}
\end{equation}
The symbols $d$ in this formula mean $\delta$ for singular
cochains and the exterior derivative for forms. The integral
boundary nullhomotopy $b$ makes $(x_{\Z},b)$ a relative
cocycle. Its class is the specified
$z\in H^4(W,B;\Z)$, with $jz=\lambda_W+c_W$; here $j$
is the map to $H^4(W;\Z)$ that forgets the boundary trivialization.

The boundary refinement $(b,k,h)$ obeys
\[
 \delta b=i^*x_{\Z},\qquad
 \mathcal I h-b-\delta k=i^*x_{\R},\qquad
 dh=i^*X.
\]
Together with the bulk cocycle equation these imply
\begin{align}
 (\mathcal I X,\mathcal I h)-(x_{\Z},b)
 &=d_{\rm rel}(x_{\R},-k)\nonumber\\
 &=(\delta x_{\R},i^*x_{\R}+\delta k).
 \label{eq:differential-data:real-image}
\end{align}
The minus sign on $k$ is forced by the second component of the
cone differential. Thus $(X,h)$ is a relative de Rham representative
of the real image of this particular integral class, rather than a
real lift chosen independently of $b$.

Multiplication must respect the comparison between forms and
singular cochains. Write the differential cocycle for $c_W$ as
$\check c_W=(c_{\Z},c_{\R},c_W)$. In the degree-four product
used here, Redden's formula includes a real-cochain correction
\begin{equation}
 \check X\cdot\check c_W=
 \bigl(x_{\Z}\smile c_{\Z},\,
 x_{\Z}\smile c_{\R}+x_{\R}\smile\mathcal I c_W
       +\mathcal B(X,c_W),\,X\wedge c_W\bigr).
 \label{eq:differential-data:product}
\end{equation}
Here $\mathcal B$ is the natural chain homotopy comparing cup
products of integrated forms with integration of wedge products
\cite[\S3.4, equation (3.21)]{Redden}. Integration of
forms is not a strict equality between these two products at the
cochain level. The correction makes the differential satisfy the
graded product rule.

In particular, put $c_{\Z,B}=i^*c_{\Z}$ and
$c_{\R,B}=i^*c_{\R}$. The boundary product has all three
components
\[
 (b,k,h)\cdot\check c_W|_B
 =\bigl(b\smile c_{\Z,B},\,
        -b\smile c_{\R,B}+k\smile\mathcal I c
                 +\mathcal B(h,c),\,h\wedge c\bigr).
\]
The minus sign in its real component comes from the odd degree
of the trivialization. Applying the graded product rule and
$d\check c_W=0$ shows that its differential is the boundary
restriction of \eqref{eq:differential-data:product}; this is Redden's
compatibility of trivializations with products
\cite[Proposition~3.23]{Redden}. The real component therefore
participates in the comparison between the integral product and its
form representative. After passage to relative
cohomology, this supplies the product of $z$ with the absolute
integral class $c_2(\widetilde U)$.

Relative integration uses the same boundary sign as
\eqref{eq:differential-data:cone}:
$(\alpha,\beta)$ pairs with the relative fundamental class $[W,B]$ as
$\int_W\alpha-\int_B\beta$. Indeed a relative exact pair
$(du,i^*u-dv)$ has zero integral by Stokes and the fact that $B$
is closed. Applying this convention to the product gives
\begin{equation}
 \Krel=\left\langle c_2(\widetilde U)\smile z,[W,B]\right\rangle
 =\int_W c_W(\lambda_W+c_W)-\int_B h\wedge c
 \in\Z.
 \label{eq:differential-data:K}
\end{equation}
The displayed integrality comes from the integral relative
cup-product pairing. The form identity computes its real image;
it does not supply integrality on its own. No extension of
$\check h$ as a trivialization over $W$ is required.

If one retained only $dh=\lambda+c$, the pair $(X,h)$ would be
closed in the relative form complex, but its identification with
$(x_{\Z},b)$ in \eqref{eq:differential-data:real-image} would be
missing. Replacing $h$ by $h+\rho$ for a closed real three-form
preserves that curvature equation and changes the displayed
relative integral by $-\int_B\rho\wedge c$. There is no integral
quantization condition on such an arbitrary real change. In the
actual background it is the full trivialization equations and the
specified morphisms that control which changes are allowed.

Equation~\eqref{eq:differential-data:K} is the integer entering
\eqref{eq:relative-invariant:K} and the response
$\phi(S(V),\check s_M)=\Krel/3$ in
\eqref{eq:relative-invariant:response}. Its boundary subtraction
contains exactly the $h\wedge c$ term whose local contribution is
$a_N=\int_B h\wedge c/3$ in
Section~\ref{sec:disk-construction}. This returns the cochain
construction to the scalar with its fixed sign $\phi=-a-\mu$,
where $a=a_N$ and $\mu$ is the closed real eta contribution
defined in Section~\ref{sec:response-data}.
The rational lift and factor $1/24$ needed for the Crowley--Goette
comparison are separate steps in
Section~\ref{sec:relative-invariant}; they do not replace the
integral datum used here.

\section{Real eta invariants and Pfaffian conventions}
\label{app:real-eta-conventions}

The real fermion coefficient accounts for the factor $1/2$ in the
anomaly polynomial. The sign of the closed eta phase additionally
depends on the chirality and the anomaly line, while first-slot
conjugation in the Hermitian reference pairing determines the local
sign in $\phi=-a-\mu$. This appendix supplies
the character expansion and Atiyah--Patodi--Singer (APS) spectral
conventions used in Sections~\ref{sec:response-data}
and~\ref{sec:aps-evaluation}, then
translates them into the closed-operator comparison of
Section~\ref{sec:heterotic-comparison}. We retain actual real bundles,
their Spin lifts and connections, product collars, and kernel terms.
The full real anomaly polynomial, real APS period and closed
fermion-phase comparison follow the mathematical companion
\cite[\S2, \S9]{C1Working}.
The real boundary-state construction below fixes its spectral-cut
transitions, graded duality and actual seam maps, thereby supplying
the contraction used by the normalized section and reference pairing.

\subsection{The vector and half-spin characters}

For a rank-sixteen real Spin bundle $E$, write
$p=p_1(E)$, $q=p_2(E)$ and $r=p_3(E)$, and use primes for $E'$.
If $x_1,\ldots,x_8$ are its formal Pontryagin roots, the vector
character is $\sum_i(e^{x_i}+e^{-x_i})$. The half-spin character is
$128\prod_i\cosh(x_i/2)$ plus a chirality-dependent product of
hyperbolic sines. Each $x_i$ has form degree two, so the latter product
first contributes in degree sixteen. Consequently the choice of
half-spin chirality does not alter the following degree-twelve
expansion, although its actual representation is fixed by the gauge
embedding:
\begin{align}
 \ch(E_{\C})
 &=16+p+\frac{p^2-2q}{12}
       +\frac{p^3-3pq+3r}{360},
 \label{eq:real-eta-conventions:vector-character}\\
 \ch(S^+_{\C}(E))
 &=128+16p+\frac{p^2+4q}{3}
       +\frac{p^3}{360}+\frac{pq}{30}+\frac{2r}{15}.
 \label{eq:real-eta-conventions:spin-character}
\end{align}
Here and below character equalities are truncated through degree twelve.
The roots use the characteristic-class normalization of
Section~\ref{sec:response-data}; in Tachikawa--Zhang's curvature notation,
$1/(2\pi)$ is absorbed into $F$ and $\ch=\tr e^{iF}$
\cite[\S2, \S4.3]{TZ}.

Multiply the two vector characters and subtract the two half-spin
characters, as required by
$\Delta_{\R}=E\otimes_{\R}E'-S^+_{\R}(E)-S^+_{\R}(E')$.
The degree-zero cancellation is $256-128-128=0$, and the degree-four
cancellation is $16(p+p')-16p-16p'=0$. In degree eight the remaining
terms are
\[
 \ch_8(\Delta_{\C})=p^2+pp'+p'^2-4q-4q'=4S,
 \qquad S=n^2+nn'+n'^2-q-q',
\]
where $n=p/2$ and $n'=p'/2$. In degree twelve the $r,r'$ terms
cancel, while the other terms give
\begin{align}
 \ch_{12}(\Delta_{\C})
 &=\frac{p^3+p'^3}{24}
   +\frac{p^2p'+pp'^2}{12}
   -\frac{(p+p')(q+q')}{6}
 \nonumber\\
 &=\frac{n+n'}{3}S.
 \label{eq:real-eta-conventions:virtual-character}
\end{align}
Because the character has no degree-zero or degree-four term, only
$\widehat A_4=-p_1(T)/24=-\lambda(T)/12$ contributes alongside
$\ch_{12}$. Thus
\begin{equation}
 -\frac12[\widehat A(T)\ch(\Delta_{\C})]_{12}
 =-\frac12\left(\frac{n+n'-\lambda(T)}{3}S\right)
 =X_4Y_8,
 \label{eq:real-eta-conventions:full-polynomial}
\end{equation}
with $X_4=\lambda(T)-n-n'$ and $Y_8=S/6$. Both gauge $p_2$
terms survive. In particular, replacing $Y_8$ by a polynomial involving
only $n,n'$ would change the coefficient system being evaluated.

For the sphere-bundle gauge $E=2V_{\R}\oplus\R^8$, let
$c=c_2(U)$ for the underlying complex rank-two bundle $U$ of $V$.
Since $p(V_{\R})=1-2c+c^2$, squaring this total Pontryagin class
gives $n=-2c$ and $p_2(E)=6c^2$. With $E'$ trivial,
$Y_8=-c^2/3$. This is the specialization used on the disk and its
closed double. A single copy $V_{\R}\oplus\R^{12}$ instead has
$n=-c$, $p_2=c^2$, and $Y_8=0$; it cannot supply the same density.
This distinction is also explained by the mathematical companion's
comparison with the published specialization
\cite[\S4.3, equation (77)]{TZ}.

\subsection{Real representatives and the APS period}

For an actual real twist $R$ on a closed spin eleven-manifold $Z$, let
$D_{Z,R_{\C}}$ be the ordinary self-adjoint Levi--Civita Dirac
operator and set
\begin{equation}
 h_R=\dim_{\C}\ker D_{Z,R_{\C}},\qquad
 \xi_R=\frac{\eta_{\rm APS}(D_{Z,R_{\C}})+h_R}{2}.
 \label{eq:real-eta-conventions:reduced-eta}
\end{equation}
The APS eta invariant excludes zero eigenvalues. For a virtual real
coefficient, add or subtract these real-valued representatives, including
their kernel dimensions, before defining
\begin{equation}
 \mu_I(Z)=-\frac12\xi_{\Delta}(Z)\pmod\Z.
 \label{eq:real-eta-conventions:real-value}
\end{equation}
This construction does not divide an unspecified element of $\R/\Z$
by two. Such a division would leave two possible answers; the real
spectral system supplies the refinement that selects the phase.

In dimension $11\equiv3\pmod8$, a real-twisted Spin operator has a
compatible antilinear symmetry whose square is $-1$ and which commutes
with the operator. Its eigenspaces, including the kernel, therefore have
even complex dimension. Freed--Moore construct the resulting half-eta
function in Proposition~1.28 \cite{FM}; Witten--Yonekura display the
antilinear symmetry in equations (2.47)--(2.51) \cite{WY}. At a crossing
of a nonzero eigenvalue through zero the eta change has the even
multiplicity required to preserve the real phase. At the crossing itself
the kernel contribution in \eqref{eq:real-eta-conventions:reduced-eta}
is essential.

Orientation reversal sends the closed operator to $-D$ while preserving
its kernel. Hence
\begin{equation}
 \xi(-D)=-\xi(D)+h_R,\qquad
 \mu_I(-Z)=-\mu_I(Z)\pmod\Z,
 \label{eq:real-eta-conventions:orientation}
\end{equation}
where the second identity uses even $h_R$ and extends by virtual
subtraction. This is a statement about closed real-twisted operators;
boundary-line reversal additionally involves the graded identifications
discussed below.

For a compact spin twelve-manifold $Q$ with the actual real twists
extended and product metric and connection near its boundary, twisted
APS gives
\begin{equation}
 \xi_{\Delta}(\partial Q)
 =\int_Q[\widehat A(TQ)\ch(\Delta_{\C})]_{12}
   -\Ind_{\rm APS}D^+_{\Delta_{\C}}.
 \label{eq:real-eta-conventions:aps}
\end{equation}
The index formula is \cite[Theorem (4.2), formula (4.3)]{APS}.
In dimension $12\equiv4\pmod8$, the real-twisted chiral Spin
representations have parallel quaternionic structures compatible with
the Dirac operator and the APS spectral projections, including their
zero modes. Both chiral APS kernels have even complex dimension; their
difference, and then its virtual subtraction, is even. Multiplication
of \eqref{eq:real-eta-conventions:aps} by $-1/2$ therefore yields
\begin{equation}
 \mu_I(\partial Q)=\int_Q I_{12}\pmod\Z.
 \label{eq:real-eta-conventions:period}
\end{equation}
This evenness is the reason the half-index ambiguity is integral
\cite[(3.8)--(3.9)]{WY}. Ordinary APS requires a spin/gauge filling,
not an extension of the differential $\mathcal H$ trivialization.
It thus applies to $\mathcal Q$ in Section~\ref{sec:aps-evaluation},
even though $\int_FX_4=2$ prevents that filling from being an
$\mathcal H$-filling.

\subsection{The boundary line and the two operator conventions}

Fix positive-index chirality by
$I(R)=\tfrac12[\widehat A\ch(R_{\C})]_{12}$, and denote its
boundary fermion Pfaffian line by $\mathcal P_R$. Freed--Moore's
real state has phase $\exp(+\pi i\xi_R)$ and lies in
$\mathcal P_R^*$ \cite[(3.11)]{FM}. The positive fermion functional
integral lies in $\mathcal P_R$ \cite[(4.33)--(4.34)]{FM}.
Only this line assignment is used from the latter passage, not its
subsequent conjectural M-theory cancellation. In Tachikawa--Zhang's
prescription the fermion functional integral and its anomaly state
lie in the same line
\cite[\S4.2]{TZ}. Accordingly its closed fermion-anomaly phase is
the dual of the Freed--Moore state:
\begin{equation}
 \chi_R(Z)=\exp(-\pi i\xi_R(Z)).
 \label{eq:real-eta-conventions:dual-phase}
\end{equation}

The subscripts $\mathrm{FM}$ and $\mathrm{WY}$ below distinguish the
Freed--Moore and Witten--Yonekura operator conventions, respectively.
Witten--Yonekura express inflow as $\exp(-i\pi\eta_{\rm WY}/2)$
\cite[(2.10), (2.52)]{WY}. Its apparent minus sign cannot be
compared with \eqref{eq:real-eta-conventions:dual-phase} without
identifying the operators. Use Freed--Moore's graded Clifford module
$\mathscr M=\mathscr M^0\oplus\mathscr M^1$, with geometric Dirac
operator $D=c\circ\nabla$. In odd dimension both graded spinor
summands are copies of the ungraded spinor bundle, tensored with the
coefficient $R_{\C}$. The operator $D$ exchanges the summands.
Here $c(v)$ denotes skew Clifford multiplication, distinct from the
characteristic class $c=c_2(U)$ above. Let $\omega$ be the normalized
odd volume action and $\omega_\partial$ the boundary volume action.
For the outward unit normal $\nu$ on the
product collar, their conventions give
\begin{equation}
 \omega=c(\nu)\omega_\partial,\qquad
 \omega^2=1,\qquad c(\nu)^2=\omega_\partial^2=-1.
 \label{eq:real-eta-conventions:collar-volume}
\end{equation}
The volume action commutes with $c(v)$. On the even Clifford-module
summand $\mathscr M^0$, define the Hermitian Witten--Yonekura matrices by
\begin{equation}
 \gamma_{\rm WY}(v)=i\omega c(v),\qquad
 \gamma_{\rm WY}(\nu)=-i\omega_\partial.
 \label{eq:real-eta-conventions:clifford-dictionary}
\end{equation}
They satisfy the positive Clifford relation. Witten--Yonekura's $+1$
normal mode therefore agrees with Freed--Moore's $+i$ boundary
subspace. Composing $D$ with $\omega$ returns to $\mathscr M^0$,
giving a self-adjoint operator on one copy of the twisted odd spinors.
The two closed odd operators have opposite signs:
\begin{equation}
 d_{\rm FM}=\omega D|_{\mathscr M^0},\qquad
 D_{\rm WY}=i\gamma_{\rm WY}^a\nabla_a=-d_{\rm FM}.
 \label{eq:real-eta-conventions:operator-dictionary}
\end{equation}
The modified tangential operator in Witten--Yonekura is also the
negative of Freed--Moore's; boundary grading intertwines these
tangential operators without
reversing the graded index. Thus agreement of the displayed boundary
chirality does not imply equality of the closed eta representatives.

Witten--Yonekura count zero modes with positive sign
\cite[(2.19) and following paragraph]{WY}. For
$h_R=\dim_{\C}\ker d_{\rm FM}$ their convention
therefore gives
\begin{align}
 \eta_{\rm WY}
 &=-\eta_{\rm APS}(d_{\rm FM})+h_R=-2\xi_R+2h_R,
 \nonumber\\
 \exp(-i\pi\eta_{\rm WY}/2)&=\exp(+\pi i\xi_R).
 \label{eq:real-eta-conventions:wy-kernels}
\end{align}
The last equality uses even $h_R$. Witten--Yonekura inflow is thus the
Freed--Moore inverse-line state, whose dual is
\eqref{eq:real-eta-conventions:dual-phase}. The familiar statement
$\eta_{\rm WY}=2\xi$ holds for Witten--Yonekura's own odd operator;
using it for $d_{\rm FM}$ would erase the sign in
\eqref{eq:real-eta-conventions:operator-dictionary}.

\subsection{The closed coefficient and finite tangent transport}

On the actual common cover $G=\Spin(16)\times\Spin(16)$, the
real associated bundles decompose as
\begin{align}
 \operatorname{ad}(\Spin(32))|_G
 &=\Lambda^2E\oplus(E\otimes_{\R}E')\oplus\Lambda^2E',
 \nonumber\\
 \operatorname{ad}(E_8\times E_8)|_G
 &=\Lambda^2E\oplus S^+_{\R}(E)
   \oplus\Lambda^2E'\oplus S^+_{\R}(E').
 \label{eq:real-eta-conventions:branching}
\end{align}
The first decomposition is the skew-matrix block decomposition. In the
second, the $112$ roots $\pm e_i\pm e_j$ and eight Cartan
directions give the $D_8$ adjoint; the other $128$ roots with
half-integral coordinates and even sign parity give the embedding's
real half-spin representation. Invariant inner products make these
equivariant decompositions orthogonal, and equivariance makes them
parallel for the induced connections. Consequently they identify
actual closed Dirac systems, including holonomy and kernels. Under the
fermion prescription of Section~\ref{sec:heterotic-comparison}, the
common adjoint, gravitational and ghost factors cancel with identical
inverse factors. The remaining coefficient is $\Delta_{\R}$ up to
the common gaugino chirality.

Equation~\eqref{eq:real-eta-conventions:full-polynomial} gives
$I(\Delta_{\R})=-X_4Y_8$, while the ordered published difference
has $+X_4Y_8$ \cite[(74)--(76)]{TZ}. This nonzero universal
polynomial fixes the remaining binary chirality to $-\Delta_{\R}$
after the operator system has been identified. It does not exclude an
independent flat anomaly sector; the stated physical prescription
excludes such an addition. Direct sums add eta invariants and kernel
dimensions, and virtual subtraction inverts their phases, so
\begin{equation}
 I_{\rm diff}(Z)=\chi_{-\Delta_{\R}}(Z)
 =\exp(+\pi i\xi_\Delta(Z))
 =\exp(-2\pi i\mu_I(Z)).
 \label{eq:real-eta-conventions:closed-comparison}
\end{equation}
This is a comparison of the prescribed closed systems, not an
equivalence of arbitrary anomaly theories or their families.

The tangent connection used in the form calculation may be a specified
stable metric Spin connection $A_*$ rather than Levi--Civita. Keep
the gauge connections fixed and put
\[
 \alpha=\CS_\lambda(A_{\LC},A_*),\qquad
 d\alpha=\lambda(A_*)-\lambda(A_{\LC}),\qquad
 H_*=H_{\LC}+\alpha.
\]
The last equality transports the existing differential trivialization
with its integral class and boundary morphism. Since $Y_8$ is
gauge-only, the full index transgression is $\alpha Y_8$. On
$Z=\overline N\cup_M W_{\rm ref}$, the collar paths match and
$Y_8$ vanishes on the trivial-gauge reference. Hence
\begin{align}
 \mu_*(Z)-\mu_{\LC}(Z)&=\int_Z\alpha Y_8
                         =-\int_N\alpha Y_8\pmod\Z,
 \nonumber\\
 a_*-a_{\LC}&=\int_N\alpha Y_8,
 \qquad (a+\mu)_*=(a+\mu)_{\LC}\pmod\Z.
 \label{eq:real-eta-conventions:transport}
\end{align}
The stable value is an ordinary spectral invariant plus a finite
transgression, not the spectrum of a proposed torsionful operator.
Neither term in the last equality is separately invariant. In
particular, changing only the eta sign produces $-a+\mu$, whereas
complex conjugation produces $a+\mu$; these agree only when
$2a\in\Z$, a condition absent from the general construction.

\subsection{Real boundary states and graded gluing}
\label{app:real-eta-conventions:real-gluing}

We now identify the real boundary state, including its phase and metric,
before using it in a contraction. Work first with an honest real
coefficient $R$ on a compact spin eleven-manifold $Y$ with product
boundary data. Write $A=D^{\partial}_{\rm FM}$ for the tangential
operator and $C$ for the parallel quaternionic structure. This $C$
is distinct from normal Clifford multiplication. On the boundary,
$C$ exchanges the $+i$ and $-i$ eigenspaces of
$\omega_\partial$; neither chiral space is itself quaternionic
\cite[\S3.1, (3.2)--(3.7)]{FM}.

\subsubsection{Allowed domains and real eta coordinates}

The product collar has
$D=c(\nu)(\partial_\nu-A)$. Substituting
\eqref{eq:real-eta-conventions:clifford-dictionary} into the
Witten--Yonekura tangential operator gives
$D_{W,\rm WY}=-A$, as well as $D_{\rm WY}=-d_{\rm FM}$.
Witten--Yonekura allow positive tangential eigenvalues
\cite[\S2.2, p.~11]{WY}; these are the negative eigenvalues of $A$.
Let $\psi_\pm$ denote the positive and negative chiral parts of
the boundary restriction. Writing $A^+$ for the map between these
chiral spaces, their allowed high modes therefore satisfy precisely
Freed--Moore's boundary condition
\[
 \psi_-+\frac{A^+}{|A|}\psi_+=0
 \qquad\text{on the nonzero high modes},
 \qquad |A|=(A^2)^{1/2}.
\]
This identifies the allowed subspace, rather than changing a sign on
the basis of a Hamiltonian description of the modes that are projected
out \cite[(3.10)]{FM}.

For the low modes choose an orthonormal positive chiral basis $e_i$
and the negative basis $f_i=Ce_i$. Witten--Yonekura's low-mode
condition \cite[(2.65), (2.75)]{WY} can be written using independent
coefficients $a^i,b_i$ as
\[
 \psi_+=\sum_i a^i e_i,\qquad
 \psi_-=-i\sum_i b_i f_i,\qquad
 b_i=\sum_jK_{ij}a^j,
 \quad K^T=-K,\quad K^\dagger K=1.
\]
The coefficients $b_i$ are independent complex variables before the
boundary condition is imposed; they are not complex conjugates of
$a^i$ on a general Euclidean filling. The corresponding
Freed--Moore low-mode map is $I=iK$, with graph
$\psi_-=-I\psi_+$. It is skew and unitary for the boundary
symplectic pairing
\[
 \langle\!\langle x,y\rangle\!\rangle=\int_{\partial Y}\langle Cx,y\rangle\,d\mathrm{vol}_{\partial Y},
\]
where the Hermitian pairing is conjugate-linear in its first slot.
This bilinear form pairs the two chiral spaces nondegenerately.
The low map exists because the total boundary mod-two index of the
real-twisted filling vanishes; separate boundary
components need not have even chiral kernel dimension.

In these coordinates $C(u,v)=(-\bar v,\bar u)$ and
$I\bar I=-1$. Thus $C$ preserves $v=-Iu$, and it also preserves
the high-mode graph. Since it commutes with $d_{\rm FM}$, the
boundary-conditioned bulk eigenspaces, including its kernel, are
quaternionic. Equation~\eqref{eq:real-eta-conventions:wy-kernels}
therefore holds on these identical boundary domains:
\[
 \tau_R(Y;a,I)=\exp\bigl(\pi i\xi_R(Y;a,I)\bigr)
             =\exp\bigl(-i\pi\eta_{\rm WY}(Y;a,I)/2\bigr).
\]
Here $(a,I)$ specifies the spectral cut and low-mode map defined
below, and $\xi_R(Y;a,I)$ includes the kernel term of the
boundary-conditioned self-adjoint operator. These are actual real
half-eta coordinates. Freed--Moore's
construction assembles them into a unit state in
$\mathcal P_R(\partial Y)^*$
\cite[(3.11)--(3.12)]{FM}.

\subsubsection{Pfaffian frames and spectral cuts}

Choose $a>0$ outside the spectrum of $A^2$, and let $E_a$ be its
finite-dimensional $+i$ chiral space spanned by eigenvectors of
$A^2$ with eigenvalue below $a$.
The boundary pairing identifies the opposite chiral space with
$E_a^*$, so the Pfaffian line in this chart is
$\mathcal P_{R,a}=\det E_a^*$. For $b>a$ also outside the spectrum,
write $F_{a,b}=E_b\ominus E_a$ and regard
$B=A^+|_{F_{a,b}}$ as an invertible skew form under that pairing.
Skew invertibility gives $\dim_\C F_{a,b}=2m$ for an integer $m$;
its polar map and Pfaffian obey
\begin{equation}
 Q_{a,b}=B(B^\dagger B)^{-1/2},\qquad
 q_{a,b}=\operatorname{Pf}(B)
        =r_{a,b}\operatorname{Pf}(Q_{a,b}),\qquad
 r_{a,b}=\det(B^\dagger B)^{1/4}>0.
 \label{eq:real-eta-conventions:cut-pfaffian}
\end{equation}
To see the last equality, put $B$ in unitary skew normal form.
Each two-dimensional block has one positive singular value repeated
twice; dividing that block by the singular value leaves its unitary
skew phase. Multiplying the block Pfaffians gives the displayed
positive norm and retains every complex phase. The fourth root is
of a positive determinant, so it selects no phase branch.

The transition from cut $a$ to cut $b$ is
$p\mapsto p\otimes q_{a,b}$ in the prescribed determinant order.
Let $\rho_a>0$ be the high-spectrum Pfaffian norm factor,
\[
 \rho_a=
 \det\nolimits_\zeta
 \left(A^-A^+\big|_{E_a^\perp}\right)^{1/4},
 \qquad \rho_a=r_{a,b}\rho_b.
\]
Here $A^-=(A^+)^\dagger$, and the determinant is zeta regularized
on the positive chiral complement. This is the Pfaffian metric
correction of \cite[(1.5)--(1.10), (1.23)--(1.25)]{FM}.
For a skew-unitary low map $I$, the unit Pfaffian frame is
$u_a(I)=\operatorname{Pf}(I)/\rho_a$. The scalar
$\tau_R(Y;a,I)$ represents the inverse-line state in the dual unit
frame $u_a(I)^*$. Stabilization by the newly included block gives
\[
 u_a(I)\otimes q_{a,b}=u_b(I\oplus Q_{a,b}).
\]
Thus the unit frame and the eta coordinate are unchanged when the
cut is enlarged without changing the allowed domain. For a different
low map $I'$ at the larger cut the inverse-line coordinates satisfy
\[
 \frac{\tau_R(Y;b,I')}{\tau_R(Y;a,I)}
 =\frac{\operatorname{Pf}(I')}
        {\operatorname{Pf}(I)\otimes\operatorname{Pf}(Q_{a,b})}.
\]
The quotient uses the ordered determinant identification above.
For three cuts, the block-product identity
$q_{a,c}=q_{a,b}\otimes q_{b,c}$ supplies the cocycle.

The phase in this transition is visible in Witten--Yonekura's
nonzero-mode calculation \cite[(2.60)--(2.63)]{WY}.
With $\epsilon=\left(\begin{smallmatrix}0&1\\-1&0\end{smallmatrix}\right)$,
a kinetic block is $i\lambda\epsilon$, $\lambda>0$.
Since $D_{W,\rm WY}=-A$, the corresponding block here is
$B=-i\lambda\epsilon$, whose Pfaffian is $-i\lambda$.
Its norm is $\lambda$ and its polar Pfaffian is $-i$.
The phase is therefore retained even though the regulated fermion
magnitude contributes only the positive factor $\lambda$.

A basis change $e'=eU$, $f'=f\bar U$ that keeps the standard $K$
in the new coordinates changes the physical low map to
$I_U=\bar U I U^\dagger$. Hence
\[
 \frac{\operatorname{Pf}(I_U)}{\operatorname{Pf}(I)}
 =\det(U)^{-1}.
\]
This is exactly the half-eta transformation described after
\cite[(2.86)]{WY}; the ordered zero-mode insertion transforms by
$\det U$. Two such congruences giving the same skew form differ
by a compact symplectic transformation of determinant one.
Neither this transition nor the stabilization discards the factor
$\operatorname{Pf}(iK)=i^{\dim E_a/2}\operatorname{Pf}(K)$.

At a kernel crossing use a fixed positive cut containing the crossing
modes. Then $E_a$ has locally constant rank, $\rho_a$ stays smooth
and positive, and $\operatorname{Pf}(I)$ is nonzero. There is no
division by the low-mode Pfaffian of $A^+$, which can vanish.
For the boundary-conditioned bulk operator, a quaternionic eigenvalue
pair crossing zero changes ordinary eta by four and reduced $\xi$
by two; $\exp(\pi i\xi)$ remains continuous, including at zero
because the kernel term is retained. The analytic family construction
in \cite[Proposition~1.28, \S3.2]{FM} supplies its smooth unit state.
Witten--Yonekura's nonzero APS overlap (2.80) and their discussion
after (3.5) give the corresponding zero-mode extension
\cite[\S\S2.5, 3.1]{WY}.

\subsubsection{Ordered duality and Hermitian contraction}

For each ordered boundary component $M_j$, let
$p_j=\dim_\C\ker A_j^+\bmod2$ be its Pfaffian-line parity.
Interchanging two zero-mode blocks of dimensions $b_r,b_s$ gives
$(-1)^{b_rb_s}$. Reversing the order of the dual factors therefore
gives $(-1)^{\sum_{r<s}p_rp_s}$. Two odd components already
produce a minus sign even when total parity is even.
A change of one component's Spin identification by $-1$ acts by
$U=-1$ on that kernel and contributes $(-1)^{p_j}$.
The gauge identification at the seam is likewise specified, rather
than inferred from characteristic classes
\cite[p.~30 and footnotes~16, 23]{WY}.

These are the actual real component degrees and seam maps. The
complex determinant-line gluing of Dai--Freed is a graded
supertrace \cite[Proposition~2.15, Theorem~2.20]{DF}.
Its erratum corrects the odd collar operator to
$B_{\partial X}=J^{-1}D_{\partial X}$, where $J$ is normal Clifford
multiplication, and corrects the boundary and cylinder formulas
\cite[(1.2$'$), (4.8$'$)]{DFerr}. Its orientation pairing has
$(-1)^{\binom{k}{2}}$, with $k$ the number of odd-index components,
from reversal of the dual tensor order \cite[p.~2344]{DFerr}.
That ordinary complex-index grade must not replace $p_j$:
the real structure exchanges the chiral kernels, so their complex
index can vanish while $p_j$ is odd. We use the matched real
Witten--Yonekura and Freed--Moore domains and coordinates above,
with these ordered duality and seam conventions, rather than a
square root of the complex gluing formula.

Choose the same allowed APS domain on the common boundary of two
fillings. Let $|Y\rangle$ and $|Y'\rangle$ be the corresponding
regulated massive-fermion states, proportional to the normalized
ground state $|\Omega\rangle$. Witten--Yonekura express their
physical overlap \cite[(3.3)--(3.4)]{WY} as
\[
 \langle Y'\mid Y\rangle
 =\frac{\langle Y'\mid\mathrm{APS}\rangle
         \langle\mathrm{APS}\mid Y\rangle}
        {|\langle\mathrm{APS}\mid\Omega\rangle|^2}.
\]
The denominator is positive, and their (2.80) keeps it nonzero
with zero modes. The phase is the ratio of the same APS eta
coordinates, hence the closed glued phase by their (3.2).
The unit-frame identification just proved identifies this ratio
with the Hermitian contraction of the real inverse-Pfaffian states.
A change of cut or frame multiplies both coordinates by the same
unit scalar and cancels in that contraction. The orientation-dual
identification uses reversed dual tensor order; retaining the
original order instead requires the explicit Koszul factor above.
All comparisons use the same actual Spin and gauge seam maps.

Finally take the ordered tensor product of the honest real summand
states and their duals for $\Delta_\R$. Write $\tau_{\Delta,\LC}(Y)$
for this unit state. Transport it to the specified stable connection by
\begin{equation}
 \tau_{\Delta,*}(Y)
 =\exp\left[-2\pi i\int_Y
          T_{I_{12}}(A_{\LC},A_*)\right]\tau_{\Delta,\LC}(Y).
 \label{eq:real-eta-conventions:transported-state}
\end{equation}
The boundary lines are identified along the prescribed connection
homotopy, with its endpoint maps. Matched collar paths make this
finite factor additive under gluing. On a closed manifold its value
is $\exp(-2\pi i\mu_*)$, whereas the untransported state has
value $\exp(+\pi i\xi_\Delta)=\exp(-2\pi i\mu_I)$.
Consequently, for the specified filling and spin reference,
\begin{equation}
 \bigl\langle\tau_{\Delta,*}(N),
                \tau_{\Delta,*}(W_{\rm ref})\bigr\rangle
 =\exp\bigl[-2\pi i\mu_*(\overline N\cup_MW_{\rm ref})\bigr].
 \label{eq:real-eta-conventions:state-gluing}
\end{equation}
The first slot is conjugate-linear, so its filling is orientation
reversed. This establishes the real contraction needed in
Section~\ref{sec:response-data}, where ordered graded evaluation
and capped gluing give the required bordism composition, normalized
section and reference formula. Combining that section's local factor
with \eqref{eq:real-eta-conventions:state-gluing} gives
$\mathcal O(M)=\exp[-2\pi i(a_N+\mu_*)]$, with the local integral
retained. Section~\ref{sec:heterotic-comparison} uses this contraction
for its prescribed scalar; no equivalence of complete anomaly
functors is required. The reference is a spin filling and need not
extend the differential $\mathcal H$ structure.

\section{Lie framings and finite transport on \texorpdfstring{$\Sp(2)$}{Sp(2)}}
\label{app:sp2-framings}

The relative pairing in Section~\ref{sec:sp2-calibration} gives
$\Krel=-2$ for the quaternionic sphere bundle over $S^7$. Identifying
that answer with the Lie-framed response requires more than the
underlying diffeomorphism $S(V)\cong\Sp(2)$: the stable tangent
frame, its Spin lift, the framed gauge connections and the integral
differential String trivialization must agree. The tautological frame
on the sphere bundle gives the left Lie framing. A full Spin gauge
connection supplies the right-framed positive-unit datum, and inversion
relates their complete differential backgrounds
\cite[\S6]{C1Working}. These identifications connect the actual-line
evaluation to the known third-root response
\cite[\S3.2, equation~(63)]{TZ}. Write $s_L,s_R$ for the integral
String structures determined by the left and right Lie framings.

Throughout, $B=S(\HH^2)$ has its outward orientation,
$P=\Sp(2)\to B$ sends $p$ to $pe_2$, and the right stabilizer is
$K=\Sp(1)$ embedded by $u\mapsto\operatorname{diag}(u,1)$.
Here $e_1,e_2$ are the standard coordinate vectors of $\HH^2$.
The actual quaternionic line is $V_y=y^{\perp_{\HH}}$;
$N=D(V)$, with projection $\pi:N\to B$, is oriented by base then
real quaternionic fibre, and
$M=\partial N=S(V)$ has the outward boundary orientation.
The real fibre orientation is $(1,i,j,k)$, whereas the complex
structure is right multiplication by $i$. Thus
$e(V_{\R})=-c_2(U)$, with $U$ the corresponding complex rank-two
bundle, and $\lambda(V_{\R})=-c_2(U)$ with $\lambda=p_1/2$.
The stabilizer $K$ and the
relative integer $\Krel$ have different roles.

\subsection{The tautological frame and the integral framing shift}

At $y\in S(\HH^2)$, the decomposition
$T_yB=y^{\perp_{\HH}}\oplus y\operatorname{Im}\HH$ has three
fixed directions $(yi,yj,yk)$. A connection on $V$ splits the
tangent sequence of its disk bundle, giving actual oriented
identifications
\begin{equation}
 TB=V_{\R}\oplus\R^3,\qquad
 TN\cong\pi^*(2V_{\R}\oplus\R^3).
 \label{eq:sp2-framings:tangent-splitting}
\end{equation}
With $Q=\pi^*P$ and its associated real quaternionic line
$Q_{\HH}$, these yield
\begin{equation}
 E_L=2Q_{\HH}\oplus\R^8\cong TN\oplus\R^5,
 \qquad E'_L=\R^{16}.
 \label{eq:sp2-framings:left-bundles}
\end{equation}
The representation $u\mapsto\operatorname{diag}(L_u,L_u,I_8)$,
where $L_u$ is left quaternionic multiplication, lifts to
$\Spin(16)$ uniquely after fixing the identity because $K$ is simply
connected. These are therefore actual Spin gauge bundles. Since $N$
retracts onto $S^7$, both $H^1(N;\Z/2)$ and $H^2(N;\Z/2)$ vanish:
there is no obstruction or choice of Spin structure in these
identifications.

The unit-vector identification
$[p,u]\mapsto p\operatorname{diag}(u,1)$ identifies $M$ with $P$.
On this boundary, $Q$ has the tautological section
$\tau(p)=(p,p)$. To compare its frame with the left Lie frame,
choose the principal connection
$a=\operatorname{pr}_{\mathfrak k}(p^{-1}dp)$, where the projection
is onto the stabilizer Lie algebra, and write
\[
 X=\begin{pmatrix}\alpha&\zeta\\-\bar\zeta&\gamma\end{pmatrix}
 \in\mathfrak{sp}(2),\qquad
 \alpha,\gamma\in\operatorname{Im}\HH,\quad\zeta\in\HH.
\]
The vector $pX$ projects to $p(\zeta,\gamma)$ at $pe_2$.
Its horizontal $V$ component is $pe_1\zeta$, and its vertical
covariant derivative is $pe_1\alpha$. The outward normal supplies
the remaining real direction $pe_1$. Two copies of the
$\tau$-frame, together with the three fixed directions, consequently
give the stable left Lie frame up to a constant oriented basis change.
Passing the outward normal across the seven base directions supplies
the boundary minus sign; the resulting boundary order is
$(\alpha,\zeta,\gamma)$. Fix one Spin lift of this frame on the
simply connected group $M$ and retain it in every boundary gluing.
The diagonal identification in
\eqref{eq:sp2-framings:left-bundles} then gives an integral
nullhomotopy of $\lambda(TN)-\lambda(E_L)$ whose boundary
String structure is precisely $s_L$. At the differential level the
boundary identification also uses the tautological gauge morphism
$\check\gamma_\tau$ of
\eqref{eq:differential-data:induced-String}. Its three-form is zero in
the framed flat model, but subtracting $2\check\gamma_\tau$ from the
restricted disk trivialization retains the two specified gauge
identifications.

The difference from the right frame can be measured integrally.
Let $U_{\Th}\in H^4(N,M;\Z)$ be the Thom class with disk-fibre
integral $+1$. Because $H^3(N;\Z)=H^4(N;\Z)=0$, the connecting
map is an isomorphism. Define $\nu\in H^3(M;\Z)$ by
$\delta\nu=U_{\Th}$; its stabilizer-$S^3$ period is $+1$.
The relative classes in the specified frames are
\begin{equation}
 \begin{aligned}
 c_2(Q,\tau)&=-U_{\Th},&
 \lambda(TN,f_L)&=2U_{\Th},\\
 \lambda(TN,f_R)&=-U_{\Th},&
 s_R-s_L&=3\nu.
 \end{aligned}
 \label{eq:sp2-framings:relative-classes}
\end{equation}
Here $f_L,f_R$ are the left and right stable boundary frames,
including the outward-normal convention.
For the first equality, the tautological unit vector extends on a
disk fibre by $x\mapsto x$, with one positive zero. Its relative
Euler class is $U_{\Th}$, and the opposite complex orientation
gives the minus sign in $c_2$: the complex basis $(1,j)$ for right
multiplication by $i$ induces the real basis $(1,i,j,-k)$.
Each left multiplication block
therefore contributes $+U_{\Th}$ to $\lambda$; the left stable
transition has two such blocks.

On the stabilizer, a right-invariant vector
$Xp$ at $p=\operatorname{diag}(u,1)$ has base projection
$(\zeta,\gamma)$ independent of $u$ and vertical component
$\alpha u$. Including the normal, the transition is
$R_u\oplus I_7$, with $R_u$ denoting right multiplication.
Quaternionic conjugation $C_{\HH}$ satisfies
$R_u=C_{\HH}L_{u^{-1}}C_{\HH}$. Constant conjugation preserves
the trace polynomial, whereas inversion on $S^3$ has degree $-1$.
This gives the relative right-framed class $-U_{\Th}$.
The unique ordinary String structure on $N$ restricts to
$s_{\mathrm{bnd}}=s_L+2\nu=s_R-\nu$, proving the last equality
in \eqref{eq:sp2-framings:relative-classes}. It compares the
structures over the identity map; inversion below is a different
comparison.

\subsection{Smooth left connections and the selected local integral}

The left construction needs a smooth bulk gauge connection that is
flat in the tautological frame on a whole collar. In a local base
gauge, write the fibre coordinate as $x=ru$ and choose
$f(r)=r^2g(r^2)$ near zero, with $g$ smooth and $f=1$ on
the collar. Set
\begin{equation}
 A=(1-f)a-f\,du\,u^{-1}.
 \label{eq:sp2-framings:radial-connection}
\end{equation}
The identity $r^2du\,u^{-1}=\operatorname{Im}(dx\,\bar x)$
proves smoothness at the centre. Under a base change $p\mapsto pk$,
the coordinate becomes $k^{-1}u$, and $A$ obeys the ordinary
connection transformation law. In the $\tau$-frame it is
$(1-f)(u^{-1}au+u^{-1}du)$, hence zero throughout the collar.
The fixed Spin lift makes this the trivial Spin connection too; flatness
alone would not specify that lift or the boundary gauge frame.

Choose a radial metric $dr^2+\ell(r)^2g_{S^3}$, with $\ell=r$ near
zero and $\ell$ constant near the boundary. On the round base take
$A_B^*=a_V\oplus d_{\R^3}$, for which
$\lambda(A_B^*)=-c$, where $c=c_2(P,a)$ denotes the
characteristic form and $a_V$ is induced by $a$. The vertical
connection is fibre Levi--Civita
vertically and uses horizontal $K$-transport otherwise. Its sum
with $\pi^*A_B^*$ is the product tangent connection $A_*$,
preserving the outward normal on the collar. A positive radial
isometry to \eqref{eq:sp2-framings:tangent-splitting} preserves
the frame homotopy and identifies the intermediate connection as
$A_0=\pi^*(A_B^*\oplus a_V)$, with the trivial stabilizing summands
understood. Thus $\lambda(A_0)=-2\pi^*c$ and
$\lambda(E_L,A)=-2C$. This intermediate connection need not preserve
the normal and is not used for gluing.

Transport on the same identified stable Spin bundle first from the
gauge connection to $A_0$ and then to $A_*$. With
$C=c_2(Q,A)$, the curving of the diagonal trivialization is
\begin{equation}
 \begin{aligned}
 H_L&=2T_q+K_3,&
 T_q&=T_{c_2}(\pi^*a,A),&
 K_3&=T_\lambda(A_0,A_*),\\
 dH_L&=2(C-\pi^*c)+\lambda(A_*)+2\pi^*c
      =\lambda(A_*)+2C.
 \end{aligned}
 \label{eq:sp2-framings:left-curving}
\end{equation}
For an invariant polynomial $\mathcal P$, the transgression convention
is $dT_{\mathcal P}(A_0,A_1)=\mathcal P(A_1)-\mathcal P(A_0)$.
This path retains the integral nullhomotopy, not only the displayed
curvature equation. In the cochain notation of
\eqref{eq:response-data:trivialization}, choose its integral
trivializing cochain $b$. Then
$\mathcal I(H_L)-b-h_X$ is a closed real three-cochain.
Since $H^3(N;\R)=0$, it is $\delta k$ for a real two-cochain
$k$. The path and collar identifications retain the specified
boundary morphisms, so all three trivialization equations hold.
Redden's torsor description explains why the integral nullhomotopy
and its geometric refinement must both be retained
\cite[Propositions~3.11--3.12, Theorem~3.15]{Redden}.

At $A_B^*$, the degree-four differential character
$\check\lambda+\check c$ is flat. Its flat class vanishes because
$H^3(S^7;\R/\Z)=0$, and the integral base trivialization is
unique. Choose its geometric representative with base three-form
curving $h=0$, keeping
the characteristic morphisms and cochains. The matched base
connections at both ends of the transgression give exact
fibre-integrated form identities \cite[\S3, \S6]{C1Working}:
\begin{equation}
 \pi_{N*}C^2=-c,\qquad
 \pi_{N*}(T_qC^2)=\pi_{N*}(K_3C^2)=0,
 \qquad a_{N,*}=\int_N H_LY_8=0.
 \label{eq:sp2-framings:selected-local-term}
\end{equation}
Here $\pi_{N*}$ integrates over the oriented disk fibres, and
$Y_8=-C^2/3$ includes the full contribution
$p_2(E_L)=6C^2$. In the universal $K$-equivariant disk, the Thom
relation $U_{\Th}^2=-cU_{\Th}$ and the relative equivariant
Chern class $C_K=-U_{\Th}$ give
$\int_{D^4}C_K^2=-c$. The fibre integral lies in the Cartan
complex of a point: the $K$-invariant polynomials in a formal
Lie-algebra variable of degree two, with zero differential.
The polynomial identity therefore becomes an equality of forms after
association with $a$. The integrated products of the degree-three equivariant
transgressions with $C_K^2$ have degree seven. The odd-degree part
of that complex vanishes, proving the two zero pushforwards;
the framed curvature is zero on the collar, so no boundary term
is omitted. The vanishing concerns this chosen representative;
the general local term remains $\int_Bhc/3$.
For a finite change from the Levi--Civita base connection
$A_B^{\LC}$ to $A_B^*$, set
$\alpha=T_\lambda(A_B^{\LC},A_B^*)$ and transport the same
trivialization by $h_*=h_{\LC}+\alpha$, with its morphism.
For the associated mixed double of Section~\ref{sec:aps-evaluation},
$Z=\overline N\cup_M W_{\rm ref}$ reverses the disk and has trivial
gauge curvature on the reference half. Writing
$\pi_Z:Z\to B$ for its projection, one therefore has
$\pi_{Z*}C^2=c$. The index transgression is
$-\pi_Z^*\alpha C^2/3$, so
\begin{equation}
 a_{N,*}-a_{N,\LC}=\frac13\int_B\alpha c,\qquad
 \mu_* -\mu_{\LC}=-\frac13\int_B\alpha c\pmod\Z.
 \label{eq:sp2-framings:base-transport}
\end{equation}
Here both closed real eta values include the finite tangent correction
of \eqref{eq:real-eta-conventions:transport}; they are not spectra of
an additional torsionful operator. The changes cancel in
$\phi=-a_N-\mu$. Thus the zero local term is a useful
representative of the left datum evaluated in
\eqref{eq:sp2-calibration:phase}, rather than an assumption that
local terms always vanish.

\subsection{A right tangent connection with positive unit period}

For the right filling, allow a full $\Spin(16)$ gauge connection
on $E_R=TN\oplus\R^5$, with $E'_R=\R^{16}$. Requiring the
two-quaternionic-block reduction used on the left would impose an
extra condition on this construction. The tangent connection can
instead be controlled separately, so that its Pontryagin form vanishes
pointwise on each disk fibre.

Choose metric connections $D^B,D^V$ on $TB,V_{\R}$ and the
horizontal distribution of $D^V$. Away from the zero section put
$n=x/|x|$ and
\[
 \mathcal K_n=n(D^Vn)^*-(D^Vn)n^*,\qquad
 \widehat D^V=D^V+\kappa\mathcal K_n,
\]
where the star denotes the metric adjoint and the radial cutoff
$\kappa$ is zero near the centre and one on the boundary
collar. The cutoff ensures smoothness, while
$\mathcal K_nn=-D^Vn$ makes the normal parallel on the collar.
For the radial metric with $\ell=r$ near zero and $\ell=1$ on the
collar, the positive oriented isometry $J$ to
$\pi^*(TB\oplus V_{\R})$ fixes horizontal vectors, sends
$\partial_r$ to $n$ and an angular vector $rv$ to $\ell(r)v$.
It is smooth at the centre and independent of the radial coordinate
on the collar. Consequently
\[
 A_T=J^{-1}(\pi^*D^B\oplus\widehat D^V)J
\]
is a smooth metric Spin connection with product, normal-preserving
boundary behaviour.

On a fixed fibre $D^V=d$, and the curvature entries in a fixed
orthonormal frame are
\[
 F_{ij}=d\kappa\wedge(n_i\,dn_j-dn_i\,n_j)
       +(2\kappa-\kappa^2)dn_i\wedge dn_j.
\]
Rotating $n$ to $e_0$ at the point of calculation makes $dn_0=0$ and
gives the nonzero entries as
\[
 F_{0a}=\kappa' dr\wedge dn_a,\qquad
 F_{ab}=(2\kappa-\kappa^2)dn_a\wedge dn_b,
 \qquad a,b\in\{1,2,3\}.
\]
Each entry is decomposable, so
$\tr_{\R}(F^2)=-2\sum_{i<j}F_{ij}\wedge F_{ij}=0$.
The base connection has no fibre curvature, and conjugation by $J$
preserves the trace. Thus $p_1(A_T)$ vanishes on every fibre even
though the connection need not be fibrewise flat.

Positively orthonormalize the right Lie frame without changing its
oriented Spin/String homotopy class. Extend its flat boundary gauge
connection to $A_G$ on the actual bundle $E_R$, constant on a
collar; the affine space of connections permits such an extension.
The diagonal Spin identification fixes the integral refinement of
\begin{equation}
 H_R=T_\lambda(A_G,A_T\oplus d_{\R^5}),\qquad
 dH_R=\lambda(A_T)-\lambda(E_R,A_G).
 \label{eq:sp2-framings:right-curving}
\end{equation}
The integral nullhomotopy and the real cochain completing this
transgression are obtained as in
\eqref{eq:sp2-framings:left-curving}, with their endpoint morphisms.
Its boundary is $s_R$. The right relative class in
\eqref{eq:sp2-framings:relative-classes} has disk integral $-1$.
Since the gauge connection is flat in that prescribed boundary frame,
Stokes gives the real equality
\begin{equation}
 \int_{S^3}H_R
 =\int_{D^4}\bigl(\lambda(A_T)-\lambda(A_G)\bigr)
 =0-(-1)=1.
 \label{eq:sp2-framings:right-period}
\end{equation}
The relative class is independent of a connection change with its
fixed boundary trivialization; a boundary Chern--Simons term accounts
for a change of representative. This argument uses the full Spin
polynomial of \eqref{eq:response-data:polynomial}. The positive
unit period is special to this right construction; the outward fibre
period of the general induced sphere-bundle datum is $-2$.

\subsection{Inversion and the differential String cylinder}

Inversion $\iota(p)=p^{-1}$ exchanges the Lie frames through
\[
 d\iota_p(pX)=-Xp^{-1}.
\]
The constant matrix $-I_{10}$ has positive determinant and is
connected to $I_{10}$ in $\SO(10)$. Keeping the outward normal
fixed and fixing a Spin lift at one point therefore gives an oriented
String identification
\begin{equation}
 \iota^*s_R=s_L.
 \label{eq:sp2-framings:inversion}
\end{equation}
This pullback relation coexists with the nonzero difference
$s_R-s_L=3\nu$ over the identity.

Pull back the complete right metric, connections and refinement.
Interpolate tangent metrics through positive bundle identifications,
transporting the refinement by the corresponding Chern--Simons
transgression. At a common tangent connection the integral lifts
agree. Their remaining difference is a degree-three differential
character with zero characteristic class, hence is represented by a
global two-form $\beta$ through
\eqref{eq:differential-data:zero-characteristic}. This fixes the
remaining geometric difference without changing the integral lift.
On a product-ended cylinder with interval
coordinate $s$, choose $\chi$ constant at zero and one near its
respective ends. If $H_0$ is the initial endpoint curving pulled
back to the product, the required three-form is
\begin{equation}
 H_{\mathrm{cyl}}=H_0+d(\chi\beta)
 =H_0+\chi\,d\beta+\chi' ds\wedge\beta.
 \label{eq:sp2-framings:string-cylinder}
\end{equation}
The integral lift and endpoint cochain isomorphisms are retained.
The interval term makes the added form exact on the cylinder and
preserves the differential cocycle equation; interpolating only the
endpoint three-forms would omit that term.

Both gauge connections on these cylinders are trivial in their
specified frames. Their real virtual coefficient cancels as
$256-128-128=0$, and $Y_8=0$. The reference-and-boundary
transport property of Section~\ref{sec:response-data}, with the real
line conventions of Appendix~\ref{app:real-eta-conventions}, therefore
identifies the two values \cite[\S2]{C1Working}. This complete
transport, including the Spin lift and cochain morphisms, returns
\eqref{eq:sp2-calibration:phase} to the right-framed positive-unit
datum. Equality of bare periods would not supply it, and periods need
not remain fixed during a change of tangent connection.

Finally, reversing orientation together with all associated boundary,
differential and reference data reverses the local integral and the
closed real eta contribution. Thus $\phi(-M)=-\phi(M)$ and the
response is complex conjugated, giving $e^{-2\pi i/3}$ for the
reversed calibrated datum. Resetting its period independently to
$+1$ would change the input. With the data retained as above,
Section~\ref{sec:sp2-calibration} obtains the same
$e^{2\pi i/3}$ response on the two specified Lie-framed data.

\section{Associated families and the representation-valued index}
\label{app:family-reduction}

The eleven-dimensional closed eta term can also be evaluated by reducing
along the four-sphere fibres of the mixed double. This gives an analytic
route to the intrinsic seven-dimensional expression in
Section~\ref{sec:relative-invariant}, separate from the twelve-dimensional
Atiyah--Patodi--Singer (APS) evaluation in
Section~\ref{sec:aps-evaluation}. The reduction retains
the representation carried by the vertical zero modes and the connection
induced on their bundle. It also retains a finite tangent transgression
when the prescribed connection differs from the natural split connection.
These two pieces are needed to recover the same differential-index value.

The associated-family reduction of Li and Sun
\cite[Appendices~A--B]{C1Working} computes this term from an equivariant
four-sphere index and its induced connection. The vertical zero modes
specify the coefficient of an auxiliary base Dirac operator; they are
not additional dynamical degrees of freedom. The vanishing eta form
and finite connection transgression then identify the invariant computed
by that operator. The conditional family in the final subsection instead
concerns the three-sphere boundary of the disk fibres, whose
spectral-section problem is separate from this closed calculation.

\subsection{The real coefficient becomes an integer eta combination}

Use the closed mixed double
$Z=\overline N\cup_M W_{\mathrm{ref}}=P\times_K F$, with
$K=\Sp(1)$ and $F=S(\HH\oplus\R)\cong\HH P^1$, constructed
in Section~\ref{sec:aps-evaluation}. Its Spin structure and its coefficient
line are the actual glued ones. The line on the lower disk has the
left-multiplication action, whereas its frame on the upper disk has trivial
action; the tautological boundary morphism glues them. In projective
coordinates these are the restrictions of the tautological line on
$\HH P^1$. The orientation is fixed by
$\int_F c_2(U_F)=1$, where $U_F$ denotes its complex rank-two bundle.
The invariant radial connection is fixed throughout the reduction. Thus
the comparison uses the bundle, action, Spin lift and connection, rather
than only a fibre characteristic number.

Restrict the actual real coefficient of Section~\ref{sec:response-data}
to this $K$ action. Let $U_Z$ be the complex rank-two bundle underlying
the glued quaternionic line on $Z$. In the representation identities,
$U$ denotes the defining complex rank-two representation of $K$ and
$\Ad_{\R}$ its real adjoint representation. The restricted representations are
\begin{align}
 \Delta_{\R}|_K&=16\HH-8\Ad_{\R}-40\R,
 &\Delta_{\C}|_K&=32U_Z-8\Ad(U_Z)-40.
 \label{eq:family-reduction:restricted-coefficient}
\end{align}
One can see the multiplicities directly. The first vector representation
complexifies to $4U+8$, and its tensor product with the trivial second
vector is $64U+128$. For torus parameters $(x,x,x,x,0,0,0,0)$ the first
half-spin character is
\begin{equation}
 128\cosh(x/2)^4
 =8(e^{2x}+1+e^{-2x})+32(e^x+e^{-x})+40.
 \label{eq:family-reduction:half-spin-character}
\end{equation}
It is therefore $8\Ad(U)+32U+40$; the other half-spin character is
$128$ trivial copies. Subtraction gives the complex identity in
\eqref{eq:family-reduction:restricted-coefficient}. Since
$\HH_{\C}=U\oplus U$ and the adjoint and trivial representations
are real, this also gives the displayed actual virtual real identity.
Unitary equivariant intertwiners induce parallel bundle identifications
for the associated connections, including their specified boundary gluing.

The real normalization $\mu_I=-\xi(D_{\Delta_{\C}})/2$ can consequently
be rewritten as
\begin{equation}
 \mu_{\LC}(Z)
 =-16\xi(D_{Z,U_Z})+4\xi(D_{Z,\Ad(U_Z)})+20\xi(D_{Z,1})
 \pmod\Z.
 \label{eq:family-reduction:integer-eta}
\end{equation}
Here every $\xi$ includes the kernel term,
$\xi(D)=(\eta_{\mathrm{APS}}(D)+\dim_{\C}\ker D)/2$.
The real-valued representatives were combined before reduction modulo
$\Z$. All subsequent quotient-valued family identities are used with
the integer coefficients $-16,4,20$; none is divided by two. The stable
tangent connection will be introduced through its transgression after
the ordinary Levi--Civita operators have been evaluated.

\subsection{The equivariant four-sphere index}

The vertical index specifies which bundles twist the base Dirac operator.
Its rank would specify only the net number of modes, losing their
$K$ action. To determine this action, first compute for the homogeneous
$G=\Sp(2)$ action on $\HH P^1$, and then restrict to the subgroup
$K=\operatorname{diag}(\Sp(1),1)$. Write $\mathbf4$ for the
defining complex representation of $\Sp(2)$, $\mathbf1$ for its trivial
representation, and use undecorated integers for trivial multiplicities.
The three complex indices are
\begin{equation}
 \Ind_G D_{U_F}=-\mathbf1,\qquad
 \Ind_G D_{\Ad(U_F)}=-\mathbf4,\qquad
 \Ind_G D_1=0.
 \label{eq:family-reduction:fibre-indices}
\end{equation}

The chirality and orientation fix the signs in this formula. Let
$Q_F=U_F^\perp$ in the equivariantly trivial quaternionic two-plane
bundle. At $[1:0]$, the stabilizer
$\Sp(1)_a\times\Sp(1)_b$ acts on the tangent quaternion by
$q\mapsto bqa^{-1}$. This is the $\Spin(4)$ lift of the tangent
action and supplies an honest equivariant Spin bundle. The spin structure
of $S^4$ is unique. Its two half-spin bundles are $U_F$ and $Q_F$;
the identity
\begin{equation}
 c_2(S^-)-c_2(S^+)=e(TF),\qquad
 c_2(Q_F)=-c_2(U_F),\qquad \int_F e(TF)=2
 \label{eq:family-reduction:chirality}
\end{equation}
then identifies $S^-=U_F$ and $S^+=Q_F$ in the chosen orientation.

For a regular torus element $t=\operatorname{diag}(a,b)$, set
$A=a+a^{-1}$ and $B=b+b^{-1}$. If $a\ne b,b^{-1}$, its only
fixed quaternionic lines are $p=[1:0]$ and $q=[0:1]$. A line with
both coordinates nonzero would require $a$ and $b$ to be conjugate unit
quaternions, excluded by this condition. The tangent weights give
\begin{equation}
 \det_{\R}(1-t|_{TF})
 =(1-ab)(1-(ab)^{-1})(1-a/b)(1-b/a)=(A-B)^2.
 \label{eq:family-reduction:fixed-point-denominator}
\end{equation}
The coefficient and spin supercharacters at these points are
\[
\begin{array}{c|ccccc}
 &U_F&Q_F&\Ad(U_F)&1&S^+-S^-\\ \hline
 p&A&B&A^2-1&1&B-A\\
 q&B&A&B^2-1&1&A-B
\end{array}
\]
The elliptic Dirac complex is geometric under the compact-group action,
and the fixed points are transverse by
\eqref{eq:family-reduction:fixed-point-denominator}. Thus the
Atiyah--Bott fixed-point formula applies
\cite[\S2, Theorem~A, (2.8)--(2.9)]{AB}. To match its pullback
convention, use the base map $t^{-1}$ and the fibre lift induced by $t$;
the real determinant in the denominator is unchanged. For coefficient
characters $f(a),f(b)$, the sum of its two contributions is
\begin{equation}
 \frac{(B-A)f(a)+(A-B)f(b)}{(A-B)^2}
 =\frac{f(b)-f(a)}{A-B}.
 \label{eq:family-reduction:fixed-point-character}
\end{equation}
The three choices $f(a)=A,A^2-1,1$ give $-1,-(A+B),0$.
These are the characters in \eqref{eq:family-reduction:fibre-indices}.
Continuity extends equality from the regular torus to the whole torus,
and conjugation extends it to $G$; virtual characters determine the
representation ring.

The ordinary twisted index provides a separate sign and rank check.
Since $p_1(TS^4)=0$, the degree-four terms
$\ch(U_F)=2-c_2(U_F)+\cdots$ and
$\ch(\Ad(U_F))=3-4c_2(U_F)+\cdots$ give ranks $-1,-4,0$
\cite[\S5, Theorem~(5.3), Remark~2]{AS}. They do not determine
the representation. In particular, the central element $-I$ acts by
$-1$ on the spinors and by $+1$ on the adjoint twist. The virtual
adjoint-twist index has character $+4$ there, as does $-\mathbf4$;
$-4\mathbf1$ would have character $-4$. Replacing one by the other
would therefore change the bundle entering the base eta invariant.

Combining \eqref{eq:family-reduction:fibre-indices} with the integer
coefficients in \eqref{eq:family-reduction:integer-eta} gives
\begin{equation}
 R_{\eff}=16-4\mathbf4,\qquad
 R_{\eff}|_K=8-4U,
 \qquad \mathbf4|_K=U+2.
 \label{eq:family-reduction:effective-representation}
\end{equation}
The homogeneous metric and connection can be deformed to the invariant
radial data through $K$-invariant positive metrics and unitary connections.
The equivariant elliptic index is unchanged. This identifies its virtual
representation, but the connection on the actual graded kernels still
has to be identified before using
\eqref{eq:family-reduction:effective-representation} analytically.

The rank-zero result is particularly instructive here. Both
$16-4\mathbf4$ and its restriction $8-4U$ have virtual rank zero,
but the latter has degree-four Chern character $4c$, where $c=c_2(U)$. It retains the
quaternionic background in the reduced calculation. A rank-only
replacement would produce the zero coefficient and erase that
dependence. Conversely, the representation-valued index is a difference
of graded kernels, not a statement that the individual kernels vanish
or that nonzero vertical eigenvalues may be omitted from an arbitrary
real eta formula. The associated-family argument determines precisely
the quotient-valued invariant to which this virtual coefficient applies.

\subsection{Parallel vertical kernels and the vanishing eta form}

The relevant connection statement holds for an associated family more
generally. Let a compact connected Lie group $K$ act isometrically on a
closed spin manifold $F$ of positive even dimension, with an honest Spin lift.
Fix an equivariant Hermitian twisting bundle $\mathcal E_F$ with invariant
unitary connection. For a principal right bundle $(P,a)$ on a closed
odd-dimensional spin manifold $B$, let $\pi:Z\to B$ be the associated
bundle projection and use
\begin{equation}
 Z=P\times_K F,\qquad \mathcal E_Z=P\times_K\mathcal E_F,
 \qquad (pk,f)\sim(p,kf),
 \quad g_\varepsilon=g_F+\varepsilon^{-2}\pi^*g_B.
 \label{eq:family-reduction:associated-data}
\end{equation}
The horizontal distribution is induced by $a$. Horizontal transport on
the twist is the prescribed $K$ action, and the tangent Spin structure
is the product one for $TZ=T^VZ\oplus\pi^*TB$, where $T^VZ$
is the tangent bundle along the fibres. At unit horizontal scale the
Clifford convention is
$c_Z(v)=c_F(v)\otimes1$ and
$c_Z(X^H)=\Gamma_F\otimes c_B(X)$, where $\Gamma_F$ is vertical
chirality, with eigenspaces $S_F^\pm$, and $X^H$ is the horizontal
lift of a base tangent vector $X$.

Let $R(K)$ denote the complex representation ring of $K$. For the fixed
fibre operator define
\begin{equation}
 \mathsf V^\pm=\ker(D_F|_{S_F^\pm\otimes\mathcal E_F}),
 \qquad \mathcal V^\pm=P\times_K\mathsf V^\pm,
 \qquad R=\mathsf V^+-\mathsf V^-\in R(K).
 \label{eq:family-reduction:kernel-representations}
\end{equation}
Horizontal transport is by spin isometries preserving the twisting
connection and vertical volume. The fibres are totally geodesic, with
zero mean curvature. The vertical connection is fibre Levi--Civita on
vertical arguments and satisfies
$\nabla^V_{X^H}U=[X^H,U]$ on horizontal arguments, for a vertical
vector field $U$. Write $\rho$ for the $K$ action on fibre spinors.
The induced connection on their Hilbert bundle is consequently
\begin{equation}
 \nabla^{\mathscr H}=d+\rho_*(a),\qquad
 [\nabla^{\mathscr H},D_F]=0.
 \label{eq:family-reduction:parallel-kernel}
\end{equation}
The second identity follows from equivariance of the fixed operator.
Thus the kernel projections are parallel, and their projected connections
are exactly the associated connections on $\mathcal V^\pm$.
These are the projected kernel connections entering the analytic index
\cite[\S3.1, (3.7)--(3.10), p.~919]{FL}.
The vertical kernels form smooth bundles of constant rank. This is a
geometric consequence of the specified associated transport, rather
than a separate kernel-stability assumption for the total operator.

A section of $\mathcal V^\pm$ is transported by the representation
of $K$ on the fixed kernel of $D_F$. Its covariant derivative is therefore
induced by the original connection $a$, and the parallel kernel projection
leaves this derivative unchanged. An equivariant intertwiner consequently
identifies the twisted base operators with their connections. Equality of
characteristic classes alone would not give this spectral identification.

The same geometry makes the Bismut--Cheeger eta form identically zero.
Set $\mathcal T(X,Y)=-\operatorname{pr}_V[X^H,Y^H]$ and
$\mathcal C=c_F(\mathcal T)$. For the right-bundle convention above,
$\mathcal T=(\Omega_a)_F$, where $\Omega_a$ is the curvature of $a$
and the subscript $F$ denotes its induced vertical vector field. Equivariance of
$b\mapsto c_F(b_F)$ sends the Bianchi identity to
$d^{\nabla^{\mathscr H}}\mathcal C=0$. This is an exterior covariant
derivative identity, not directional covariant constancy of the curvature.
With the geometric superconnection convention of
\cite[\S1.1, pp.~271--273]{Dai},
\begin{align}
 \mathbb A_t&=\sqrt tD_F+\nabla^{\mathscr H}
                         -\frac{\mathcal C}{4\sqrt t},\\
 \mathbb A_t^2&=tD_F^2+(\nabla^{\mathscr H})^2
       -\frac14(D_F\mathcal C+\mathcal C D_F)
       +\frac{\mathcal C^2}{16t},
 \qquad t>0.
 \label{eq:family-reduction:superconnection}
\end{align}
Every coefficient of the square is even in vertical chirality and has
even form degree. In contrast,
$\partial_t\mathbb A_t=D_F/(2\sqrt t)+\mathcal C/(8t^{3/2})$
has chirality-odd coefficients. The heat operator is smoothing for
$t>0$, so the actual fibrewise supertrace vanishes:
\begin{equation}
 \Str\bigl((\partial_t\mathbb A_t)e^{-\mathbb A_t^2}\bigr)=0,
 \qquad \widetilde\eta=0.
 \label{eq:family-reduction:zero-eta-form}
\end{equation}
The vanishing holds in every degree of the defining eta-form integral.
It is stronger than a cohomological vanishing and uses the natural
horizontal connection. Isotriviality of the fibres by itself would not
give this conclusion for arbitrary horizontal data.

In particular, the proof did not assume that $a$ is flat. Its curvature
appears in $\mathcal C$ and in $(\nabla^{\mathscr H})^2$, but the
chirality of the heat-supertrace integrand still forces its vanishing.
Nor did the proof use a vanishing odd cohomology group on the base.
The eta form is zero as the actual differential form determined by this
geometry, so its pairing in the adiabatic formula vanishes before any
cohomological simplification. This distinction matters when the answer
is subsequently compared at specified finite connections.

\subsection{The reduced limit and the finite tangent correction}

The smooth-kernel reduced adiabatic formula of Freed--Lott
\cite[Theorem~6.2, (6.15), p.~938]{FL} now gives
\begin{equation}
 \lim_{\varepsilon\downarrow0}[\xi(D_{Z,g_\varepsilon})]
 =[\xi(B,R)]+\left[\int_B\widehat A(TB)\widetilde\eta\right]
 =[\xi(B,R)]\quad\text{in }\R/\Z,
 \label{eq:family-reduction:reduced-limit}
\end{equation}
where $\xi(B,R)=\xi(D_{B,\mathcal V^+})-
\xi(D_{B,\mathcal V^-})$. The Spin case is the cited Spin$^c$
formula with trivial determinant connection. Its fibre-shrinking metric
is $\varepsilon^2g_\varepsilon=
\varepsilon^2g_F+\pi^*g_B$, after a positive parameter change.
A uniform positive rescaling multiplies all nonzero eigenvalues by the
same positive factor. The eta function acquires the corresponding
factor to the power of its argument, which is one at zero, and the
kernel dimension is unchanged. Thus the two metric conventions have
the same reduced invariant. Compactness, the even-dimensional fibres, their Spin lift,
and the smooth associated kernels supply the required hypotheses.
Invertibility of the effective base operator and eventual constancy of
the total-space kernel are not required for this quotient-valued limit.
In particular, we do not use the stronger real-valued formula of Dai's
Theorem~0.1 with its additional total-kernel hypothesis.

This is a limiting identity, not a finite-$\varepsilon$ error estimate.
The effective operator is
$D_{B,\mathcal V^+}\oplus(-D_{B,\mathcal V^-})$; its reduced
invariant differs from $\xi(B,R)$ by
$\dim\ker D_{B,\mathcal V^-}$, an integer. Small eigenvalues and
zero modes cannot simply be dropped in a real-valued formula. For the
flat product $T^2\times S^1$ with parallel spinors and trivial twist,
the graded vertical index is zero while the total reduced eta is $1$.
This product is consistent with \eqref{eq:family-reduction:reduced-limit},
but rules out an unqualified real equality. The limit supplies no
half-integral refinement of a class in $\R/\Z$.

To return to a fixed metric, one must compare the tangent connections.
Put $\nabla^{\oplus}_{\nat}=\nabla^V\oplus\pi^*\nabla^{B,\LC}$,
and retain the fixed twisting connection. With
$d\CS_P(\nabla_0,\nabla_1)=P(\nabla_1)-P(\nabla_0)$, the
finite formula is
\begin{equation}
 [\xi(D_{Z,g})]=[\xi(B,R)]+
 \left[\int_Z\CS_{\widehat A}
  (\nabla^{\oplus}_{\nat},\nabla^{Z,\LC}_g)
          \ch(\nabla^{\mathcal E_Z})\right].
 \label{eq:family-reduction:finite-metric}
\end{equation}
Only the total-space dimension component is integrated. We recall why
the limiting connection in this formula is the natural split one.
In the fixed unrescaled splitting, Koszul's formula gives a smooth limit
\begin{equation}
 \nabla^{Z,\LC}_{g_\varepsilon}
       \longrightarrow\nabla^{\oplus}_{\nat}+\mathcal N,
 \label{eq:family-reduction:connection-limit}
\end{equation}
where $\mathcal N$ has only the upper block
$\pi^*TB\to T^VZ$; the opposite block is $O(\varepsilon^2)$.
The diagonal limits are the vertical and base connections already
specified. Along $\nabla^{\oplus}_{\nat}+s\mathcal N$, curvature
is upper triangular with fixed diagonal. A trace with one inserted
$\mathcal N$ therefore vanishes. The Pontryagin trace polynomials
defining each relevant degree of $\widehat A$ imply
\begin{equation}
 \CS_{\widehat A}(\nabla^{\oplus}_{\nat},
                      \nabla^{\oplus}_{\nat}+\mathcal N)=0,
 \qquad
 \CS_{\widehat A}(\nabla^{\oplus}_{\nat},
                  \nabla^{Z,\LC}_{g_\varepsilon})\longrightarrow0.
 \label{eq:family-reduction:vanishing-transgression-limit}
\end{equation}
The second convergence is smooth because the finite-degree transgression
is polynomial in connection coefficients and their derivatives. The
limit need not be metric; the same invariant trace polynomials apply
to that linear connection. The block estimate here identifies the
connection limit, without asserting a spectral remainder estimate.

For a homogeneous invariant polynomial $P$ of degree $j$, set
$\nabla_s=\nabla_0+s(\nabla_1-\nabla_0)$ and let $F_s$ be its
curvature. The transgression is
$j\int_0^1P(\nabla_1-\nabla_0,F_s,\ldots,F_s)\,ds$
\cite[Proposition~3.8]{CS}. Its triangle identity follows by integrating
the closed characteristic form of the affine connection on
$Z\times\Delta^2$ and applying Stokes with boundary
$[1,2]-[0,2]+[0,1]$. The oriented two-simplex integral $S_P$ obeys
\begin{equation}
 \CS_P(\nabla_0,\nabla_1)+\CS_P(\nabla_1,\nabla_2)
       -\CS_P(\nabla_0,\nabla_2)=dS_P.
 \label{eq:family-reduction:triangle}
\end{equation}
The two-simplex identity, rather than the infinitesimal proposition
alone, justifies inserting the intermediate connection. Exact terms
integrate to zero on closed $Z$ against the closed twisting character.

Finally, twisted APS on a product-collar cylinder with boundary
$Z_1\sqcup(-Z_0)$ gives
\begin{equation}
 [\xi(D_{Z,g_1})-\xi(D_{Z,g_0})]
 =\left[\int_Z\CS_{\widehat A}
    (\nabla^{Z,\LC}_{g_0},\nabla^{Z,\LC}_{g_1})
                    \ch(\nabla^{\mathcal E_Z})\right]
 \label{eq:family-reduction:cylinder}
\end{equation}
\cite[Theorem~(4.2), (4.3)]{APS}. The endpoint identity
$\xi(-D)=-\xi(D)+\dim\ker D$ retains the integer kernel
contribution even when an endpoint is not invertible. Set
$g_0=g_\varepsilon$, insert $\nabla^{\oplus}_{\nat}$ using
\eqref{eq:family-reduction:triangle}, and take the reduced limit.
Equations~\eqref{eq:family-reduction:reduced-limit} and
\eqref{eq:family-reduction:vanishing-transgression-limit} give
\eqref{eq:family-reduction:finite-metric}, including its sign.

For a specified split stable metric connection $\nabla_*$ on the same stabilized
tangent bundle, define the differential-index value by
\begin{align}
 \xi_*(Z,\mathcal E_Z)
 &:=[\xi(D_{Z,g})]-
 \left[\int_Z\CS_{\widehat A}(\nabla_*,\nabla^{Z,\LC}_g)
                   \ch(\nabla^{\mathcal E_Z})\right]\nonumber\\
 &=[\xi(B,R)]+\left[\int_Z\CS_{\widehat A}
   (\nabla^{\oplus}_{\nat},\nabla_*)
                      \ch(\nabla^{\mathcal E_Z})\right].
 \label{eq:family-reduction:stable-value}
\end{align}
The triangle identity and cylinder variation make it independent of the
auxiliary metric $g$. It is not the raw eta invariant of a torsionful
tangent connection. A change of twisting connection would require its
own Chern--Simons contribution; the twist was fixed here.

For example, if two auxiliary metrics are used in the first line of
\eqref{eq:family-reduction:stable-value}, their ordinary reduced eta
values differ by the cylinder integral
\eqref{eq:family-reduction:cylinder}. The difference between their
two subtracted transgressions is that same integral by
\eqref{eq:family-reduction:triangle}, modulo an exact form. These
differences cancel on the closed total space. The result still depends
on the specified stable connection and twist; independence of the
auxiliary Levi--Civita representative does not remove those background
data from the differential-index value.

\subsection{The base response and the special stable connection}

For the four-sphere family, the kernel representation identified above
has precisely the associated connection required by
\eqref{eq:family-reduction:stable-value}. Equality of virtual
representations is realized, by complete reducibility, as an equivariant
stable isomorphism after adding the same representation to both sides.
A unitary intertwiner becomes parallel under association to $(P,a)$.
Additivity of reduced eta, including the kernels, therefore permits
the effective coefficient $-4U+8$, rather than only its rank zero.

At the general natural split tangent connection, the transgression
in \eqref{eq:family-reduction:stable-value} vanishes. Applying that
formula to the three twists with the integer multiplicities gives
\begin{equation}
 \mu_*(Z)=\xi(B,-4U+8)\quad\text{in }\R/\Z.
 \label{eq:family-reduction:general-base-value}
\end{equation}
Combining it with $a_N=\int_Bh\wedge c/3$ and the scalar sign
$\phi=-a_N-\mu_*$ yields
\begin{equation}
 \phi=\xi(B,4U-8)-\frac13\int_Bh\wedge c
      =\frac{\Krel}{3}\quad\text{in }\R/\Z.
 \label{eq:family-reduction:scalar}
\end{equation}
This is the intrinsic expression and relative pairing already established
in Section~\ref{sec:relative-invariant}. The equality needs no rational
$H^3$ or $H^4$ vanishing; those restrictions enter only its
Crowley--Goette comparison. The local term remains generally nonzero.

The integral differential trivialization on the base is still present
in this formula, through its curving $h$ and the relative class defining
$\Krel$. The family theorem evaluates the closed spectral part; it
does not trivialize $\lambda+c$ over an eight-dimensional spin/gauge
filling or replace the specified integral nullhomotopy by a freely
chosen real primitive. Thus the reduction and the relative pairing have
complementary roles: the former produces the base operator and its
connection, while the latter fixes the globally quantized combination
with the disk term.

On the original $S^7$ base, the special splitting
$TB=V_{\R}\oplus\R^3$ supplies
$A_B^*=a_V\oplus d_{\R^3}$. Let
$k_3=T_\lambda(A_B^{\LC},A_B^*)$ and keep the same vertical
connection on $Z$. The index polynomial governing the eleven-dimensional
eta invariant is
$I_{12}=-(\lambda(TZ)+2C)C^2/3$, with $C=c_2(U_Z)$.
Its tangent dependence is linear in $\lambda$, so the natural-to-stable
transgression is exactly
\begin{equation}
 T_I(A_{\nat},A_*)=-\frac13\pi^*k_3\,C^2,
 \qquad \pi_{Z*}C^2=c.
 \label{eq:family-reduction:finite-correction}
\end{equation}
The pushforward is the exact form identity for the glued radial data
of Sections~\ref{sec:aps-evaluation} and~\ref{sec:sp2-calibration}.
It gives
\begin{equation}
 \mu_*(Z)=\xi_{\eff}^{\LC}(B)-\frac13\int_B k_3\wedge c
          =\xi_{\eff}^*(B),
 \qquad \xi_{\eff}^{\LC}(B)=\xi(B,-4U+8).
 \label{eq:family-reduction:special-base-value}
\end{equation}
Here $\xi_{\eff}^*(B)$ denotes the base differential-index value at
$A_B^*$, obtained from its ordinary reduced eta invariant by the
specified tangent transgression.

The equality with the stable base value can be checked from its full
eight-form, rather than inferred from the fibre index. Use roots
$x,-x,y,-y$ for $\mathbf4$, set $c_2=-(x^2+y^2)$ and
$T_4=2(x^4+y^4)$, and expand
$\ch(\mathbf4)=4-c_2+T_4/24+\cdots$. Since $R_{\eff}$ has rank
zero,
\begin{equation}
 [\widehat A\ch(R_{\eff})]_8
       =-\frac16T_4-\frac16p_1c_2
       \xrightarrow{\ K\ }-\frac13(\lambda(TB)+c)c.
 \label{eq:family-reduction:base-polynomial}
\end{equation}
The restriction uses $T_4=2c^2$ and $\lambda=p_1/2$.
Its tangent transgression is again $-k_3c/3$, proving that the
total-space and base corrections match. Transporting the same
differential trivialization gives $h_*=h_{\LC}+k_3$, up to the
retained exact-form and morphism data. Hence the disk term changes by
$+\int_Bk_3c/3$, while the closed eta term changes by its negative.
Their corrections cancel in $-a_N-\mu_*$.

For this special connection $\lambda(A_B^*)+c=0$ as a form.
The corresponding degree-four differential character is flat and lies
in $H^3(S^7;\R/\Z)=0$, so it vanishes. The integral trivialization
class is unique because $H^3(S^7;\Z)=0$, and its geometric
representative can be chosen with $h_*=0$. This recovers the selected
zero local term used in Section~\ref{sec:sp2-calibration}, with the
same finite connection transport. It is a property of that specified
datum, not a vanishing statement for all sphere-bundle backgrounds.

\subsection{A separate conditional boundary-family result}

There is a different analytic question for the $S^3$ boundary fibres
of $N=D(V)\to S^7$. Suppose one chooses a smooth complex vertical
Dirac-type realization of the difference anomaly whose Hermitian
Clifford module, even-dimensional grading, twisting data, metrics and
connections extend over the disk family and are product near its
boundary. This assumption includes the identification of that
realization with the complexified anomaly. Let $D_\partial$ be its
induced self-adjoint family on the closed boundary fibres. Its odd
analytic index takes values in $K^1(S^7)$, the odd complex $K$-theory
group of the base; this notation is distinct from the symmetry group
$K=\Sp(1)$ used above. Under these hypotheses, the odd analytic index
of the boundary family vanishes
\cite[Appendix~E, Proposition~E.1]{C1Working}:
\begin{equation}
 \Ind_{\C}(D_\partial)=0\in K^1(S^7).
 \label{eq:family-reduction:conditional-boundary-index}
\end{equation}
The extending graded Clifford module supplies an even Dirac-symbol family
on the compact disk fibres, and product structure identifies its induced
boundary symbol over the whole base. This is precisely the situation
of Melrose--Piazza's family cobordism theorem
\cite[\S\S1--2, Proposition~3]{MP}. Carvalho gives the corresponding
family-symbol extension formulation \cite{Carvalho}. Cobordism invariance
applies to an extending realization; it does not construct a realization
of the anomaly.

For a smooth self-adjoint elliptic family on closed fibres over a compact
base, vanishing of the odd analytic index is equivalent to existence of a
Melrose--Piazza spectral section \cite[\S2, Proposition~1]{MP}.
Applying this criterion to
\eqref{eq:family-reduction:conditional-boundary-index} therefore gives
a spectral section for the assumed extending realization
\cite[Appendix~E, Corollary~E.2]{C1Working}. Such a section is a smooth family
of self-adjoint pseudodifferential projections, equal to the identity on
eigenspaces sufficiently far above a spectral cutoff and to zero
sufficiently far below its negative. It is defined over the whole base,
including points where eigenvalues cross zero. The theorem establishes
existence; it selects neither a preferred section nor a spectral gap.

A characteristic-class check is available under further specified
twisting hypotheses. For a spin-Dirac realization on
$M=\Sp(2)\to S^7$, distinguish the twisting bundle $\mathcal E$
from the full Clifford module. If $\mathcal E$ comes from an
$\SU(2)$ representation or is the complexification of a real bundle,
its rational Chern character, like $\widehat A(T_\pi)$ for the vertical
tangent bundle $T_\pi$, has components
only in degrees divisible by four. The families index character is
\begin{equation}
 \ch\bigl(\Ind_{\C}(D_\partial)\bigr)
       =\int_{M/S^7}\widehat A(T_\pi)\ch(\mathcal E).
 \label{eq:family-reduction:conditional-character}
\end{equation}
Integration over three-dimensional fibres leaves positive odd degrees
congruent to one modulo four. The only positive odd rational cohomology
of $S^7$ is in degree seven, so this character vanishes. Bott periodicity
gives $K^1(S^7)\cong\Z$; rational injectivity and torsion-freeness
then give integral vanishing. This check does not identify
$\mathcal E$ with the effective coefficient of the closed four-sphere
calculation and does not replace the extending-operator hypotheses.

Neither a vanishing complex index nor an available complex spectral
section supplies a real Pfaffian refinement. A boundary-family evaluation
would still require a normalized taming, meaning a smoothing perturbation
that makes the family invertible, together with its spectral-flow dependence,
and compatible eta-form and real differential data. The closed reduction
\eqref{eq:family-reduction:general-base-value} and the direct APS
calculation require none of those additional choices. Their returned
observable remains \eqref{eq:family-reduction:scalar}, on the specified
Spin, gauge and integral differential background.

\section{Euler calibration and the two gauge embeddings}
\label{app:alternative-calibrations}

Two optional calculations illuminate different features of the response.
An Euler integral fixes the sign and normalization of the effective
seven-dimensional eta value in the $\Sp(2)$ example of
Section~\ref{sec:sp2-calibration}. A comparison of two gauge embeddings
explains the cubic consistency of the response discussed in
Section~\ref{sec:six-dimensional}. The first calculation enlarges the
base gauge bundle to a symplectic bundle on $D^8$; the second keeps the
eleven-dimensional disk and its complete boundary datum fixed. Neither
is a premise of the direct twelve-dimensional computation in
Section~\ref{sec:aps-evaluation}. The Euler calibration and the
comparison of real gauge coefficients follow Li--Sun
\cite[Lemma C.1, Proposition C.2, Remark C.3; Appendix D]{C1Working}.

\subsection{The outward relative Euler unit}

Let $B=S(\HH^2)=\partial D^8$ have the outward orientation and let
$V_y=y^{\perp_{\HH}}$, with underlying complex rank-two bundle
$U$ given by right multiplication by $i$. Although this quaternionic line does not
extend across the ball as a quaternionic line, its stabilization
does: take the defining complex rank-four symplectic bundle
$\mathcal W=D^8\times\HH^2$. On the boundary it splits as
$V\oplus\HH$, with unit section $s(y)=y$. Prescribe the connection
$a$ on $V$ and the flat connection in the complementary quaternionic
frame $(y,yi,yj,yk)$. The affine space of symplectic connections
allows a smooth extension of this boundary connection.

This boundary frame specifies a relative Euler number, rather than
an absolute characteristic class of the trivial bundle. For one
quaternionic block, the complex orientation is opposite to the real
orientation $(1,i,j,k)$; for two blocks the signs multiply to $+1$.
Thus $c_4(\mathcal W)=e(\mathcal W_{\R})$ with the real orientation
used here. The section $s(x)=x$ has a single zero with derivative the
identity. Its relative Euler number is consequently $+1$.

The Atiyah--Patodi--Singer (APS) calculation needs the corresponding
equality for the real
Chern--Weil integral, not only an integral-cohomology statement.
For two metric connections making $s$ parallel on the boundary,
choose a boundary frame beginning with $s$. Their curvature and
connection variation have zero first row and column. The polarized
Euler Pfaffian with one variation insertion therefore vanishes at
the boundary. Chern--Weil transgression and Stokes show that their
Euler integrals agree. A convenient comparison is the round
hemisphere connection, transported to the trivial real bundle by
its positive orthonormal frame. At the totally geodesic equator its
outward normal is $s(y)=y$ and is parallel. Its Euler density and
volume give
\begin{equation}
\label{eq:alternative-calibrations:euler-unit}
 \int_{D^8}c_4(\mathcal W)
 =\frac{105}{16\pi^4}\frac{16\pi^4}{105}=1.
\end{equation}
This comparison proves the integral for the prescribed symplectic
connection. The hemisphere metric is used only to evaluate the Euler
integral; the Dirac operators below use genuine product collars.

\subsection{The effective eta value and finite tangent transport}

The associated-family calculation supplies the effective complex
coefficient $R_{\eff}=-4\cdot\mathbf4+16$, where $\mathbf4$ is the
defining representation of $\Sp(2)$, with its associated connection.
Undecorated integers in complex coefficients denote trivial summands.
On the boundary $\mathbf4|_{\Sp(1)}=U+2$, so this coefficient
restricts to $-4U+8$. Appendix~\ref{app:family-reduction} identifies
this representation and its associated connection with the effective
base coefficient; a rank-zero replacement would
lose the twisted base operator. Write
$\xi(A)=(\eta_{\mathrm{APS}}(A)+\dim_{\C}\ker A)/2$.
For Levi--Civita tangent data and a product-collar spin metric on the
ball, fix an inward coordinate
$u$ and the convention for the boundary Dirac operator $A_E$
twisted by $E$:
\begin{equation}
\label{eq:alternative-calibrations:collar}
 D_E^+=c(du)(\partial_u+A_E),\qquad
 \nu=-\partial_u,\qquad c_B(v)=c_D(\nu)c_D(v).
\end{equation}
The APS condition projects out the nonnegative eigenspaces, including
the zero eigenspace. Applying the twisted APS theorem to the actual
symplectic twist and to the trivial complex line gives
\cite[Theorem (4.2), formula (4.3)]{APS}
\begin{align}
\label{eq:alternative-calibrations:aps-integer}
 \xi_R^{\LC}&=\int_{D^8}\widetilde I_8-N_R,\qquad
 \xi_R=-4\xi(A_{\mathcal W})+16\xi(A_1),\\
 N_R&=-4\Ind_{\mathrm{APS}}D_{\mathcal W}^+
       +16\Ind_{\mathrm{APS}}D_1^+\in\Z.\notag
\end{align}
The parallel trivial quaternionic line gives the operator splitting
$A_{\mathcal W}=A_V\oplus2A_1$. Hence
$\xi_R=-4\xi(A_V)+8\xi(A_1)=\xi_{\eff}$, with all kernels
included. The real normalization has already been converted into
this integer combination of complex reduced eta values. There is no
additional Majorana halving, and no congruence is divided by two.
The integer $N_R$ is retained until the final reduction modulo $\Z$;
it is not a substitute for the real parity argument in the
twelve-dimensional calculation.

Put $Z_4=\lambda+c$ and $W'_4=-c/3$, with
$\lambda=p_1/2$. On $D^8$, $c$ denotes $c_2(\mathcal W)$;
its restriction is the original $c=c_2(U)$ on $B$. For roots
$x,-x,y,-y$, the degree-four and degree-eight components of
$\ch(\mathcal W)$ are $-c$ and $T_4/24$. The rank of
$R_{\eff}$ is zero. Combining its degree-eight character with
the product of $\widehat A_4=-\lambda/12$ and its degree-four
character gives the full effective eight-form:
\begin{equation}
\label{eq:alternative-calibrations:effective-density}
 T_4=2(x^4+y^4),\qquad
 \widetilde I_8=-\frac16T_4-\frac13\lambda c,\qquad
 T_4=2c^2-4c_4(\mathcal W).
\end{equation}
Since $H^4(D^8;\R)=0$, the extended form $Z_4$ has a
primitive. Since $H^3(S^7;\R)=0$, that primitive can be
adjusted by an exact term to agree with the specified curving $h$
on the boundary. Subtracting the boundary term by Stokes cancels the
entire tangent-dependent part of the density:
\begin{align}
\label{eq:alternative-calibrations:relative-density}
 \widetilde I_8-Z_4W'_4
 &=-\frac16T_4+\frac13c^2=\frac23c_4(\mathcal W),\\
\label{eq:alternative-calibrations:effective-value}
 \xi_{\eff}^{\LC}(B)-\int_BhW'_4
 &=\frac23\pmod\Z.
\end{align}
The residual polynomial vanishes on the boundary $\Sp(1)$
reduction, and its real integral is fixed by
\eqref{eq:alternative-calibrations:euler-unit}.

For a specified stable tangent connection $A_*$, keep the gauge
connection fixed and set $k_3=T_\lambda(A_{\LC},A_*)$. The
tangent dependence $-\lambda c/3$ determines the finite correction.
Integrating Chern--Weil variation along the connection path
\cite[Proposition 3.8]{CS}, extended over the same disk with product
collars, changes the density integral by $-\int_Bk_3c/3$.
The corrected differential-index value therefore retains the same
integer $N_R$ in \eqref{eq:alternative-calibrations:aps-integer}.
In the following formula, $b$ denotes a
global real two-form encoding the exact change of curving, distinct
from the integral trivializing cochain in
Appendix~\ref{app:differential-data}:
\begin{equation}
\label{eq:alternative-calibrations:stable-correction}
 \xi_{\eff}^*=\xi_{\eff}^{\LC}-\frac13\int_Bk_3c,
 \qquad h_*=h_{\LC}+k_3+db.
\end{equation}
The change in $\int_BhW'_4$ is the same
$-\int_Bk_3c/3$, since $c$ is closed. Thus the corrected
eta-minus-local combination remains $2/3$. This uses ordinary
Levi--Civita APS operators and a differential-index correction, not
a raw spectrum for the stable tangent connection. A change of affine
comparison path changes the transgression by an exact form through
the connection-simplex identity; its integral against closed $c$
vanishes. For the selected
$\Sp(2)$ representative with $h_*=0$, it gives $\mu_*=2/3$;
the response convention $\phi=-a_N-\mu_*$ then returns
$\phi=1/3$ and $\mathcal O=e^{2\pi i/3}$ as in
Section~\ref{sec:sp2-calibration}.

The outward Euler unit must also be distinguished from the published
clutching generator. Tachikawa--Zhang normalize that generator by
$\int_{S^8}T_4/48=1$ and use the homotopy map with multiplier
$12$ \cite[equations (42), (54)--(55)]{TZ}. Degree-four products
integrate to zero on $S^8$, so this convention gives
$\int_{S^8}c_4=-12$. With the cap extending the ball and its
boundary section by a trivial cap, it is $-12$ times the geometric
relative unit above. The relative homotopy generator whose twelvefold
multiple is this clutching class therefore has Euler number $-1$.
For $R_{\eff}$ the
primitive coefficient is $48(-1/6)=-8$, giving the
value $-8/12=-2/3$ on the source generator. Our outward relative
unit is its negative, so the additive eta-minus-local value changes
sign and becomes $+2/3$. The two numbers refer to opposite
generators; the direct Euler integral fixes the value here without
using the homotopy normalization.

\subsection{Two arrangements of the same boundary gauge datum}

The embedding comparison applies to the general sphere-bundle
background of Section~\ref{sec:disk-construction}: a quaternionic
line $V$ over a closed connected spin seven-manifold $B$, its
principal $\Sp(1)$ bundle $P$, and the specified integral
differential trivialization of $\lambda(TB)+c$. No rational-cohomology
vanishing is required for this comparison. On $N=D(V)$, keep the
tangent connection, quaternionic connection and integral differential
datum $\check H_N$ fixed. The
arrangement used in the main calculation is
$E_{20}=2Q_{\HH}\oplus\R^8$, $E'_{20}=\R^{16}$, where
$Q=\pi^*P$ and $Q_{\HH}$ is its associated real quaternionic
line. Write $C=c_2(Q,A)$ at the retained connection $A$.
Compare it with
\begin{equation}
\label{eq:alternative-calibrations:gauge-arrangements}
 E_{11}=Q_{\HH}\oplus\R^{12},\qquad
 E'_{11}=Q_{\HH}\oplus\R^{12}.
\end{equation}
The labels count quaternionic copies in the two ordered factors;
they are not fluxes or current-algebra levels. Both constructions
give actual $\Spin(16)^2$ bundles with total gauge class $-2C$.

Equality of this class does not by itself identify the boundary
response. Order the total real blocks as
$(Q_1,Q_2,\R^8,\R^{16})$ in the first arrangement and
$(Q_1,\R^{12},Q_2,\R^{12})$ in the second. Moving the second
four-plane across twelve trivial directions has determinant
$(-1)^{4\cdot12}=1$. Fix its constant $\Spin(32)$ lift. This
intertwines the total connections and the combined tautological
frames. It identifies the additive gauge data, while keeping the
two anomaly coefficient systems distinct.

Retain the quaternionic and Whitney characteristic morphisms of
Appendix~\ref{app:differential-data}, in the order $(Q_1,Q_2)$.
For $\check q=\check c_2(Q,A)$, both arrangements then have
total gauge cocycle $(-\check q)+(-\check q)$ and boundary
trivialization $-2\check\gamma_\tau$, including their integral
and real cochains. The same $\check H_N$ therefore induces the
same $\check s_M=\check H_N|_M-2\check\gamma_\tau$.
No independent degree-three class is inserted. The tangent Spin
structure, normal splitting, connection and product collars also
agree. Trivialize each boundary $\Spin(16)$ factor by its own
specified frame, preserving the factor order. Each actual boundary
coefficient is then $256-128-128=0$ as a virtual real
representation. Their normalized reference elements coincide,
without a square-root or Pfaffian-sign choice. Independently chosen
nullhomotopies with equal curvature would not suffice.

The full invariant polynomials distinguish the two interiors.
Since $p(Q_{\HH})=1-2C+C^2$, the first arrangement has
$(n,n')=(-2C,0)$ and $(p_2(E),p_2(E'))=(6C^2,0)$.
The second has $(n,n')=(-C,-C)$ and
$(p_2(E),p_2(E'))=(C^2,C^2)$. Substitution into the full
$Y_8$ of \eqref{eq:response-data:polynomial} gives
\begin{equation}
\label{eq:alternative-calibrations:two-polynomials}
 Y_{20}=-\frac13C^2,\qquad
 Y_{11}=\frac16C^2,\qquad Y_{20}=-2Y_{11}.
\end{equation}
The corresponding spectral relation requires an identity of real
coefficients. In each arrangement the difference coefficient is the
tensor product of the two vector representations minus the two
positive real half-spin representations of
Section~\ref{sec:response-data}. If $u$ is the character of $U$,
the one-copy vector
and half-spin characters are $2u+12$ and $32(u+2)$; the
two-copy characters are $4u+8$ and $8(u+2)^2$. Thus the two
ordered subtractions have characters
\[
 \begin{aligned}
 16(4u+8)-8(u+2)^2-128&=-8u^2+32u-32,\\
 (2u+12)^2-64(u+2)&=4u^2-16u+16.
 \end{aligned}
\]
Using
$\Ad(U)=\operatorname{Sym}^2U$, with character $u^2-1$,
their ordered differences give
\begin{align}
\label{eq:alternative-calibrations:real-coefficients}
 \Delta_{20,\C}&=32U-8\Ad(U)-40,&
 \Delta_{11,\C}&=-16U+4\Ad(U)+20,\\
 \Delta_{20,\R}&=16\HH-8\Ad_{\R}-40\R,&
 \Delta_{11,\R}&=-8\HH+4\Ad_{\R}+20\R.\notag
\end{align}
Here $\HH_{\C}=U\oplus U$; the adjoint and trivial
representations are real. The real identity is also recovered from
$\HH\otimes_{\R}\HH=4\Ad_{\R}+4\R$. Orthogonal
compact-group intertwiners become parallel after association,
including the clutching. Thus
$\Delta_{20,\R}=-2\Delta_{11,\R}$ relates the actual virtual
real closed Dirac systems, with kernels and the same finite
connection corrections. Equality of ranks, or equality of local
polynomials alone, would not establish this spectral statement.

Additivity now yields $a_{20}=-2a_{11}$,
$\mu_{20}=-2\mu_{11}$ and
$\phi_{20}=-2\phi_{11}$ modulo $\Z$. Filling independence
for the synchronized complete boundary datum supplies the separate
equality $\phi_{20}=\phi_{11}$. Combining them gives
\begin{equation}
\label{eq:alternative-calibrations:cubic-consistency}
 3\phi=0\quad\text{in }\R/\Z,\qquad \mathcal O^3=1.
\end{equation}
This argument uses the normalized filling independence stated in
Section~\ref{sec:response-data}, including the real gluing
application discussed in Appendix~\ref{app:real-eta-conventions}.
The characteristic-class comparison does not discharge that
boundary-line identification.

For comparison, the associated-family reduction of the $(1,1)$
system has effective coefficient $2U-4$ and local term
$-\int_Bhc/6$, and gives $\phi_{11}=-\Krel/6$
\cite[Appendix D]{C1Working}. The difference from the main value
$\phi_{20}=\Krel/3$ is $\phi_{20}-\phi_{11}=\Krel/2$, which
is integral because $\Krel$ is even. On exactly the domain
$H^3(B;\Q)=H^4(B;\Q)=0$, where the Crowley--Goette invariant
$t$ is defined as in Section~\ref{sec:relative-invariant}, this is
$-4t=8t$, with $12t=\Krel/2=0$ in $\R/\Z$. These relations
preserve the actual coefficients; they do not divide an eta
congruence by two. The
embedding comparison explains cubic consistency, while the relative
pairing in the main calculation determines which of the three
response values occurs. Its reduced six-dimensional meaning remains
the anomaly response of Section~\ref{sec:six-dimensional}, without
a vacuum-counting interpretation.

\bibliographystyle{JHEP}
\bibliography{references}
\end{document}